\documentclass[preprint,aps,prc,showpacs,nofootinbib,secnumarabic,floatfix]{revtex4-1}

\usepackage{graphicx}% Include figure files
\usepackage{bm}% bold math
\usepackage{color}
\usepackage{gensymb}
\usepackage{amsmath}
\usepackage{slashed}
\usepackage{appendix}

\newcommand{\bvec}[1]{\ensuremath{\boldsymbol{#1}}}

\begin{document}

\title{Theoretical study of the $\Delta^{++}-\Delta^-$ configuration in the deuteron using antiproton beam}

\author{A.B. Larionov$^{1,2,3}$\footnote{Corresponding author.\\ 
        E-mail address: larionov@fias.uni-frankfurt.de},
        A. Gillitzer$^1$, J. Haidenbauer$^{1,4}$, M. Strikman$^5$}

\affiliation{$^1$Institut f\"ur Kernphysik, Forschungszentrum J\"ulich, D-52425 J\"ulich, Germany\\
             $^2$National Research Center "Kurchatov Institute", 
             123182 Moscow, Russia\\
             $^3$Frankfurt Institute for Advanced Studies (FIAS), 
             D-60438 Frankfurt am Main, Germany\\
             $^4$Institute for Advanced Simulation, Forschungszentrum J\"ulich, D-52425 J\"ulich, Germany\\ 
             $^5$Pennsylvania State University, University Park, PA 16802, USA}

\date{\today}

\begin{abstract}
  We study the manifestation of the $\Delta^{++}-\Delta^-$ component of the deuteron wave function in
  the exclusive reaction $\bar p d \to \pi^- \pi^- \Delta^{++}$.
  Due to the large binding energy the internal motion in the $\Delta-\Delta$ system is
  relativistic. We take this into account within the light-cone (LC) wave function formalism and, indeed, found
  large differences between calculations based on the LC and non-relativistic (NR) wave functions. We demonstrate,
  that the consistent LC treatment of the $\Delta-\Delta$ system plays the key role in the separation of the
  signal and background. Within the LC approach, the characteristic shape of the momentum distribution of the $\Delta-\Delta$
  bound system predicted by the meson-exchange model is well visible on the background of usual annihilations
  at beam momenta between 10 and 15 GeV/c.
\end{abstract}

 	%Light mesons (S=C=B=0)
      
\maketitle

\section{Introduction}
\label{intro}

One of the hottest fields of the modern nuclear physics is the study of non-nucleonic degrees-of-freedom in nuclei.
This issue is closely related to the mechanism of hadronic interactions at short distances where the partonic structure
of hadrons becomes important. As the lightest nucleus, the deuteron is an ideal object for testing theoretical models
of non-nucleonic degrees-of-freedom. In this case angular momentum and isospin conservation allow to considerably reduce
the space of possible exotic configurations and simplify the physical picture. In particular, the lightest exotic
baryonic configuration is a mixture of $\Delta^{++}-\Delta^-$ and $\Delta^{+}-\Delta^0$ states with equal probabilities.
There is a substantial difference in various theoretical predictions on the $\Delta-\Delta$ component of the deuteron.  

In the meson-exchange calculations of the deuteron \cite{Haapakoski:1974qnh,Arenhovel:1975dx,Dymarz:1990ac,Haidenbauer:1993pw},
the short-range structure of the $NN \rightarrow \Delta \Delta$ transition potential has been effectively described by inserting
the cut off (hard core radius) in the pion-exchange potential and adding $\rho$ meson exchange \cite{Haapakoski:1974qnh} or using
formfactors \cite{Arenhovel:1975dx,Dymarz:1990ac,Haidenbauer:1993pw} in meson-nucleon-delta vertices. In most calculations,
the $^7D_1$ state dominates in the $\Delta-\Delta$ wave function. The $^3S_1^{NN} \to\: ^7D_1^{\Delta\Delta}$ transition is driven
by the tensor interaction due to $\pi$ and $\rho$ exchanges which contribute with opposite signs. Thus, at short distances,
the inclusion of $\rho$ exchange has been shown to be very important, as it stabilizes the cut-off dependence of the results
\cite{Haapakoski:1974qnh}. The meson-exchange calculations typically predict that the deuteron has a $\Delta-\Delta$ component
with a probability $<1.5\%$.

The typical momenta in the $\Delta-\Delta$ wave function are $\sim 400 \div 500$ MeV/c (see Fig.~\ref{fig:DD_WF_CCF}).
This corresponds to the inter-$\Delta$ distances $\sim \pi/2k \sim 0.6 \div 0.8$ fm which are much smaller than the root-mean-square
radius of the ordinary deuteron $\sim 2$ fm. Thus, the $\Delta-\Delta$ wave function has a strong overlap with other non-nucleonic
configurations such as e.g. six quark states which one can try to model. For example, in the constituent quark model calculations
with oscillator basis \cite{Glozman:1993pm,Glozman:1994xe} the main contribution to the $\Delta-\Delta$ wave function is due to the
$s^6$ quark configuration. Thus, the $\Delta-\Delta$ configuration is described by the $0s$ oscillator state. This model predicts
the probability of the $\Delta-\Delta$ component to be $\sim 2-3\%$.

Experimentally, the $\Delta-\Delta$ component has been already discussed in previous analyses of photon \cite{Benz:1974au} and antiproton
\cite{Braun:1974fy} reactions on the deuteron. In ref. \cite{Benz:1974au}, DESY data on backward $\Delta^{++}$ production in the laboratory
(lab.) frame in the reaction $\gamma d \to \Delta^{++}+{\rm anything}$ were analyzed deducing $\sim 3\%$ $\Delta-\Delta$ admixture
in the deuteron. In ref. \cite{Braun:1974fy}, non-annihilative channels of $\bar pd$ interactions have been used for the search of the
$\Delta-\Delta$ component. The high percentage of $\sim 16\%$ of the $\Delta-\Delta$ component deduced in ref. \cite{Braun:1974fy} strongly
indicates that the background was not fully excluded in the spectator $\Delta$ decays in the backward hemisphere in the lab. frame.
An upper limit of 0.4\% on the $\Delta-\Delta$ component has been obtained in $\nu(\bar\nu) d$ interaction studies \cite{Allasia:1986kg},
where the neutrino (antineutrino) was supposed to interact with the quark content of $\Delta^-(\Delta^{++})$ leaving
the $\Delta^{++}(\Delta^-)$ as a low-momentum spectator. 

In the OBELIX@LEAR experiment \cite{Bertin:1997dw} the reaction
\begin{equation}
  \bar p d \to 2\pi^- \pi^+ p       \label{pbard_to_2pi^-pi^+p}
\end{equation}
with stopped antiprotons was used to estimate an upper limit on the annihilation probability $Y_{\bar p (\Delta^- \Delta^{++}) \to 2\pi^- \pi^+ p}$
due to the subprocess $\bar p \Delta^- \to \pi^- \pi^-$. The resulting $Y_{\bar p (\Delta^- \Delta^{++}) \to 2\pi^- \pi^+ p} \leq 6.5 \times 10^{-5}$
corresponds to a $\Delta-\Delta$ configuration probability $\leq 1\%$.
In ref. \cite{Denisov:1998sq} some enhancement in the invariant mass distribution of $\pi^- \pi^-$ pairs at 1.4-1.5 GeV from reaction
(\ref{pbard_to_2pi^-pi^+p}) visible in the OBELIX data \cite{Bertin:1997dw} was interpreted by including the $\Delta^-\Delta^{++}$ component.
However, due to the lack of statistics it is difficult to make definite conclusions on the existence of a $\Delta-\Delta$ component from
the OBELIX data.

In the present paper -- in view of the upcoming PANDA experiment -- we theoretically address the reaction channel
$\bar p d \to \pi^- \pi^- \Delta^{++}$ at $p_{\rm lab}=10\div15$ GeV/c for the kinematics with two energetic $\pi^-$ mesons in the forward
lab. hemisphere and a slow $\Delta^{++}$. The ``signal'' reaction channel is $\bar p \Delta^- \to \pi^- \pi^-$ annihilation on the virtual
$\Delta^-$ leading to a practically instantaneous (on nuclear scale) release of the spectator $\Delta^{++}$. The possible background channels
include at least two steps and, thus, are expected to be moderate. We will consider the following two possible background reactions:
(i) $\bar p n \to \pi^- \pi^0$ followed by the charge exchange (CEX) reaction $\pi^0 p \to \pi^- \Delta^{++}$ and
(ii) $\bar p n \to \pi^- \pi^- \pi^+$ followed by $\pi^+ p \to \Delta^{++}$.

The other isospin component, i.e. $\Delta^+\Delta^0$, can be studied in the $\bar p d \to \pi^- \pi^+ \Delta^0$ channel. However, background of
the types (i) and (ii) will in this case include the annihilation channels both on the proton and on the neutron, and thus presumably will be
larger. Therefore, for simplicity, we restrict ourselves in this work to the analysis of $\bar p$ annihilation on the $\Delta^{++} \Delta^-$
component.

In calculations of the signal channel we use both NR and LC descriptions of the $\Delta-\Delta$ wave function and analyze their
influence on the results. Both NR and LC descriptions are based on the same input $\Delta-\Delta$ momentum distribution provided
by calculations based on the model of ref. \cite{Haidenbauer:1993pw}, but differ in the physical meaning of the intrinsic
$\Delta-\Delta$ momentum. The elementary two-pion annihilation amplitudes are calculated in the framework of the nucleon-
and $\Delta$-exchange model. The CEX $\pi^0 p \to \pi^- \Delta^{++}$ amplitude is described by the reggeized $\rho$ exchange.
We show that LC effects are strong in case of a strongly bound $\Delta-\Delta$ configuration and crucial for the visibility of the signal
which is comparable in strength with the three-pion annihilation background in the backward lab. hemisphere.

The structure of the paper is as follows. In sect. \ref{signal} we derive the signal cross section in the NR and LC descriptions.
In sect. \ref{WF}, the wave function of the $\Delta-\Delta$ state used in the calculations is described briefly.
Section \ref{bgr} includes the formalism for the calculation of the background channels. Section \ref{results} contains numerical results. 
Finally, in sect. \ref{summary} we summarize the results and try to draw conclusions on the possibility to observe the $\Delta-\Delta$ component
of the deuteron experimentally.

The appendices contain some technical aspects. In Appendix \ref{app_LC_vs_NR} we derive the relation between
the NR and LC $\Delta-\Delta$ wave functions based on the electromagnetic formfactor of the $\Delta-\Delta$ state.
In Appendix \ref{app_piPoles} we obtain Eq.(\ref{poles}) for the poles of the pion propagator used in the calculation
of the three-pion annihilation background in sect. \ref{3pi}. The elementary amplitudes are described in Appendix \ref{elem}.

\section{Antiproton interaction with a deuteron $\Delta$-$\Delta$ configuration}
\label{signal}

We will start from the detailed NR derivation and then sketch the main steps of the LC derivation. In the latter case more details
can be found in refs. \cite{Frankfurt:1977vc,Frankfurt:1981mk}.

\begin{figure}
\begin{center}
   \includegraphics[scale = 0.60]{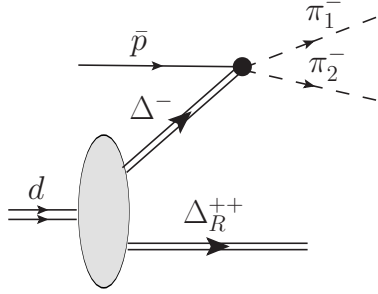}
\end{center}
\caption{\label{fig:IA} Impulse approximation graph showing the production of a pion pair in antiproton annihilation on one
  of the $\Delta$'s of a $\Delta$-$\Delta$ configuration in the deuteron.}
\end{figure}

The $S$-matrix\footnote{We use the conventions of ref. \cite{BLP} throughout the paper.}
corresponding to the Feynman diagram of Fig.~\ref{fig:IA} can be written as follows:
\begin{equation}
  S^{(0)} = \int d^3r_{\Delta_R} d^3r_{\Delta} \sum_{\lambda_\Delta} \phi_{\Delta-\Delta}^{\lambda_d}(\bm{r}_{\Delta_R},\bm{r}_{\Delta};\lambda_{\Delta_R},\lambda_\Delta)
  \frac{1}{V} \int \frac{Vd^3p_\Delta}{(2\pi)^3} e^{-i\bm{p}_{\Delta_R}\bm{r}_{\Delta_R}-i\bm{p}_{\Delta}\bm{r}_{\Delta}} S_{\pi_1\pi_2;\bar p \Delta}~, \label{S^(0)}
\end{equation}
where
\begin{equation}
  S_{\pi_1\pi_2;\bar p \Delta} = \frac{(2\pi)^4\delta^{(4)}(p_1+p_2-p_{\bar p}-p_\Delta)}{(2E_{\bar p}V 2E_{\Delta}V 2E_1V 2E_2V)^{1/2}}
  i M_{\pi_1\pi_2;\bar p \Delta}   \label{S_elem}
\end{equation}
is the $S$-matrix corresponding to the process $\bar p \Delta \to \pi_1\pi_2$. $V$ is a normalization volume.
$\phi_{\Delta-\Delta}^{\lambda_d}(\bm{r}_{\Delta_R},\bm{r}_{\Delta};\lambda_{\Delta_R},\lambda_\Delta)$ is the wave function of the $\Delta$-$\Delta$
configuration normalized according to the following condition:
\begin{equation}
  \sum_{\lambda_{\Delta_R},\lambda_\Delta} \int d^3r_{\Delta_R} d^3r_{\Delta}
  |\phi_{\Delta-\Delta}^{\lambda_d}(\bm{r}_{\Delta_R},\bm{r}_{\Delta};\lambda_{\Delta_R},\lambda_\Delta)|^2 =  P_{\Delta-\Delta}~,   \label{norma}
\end{equation}
where $P_{\Delta-\Delta}$ is the probability of a $\Delta$-$\Delta$ configuration in the deuteron.
$\lambda_d$, $\lambda_{\Delta_R}$ and $\lambda_\Delta$ are the third spin components of the deuteron, residual and struck $\Delta$'s, respectively. 

By using the center-of-mass (c.m.), $\bm{R}$, and relative, $\bm{r}$, coordinates,
\begin{equation}
  \bm{R} = \frac{m_\Delta\bm{r}_\Delta+m_{\Delta_R}\bm{r}_{\Delta_R}}{m_\Delta+m_{\Delta_R}}~,~~~\bm{r}=\bm{r}_{\Delta}-\bm{r}_{\Delta_R}~,  \label{R_and_r}
\end{equation}
one can separate the c.m. motion and relative motion in the wave function as follows: 
\begin{equation}
  \phi_{\Delta-\Delta}^{\lambda_d}(\bm{r}_{\Delta_R},\bm{r}_{\Delta};\lambda_{\Delta_R},\lambda_\Delta)
  = \frac{1}{\sqrt{V}} e^{i\bm{p}_d\bm{R}} \phi_{\Delta-\Delta}(\bm{r};\lambda_{\Delta_R},\lambda_\Delta)~.   \label{cm_split}
\end{equation}
Substituting Eqs.(\ref{S_elem}),(\ref{cm_split}) in Eq.(\ref{S^(0)}) and integrating-out the c.m. motion we have:
\begin{eqnarray}
  S^{(0)} &=& \int d^3r \sum_{\lambda_\Delta} \frac{1}{\sqrt{V}} \phi_{\Delta-\Delta}^{\lambda_d}(\bm{r};\lambda_{\Delta_R},\lambda_\Delta)
  \int d^3p_\Delta \delta^{(3)}(\bm{p}_d-\bm{p}_{\Delta_R}-\bm{p}_{\Delta}) \nonumber \\
  && \times e^{i(\bm{p}_{\Delta_R}m_\Delta-\bm{p}_{\Delta}m_{\Delta_R})\bm{r}/(m_\Delta+m_{\Delta_R})}\,
           \frac{(2\pi)^4\delta^{(4)}(p_1+p_2-p_{\bar p}-p_\Delta)}{(2E_{\bar p}V 2E_{\Delta}V 2E_1V 2E_2V)^{1/2}}
           i M_{\pi_1\pi_2;\bar p \Delta}~.         \label{S^(0)_expanded}
\end{eqnarray}
If one defines the energy of the struck $\Delta$ as
\begin{equation}
  E_\Delta = E_d - E_{\Delta_R}~,        \label{E_Delta}
\end{equation}
then, integrating-out the first $\delta$-function in Eq.(\ref{S^(0)_expanded}),  we can finally express the $S$-matrix
in the standard form,
\begin{equation}
  S^{(0)} = \frac{(2\pi)^4\delta^{(4)}(p_1+p_2+p_{\Delta_R}-p_{\bar p}-p_d)}{(2E_{\bar p}V 2E_dV 2E_{\Delta_R}V 2E_1V 2E_2V)^{1/2}}\,
  i M_{\pi_1\pi_2\Delta_R;\bar p d}^{(0)}~,    \label{S^(0)_final}
\end{equation}
where the invariant matrix element is given by the following expression:
\begin{equation}
  M_{\pi_1\pi_2\Delta_R;\bar p d}^{(0)} = \left(\frac{2E_{\Delta_R}E_d}{E_\Delta}\right)^{1/2} (2\pi)^{3/2}
  \sum_{\lambda_\Delta} \phi_{\Delta-\Delta}^{\lambda_d}(\bm{k};\lambda_{\Delta_R},\lambda_\Delta)
  M_{\pi_1\pi_2;\bar p \Delta}~,    \label{M^(0)}
\end{equation}
where
\begin{equation}
  \bm{k} = -\bm{p}_{\Delta_R}+\frac{m_{\Delta_R}\bm{p}_d}{m_\Delta+m_{\Delta_R}}   \label{k_nonrel}
\end{equation}
is the momentum of the struck $\Delta$ \footnote{Thus in the deuteron rest frame $\bm{k} = -\bm{p}_{\Delta_R}$, which
is the non-relativistic definition. See Eq.(\ref{k_int}) below for the light-cone definition.}.
The wave function in momentum space is defined as follows:
\begin{equation}
  \phi_{\Delta-\Delta}^{\lambda_d}(\bm{k};\lambda_{\Delta_R},\lambda_\Delta) = \frac{1}{(2\pi)^{3/2}}
  \int d^3r e^{-i\bm{k}\bm{r}}  \phi_{\Delta-\Delta}^{\lambda_d}(\bm{r};\lambda_{\Delta_R},\lambda_\Delta)~.   \label{phi_p}
\end{equation}  
The normalization condition for this wave function is
\begin{equation}
  \sum_{\lambda_{\Delta_R},\lambda_\Delta} \int d^3k |\phi_{\Delta-\Delta}^{\lambda_d}(\bm{k};\lambda_{\Delta_R},\lambda_\Delta)|^2 = P_{\Delta-\Delta}~.
\end{equation}
Note that based on Eq.(\ref{M^(0)}) one can obtain the relation between the deuteron vertex function $d \to \Delta \Delta$
and the wave function of the $\Delta-\Delta$ state (cf. ref. \cite{Frankfurt:1996uz}):
\begin{equation}
  \frac{i \Gamma_{d \to \Delta \Delta}(p_d,p_{\Delta_R};\lambda_d,\lambda_{\Delta_R},\lambda_\Delta)}{p_{\Delta}^2 - m_\Delta^2 +i\epsilon}
  = \left(\frac{2E_{\Delta_R}E_d}{E_\Delta}\right)^{1/2} (2\pi)^{3/2} \phi_{\Delta-\Delta}^{\lambda_d}(\bm{k};\lambda_{\Delta_R},\lambda_\Delta)~.
  \label{Gamma_d2DD}
\end{equation}

In the deuteron rest frame (lab. frame) the differential cross section of the process shown in Fig.~\ref{fig:IA} is
\begin{equation}
  d\sigma = \frac{(2\pi)^4 \delta^{(4)}(p_1+p_2+p_{\Delta_R}-p_{\bar p}-p_d)}{4 m_d p_{\rm lab}}
  \overline{|M_{\pi_1\pi_2\Delta_R;\bar p d}^{(0)}|^2} \frac{d^3p_{\Delta_R}}{(2\pi)^32E_{\Delta_R}} \frac{d^3p_1}{(2\pi)^32E_1}
  \frac{d^3p_2}{(2\pi)^32E_2}~,    \label{dsigma}
\end{equation}
where $\overline{|M_{\pi_1\pi_2\Delta_R;\bar p d}^{(0)}|^2}$ is the modulus squared of the invariant matrix element, Eq.(\ref{M^(0)}),
summed over final spins and averaged over initial spins.

For the modulus squared of the invariant matrix element in the lab. frame we have:
\begin{eqnarray}
\overline{|M_{\pi_1\pi_2\Delta_R;\bar p d}^{(0)}|^2} &=& \frac{1}{6} \sum_{\lambda_{\bar p},\lambda_d,\lambda_{\Delta_R}} |M_{\pi_1\pi_2\Delta_R;\bar p d}^{(0)}|^2 
     = \frac{2E_{\Delta_R}m_d}{E_\Delta} \frac{(2\pi)^3}{6}  \nonumber \\ 
&& \times \sum_{\lambda_{\bar p},\lambda_d,\lambda_{\Delta_R}} \sum_{\lambda_\Delta,\lambda_\Delta^\prime}
    \phi_{\Delta-\Delta}^{\lambda_d}(\bm{k};\lambda_{\Delta_R},\lambda_\Delta)
    {\phi_{\Delta-\Delta}^{\lambda_d}}^*(\bm{k};\lambda_{\Delta_R},\lambda_\Delta^\prime)
    M_{\pi_1\pi_2;\bar p \Delta} M^*_{\pi_1\pi_2;\bar p \Delta^\prime}   \nonumber \\
&=& \frac{2E_{\Delta_R}m_d}{E_\Delta} \frac{(2\pi)^3}{6} \sum_{\lambda_{\bar p},\lambda_d,\lambda_{\Delta_R},\lambda_\Delta}
    |\phi_{\Delta-\Delta}^{\lambda_d}(\bm{k};\lambda_{\Delta_R},\lambda_\Delta)|^2\,
    |M_{\pi_1\pi_2;\bar p \Delta}|^2~,    \label{M2}
\end{eqnarray}
where in the last step we neglected the interference terms between transitions with different spin projections of the struck $\Delta$. 
Neglecting the spin dependence of the transition probability $\bar p \Delta \to \pi_1\pi_2$, we can replace  $|M_{\pi_1\pi_2;\bar p \Delta}|^2$
by its averaged value over the third spin components of $\bar p$ and $\Delta$, $\overline{|M_{\pi_1\pi_2;\bar p \Delta}|^2}$.
This allows to simplify Eq.(\ref{M2}) to the following form:
\begin{equation}
  \overline{|M_{\pi_1\pi_2\Delta_R;\bar p d}^{(0)}|^2} = \frac{2E_{\Delta_R}m_d}{E_\Delta} (2\pi)^3 |\phi_{\Delta-\Delta}(\bm{k})|^2\,
  \overline{|M_{\pi_1\pi_2;\bar p \Delta}|^2}~,    \label{M2_simpl}
\end{equation}
where the deuteron-spin-averaged modulus squared of the $\Delta-\Delta$ wave function, 
\begin{equation}
  |\phi_{\Delta-\Delta}(\bm{k})|^2
  \equiv \frac{1}{3} \sum_{\lambda_d,\lambda_{\Delta_R},\lambda_\Delta} |\phi_{\Delta-\Delta}^{\lambda_d}(\bm{k};\lambda_{\Delta_R},\lambda_\Delta)|^2~,
                                             \label{phi_DD^2}
\end{equation}
describes the momentum distribution of $\Delta_R$ in the $\Delta$-$\Delta$ configuration. It is normalized as
\begin{equation}
  \int d^3k  |\phi_{\Delta-\Delta}(\bm{k})|^2 = P_{\Delta-\Delta}~.
\end{equation}

Substituting Eq.(\ref{M2_simpl}) in Eq.(\ref{dsigma}) we have
\begin{eqnarray}
  d\sigma &=& \frac{(2\pi)^4 \delta^{(4)}(p_1+p_2+p_{\Delta_R}-p_{\bar p}-p_d)}{4 p_{\rm lab} E_\Delta}\, |\phi_{\Delta-\Delta}(\bm{k})|^2\,  \nonumber \\
  && \times \overline{|M_{\pi_1\pi_2;\bar p \Delta}|^2}\, \frac{d^3p_1}{(2\pi)^32E_1} \frac{d^3p_2}{(2\pi)^32E_2} d^3p_{\Delta_R}~.     \label{dsigma_simpl}
\end{eqnarray}
This equation can be further simplified by using the elementary differential cross section
\begin{equation}
  d\sigma_{\bar p \Delta \to \pi_1\pi_2} = \frac{(2\pi)^4 \delta^{(4)}(p_1+p_2-p_{\bar p}-p_\Delta)}{4I_{\bar p \Delta}}
  \overline{|M_{\pi_1\pi_2;\bar p \Delta}|^2} \frac{d^3p_1}{(2\pi)^32E_1} \frac{d^3p_2}{(2\pi)^32E_2}~,    \label{dsig_elem}
\end{equation}
where
\begin{equation}
  I_{\bar p \Delta} =\sqrt{(p_{\bar p}p_\Delta)^2-m_{\bar p}^2 \tilde m_\Delta^2}    \label{I_barpDelta}
\end{equation}
is the M\"oller flux factor,
\begin{eqnarray}
  p_{\bar p} &=& (E_{\bar p},0,0,p_{\rm lab})                   \label{p_barp} \\ 
  p_\Delta &=& p_d-p_{\Delta_R} = (E_\Delta,\bm{k})     \label{p_Delta}
\end{eqnarray}
are the four-momenta of the incoming antiproton and of the struck $\Delta$, respectively, and
\begin{equation}
   \tilde m_\Delta^2=p_\Delta^2=(E_\Delta)^2-\bm{k}^2     \label{tildeM_Delta} 
\end{equation}
is the (off-shell) invariant mass of the struck $\Delta$.

Thus, using Eq.(\ref{dsig_elem}) we can rewrite Eq.(\ref{dsigma_simpl}) as
\begin{equation}
  d\sigma = \frac{E_{\bar p}}{p_{\rm lab}} v_{\bar p \Delta}  |\phi_{\Delta-\Delta}(\bm{k})|^2
  d\sigma_{\bar p \Delta \to \pi_1\pi_2} d^3p_{\Delta_R}~,     \label{dsigma_simpl1}
\end{equation}
where
\begin{equation}
  v_{\bar p \Delta}=\frac{I_{\bar p \Delta}}{E_{\bar p}E_\Delta}
\end{equation}
is the relative velocity of the antiproton and the struck $\Delta$. Using the invariant
\begin{equation}
  t=(p_1-p_{\bar p})^2    \label{t}
\end{equation}
 we obtain
\begin{equation}
     \frac{d^4\sigma}{d^3p_{\Delta_R} dt} = \frac{E_{\bar p}}{p_{\rm lab}} v_{\bar p \Delta}  |\phi_{\Delta-\Delta}(\bm{k})|^2
     \frac{d\sigma_{\bar p \Delta \to \pi_1\pi_2}(s^\prime,t)}{dt}~,                     \label{dsigma_over_d3pdt}
\end{equation}
where $s^\prime=(p_{\bar p}+p_d-p_{\Delta_R})^2$.
The kinematic prefactor in Eq.(\ref{dsigma_over_d3pdt}) can be explicitly calculated as follows:
\begin{equation}
  \frac{E_{\bar p}}{p_{\rm lab}} v_{\bar p \Delta} = \frac{I_{\bar p \Delta}}{p_{\rm lab}E_\Delta}
  = \frac{\sqrt{(E_{\bar p}E_\Delta - p_{\rm lab}k^z)^2-m_{\bar p}^2 \tilde m_\Delta^2}}{p_{\rm lab}E_\Delta}~,  \label{kinem_prefac}
\end{equation}
where $E_{\bar p}=\sqrt{m_{\bar p}^2+p_{\rm lab}^2}$, $E_\Delta=m_d-E_{\Delta_R}$, $E_{\Delta_R}=\sqrt{m_{\Delta_R}^2+\bm{p}_{\Delta_R}^2}$.
We recall that $\bm{k}=-\bm{p}_{\Delta_R}$ in Eqs.(\ref{dsigma_over_d3pdt}),(\ref{kinem_prefac}) since these equations are obtained
treating the deuteron non-relativistically.

One can formally express the kinematic prefactors in Eq.(\ref{dsigma_over_d3pdt}) in terms of the light cone variable
\begin{equation}
  \alpha = \frac{E_{\Delta_R}-p_{\Delta_R}^z}{m_d/2}~,  \label{alpha}
\end{equation}
as defined in the deuteron rest frame. Hence, $\alpha/2$ is the fraction of deuteron momentum carried by $\Delta_R$
in the infinite momentum frame (where $\Delta_R$ moves fast in negative $z$ direction). We have also
\begin{equation}
   2 - \alpha = \frac{E_{\Delta}-k^z}{m_d/2}~.  \label{2-alpha}
\end{equation}
In the limit of very high beam momenta such that $p_{\rm lab} \simeq E_{\bar p}$ one can 
neglect masses in the M\"oller flux factor Eq.(\ref{I_barpDelta}):
\begin{equation}
  I_{\bar p \Delta} \simeq p_{\bar p} p_\Delta \simeq p_{\rm lab} (E_\Delta - k^z)
  = p_{\rm lab} \frac{m_d}{2} (2-\alpha)~. \label{HElimit1}
\end{equation}
This leads to the relative velocity
\begin{equation}
      v_{\bar p \Delta} \simeq \frac{m_d/2}{E_\Delta} (2-\alpha)~.  \label{HElimit2}
\end{equation}
Using Eq.(\ref{HElimit2}) we can rewrite the differential cross section (\ref{dsigma_over_d3pdt}) as
\begin{equation}
    \frac{d^4\sigma}{d^3p_{\Delta_R} dt} = \frac{m_d/2}{E_\Delta} (2-\alpha) |\phi_{\Delta-\Delta}(\bm{k})|^2
    \frac{d\sigma_{\bar p \Delta \to \pi_1\pi_2}(s^\prime,t)}{dt}~,                     \label{dsigma_over_d3pdt_HE}
\end{equation}
where
\begin{equation}
  s^\prime=(p_{\bar p}+p_d-p_{\Delta_R})^2  = (p_{\bar p}+p_\Delta)^2
  \simeq 2  p_{\bar p} p_\Delta \simeq p_{\rm lab} m_d (2-\alpha)~.  \label{sprime}
\end{equation}
If we define the invariant energy squared for the antiproton collision with a nucleon at rest
\begin{equation}
  s \equiv (p_{\bar p}+p_N)^2 = 2 E_{\bar p} m_N + 2 m_N^2~,
\end{equation}
then we have
\begin{equation}
  s^\prime \simeq s(2-\alpha)~.      \label{sprime_to_s}
\end{equation}  
We stress that Eq.(\ref{dsigma_over_d3pdt_HE}) is simply the high-energy limit of Eq.(\ref{dsigma_over_d3pdt}).

The problematic feature of the derivation given above is that the contribution of the baryon-antibaryon pairs
is included in the NR wave function in an uncontrolled way. This results in the finite value of $|\phi_{\Delta-\Delta}(\bm{k})|^2$
at $\alpha > 2$. This problem can be solved within the LC formalism. It is clear that the baryon-antibaryon pairs, i.e. vacuum fluctuations,
should not contribute to the LC wave function since it is evaluated in the frame where the deuteron is fast, and thus the time scale of its
internal dynamics is slowed down \cite{Frankfurt:1977vc,Frankfurt:1981mk}.

Thus, in the LC formalism one should evaluate the graph of Fig.~\ref{fig:IA} within the non-covariant perturbation theory
(time from left to right) and perform the transformation of the result in the infinite momentum frame where another graph (not shown) with
the emission of an antidelta from the antiproton disappears.
The calculation is almost identical to that for photon absorption in ref. \cite{Frankfurt:1977vc}. Thus, we will
not repeat it here and only show the final result:
\begin{equation}
    E_{\Delta_R} \frac{d^4\sigma}{d^3p_{\Delta_R} dt} = \left|\frac{\Gamma_{d \to \Delta \Delta}(p_d,p_{\Delta_R})}{(2-\alpha)(m_d^2-M_{\Delta,\Delta_R}^2)}\right|^2
    \frac{(2-\alpha)}{(2\pi)^3} \frac{d\sigma_{\bar p \Delta \to \pi_1\pi_2}(s^\prime,t)}{dt}~.    \label{dsigma_over_d3pdt_LC_prelim}
\end{equation}
Here, $M_{\Delta,\Delta_R}^2$ is the invariant mass of the intermediate $\Delta-\Delta$ state expressed as 
\begin{equation}
  M_{\Delta,\Delta_R}^2=\frac{4(m_\Delta^2+p_{\Delta_R t}^2)}{\alpha(2-\alpha)}=4(m_\Delta^2+\bm{k}^2)~,   \label{M_DeltaDeltaR}
\end{equation}
where on the last step we inserted a new variable conveniently used in the LC formalism (cf. \cite{Frankfurt:1977vc,Frankfurt:1981mk}
and Appendix \ref{app_LC_vs_NR}), the internal momentum $\bm{k}$ defined by relations
\begin{equation}
  \alpha = 1 + \frac{k^z}{\sqrt{m_\Delta^2+\bm{k}^2}}~,~~~\bm{k}_t = - \bm{p}_{\Delta_Rt}~, \label{k_int}
\end{equation}
where $\alpha$ is related to the residual $\Delta$ momentum via Eq.(\ref{alpha}). Using Eqs.(\ref{psi_DD}),(\ref{LC_vs_NR})
of Appendix \ref{app_LC_vs_NR}, the following expression for the differential cross section can be obtained:
\begin{equation}
    E_{\Delta_R} \frac{d^4\sigma}{d^3p_{\Delta_R} dt} = \frac{m_d/2}{m_d-m_\Delta} \frac{|\phi_{\Delta-\Delta}(\bm{k})|^2}{2-\alpha} \sqrt{m_\Delta^2+\bm{k}^2}
    \frac{d\sigma_{\bar p \Delta \to \pi_1\pi_2}(s^\prime,t)}{dt}~,                     \label{dsigma_over_d3pdt_LC}
\end{equation}
where $m_\Delta$ is the physical mass of the residual $\Delta$.

\section{Wave function of the $\Delta-\Delta$ system}
\label{WF}

The $\Delta-\Delta$ component of the deuteron wave function is a superposition of the $^3S_1,~^3D_1,~^7D_1$,
and $^7G_1$ states. In our calculations we applied the wave functions of the $np$ and $\Delta-\Delta$ systems
according to the coupled-channel folded-diagram potential (CCF) model of ref. \cite{Haidenbauer:1993pw}.
This model has been primarily developed for the description of many-body systems as
the resulting two-body potential is energy-independent which substantially simplifies calculations.
The two-body observables ($NN$ phase shifts, deuteron properties) are reproduced with an accuracy comparable
to that of the (energy-dependent) full Bonn potential \cite{Machleidt:1987hj}. Indeed, the analytic expressions
for the meson-baryon vertex functions are identical to those of the Bonn potential. The numerical values
of the meson-nucleon coupling constants and cutoff masses are, however, readjusted by a best fit to the empirical
$NN$ phase shifts. The CCF model is defined in momentum space and thus we start directly from the momentum space
representation.

The $\Delta-\Delta$ wave function in momentum space, Eq.(\ref{phi_p}), can be represented in the $LS$ basis as follows:
\begin{equation}
  \phi_{\Delta-\Delta}^{\lambda_d}(\bm{k};\lambda_{\Delta_R},\lambda_\Delta)
  = \sum_{LS} (-i)^L u_{LS}(k) \sum_{M\lambda} \langle 1 \lambda_d | LM;S\lambda \rangle\,
                                    Y_{LM}(\hat{\bm{k}})\, \chi_{S\lambda}(\lambda_{\Delta_R},\lambda_\Delta)~.   \label{LS_repr}
\end{equation}
Using the orthogonality of the spin wave functions,
\begin{equation}
  \sum_{\lambda_{\Delta_R}\lambda_\Delta} \chi_{S\lambda}(\lambda_{\Delta_R},\lambda_\Delta) \chi_{S^\prime\lambda^\prime}^*(\lambda_{\Delta_R},\lambda_\Delta)
  = \delta_{SS^\prime} \delta_{\lambda\lambda^\prime}~,              \label{chi_orthog}
\end{equation}
and the properties of the spherical functions and Clebsch-Gordan coefficients (cf. \cite{Varshalovich}) leads after some algebra
to the following expression for the c.m. momentum distribution, Eq.(\ref{phi_DD^2}):
\begin{equation}
  |\phi_{\Delta-\Delta}(\bm{k})|^2 = \frac{1}{4\pi} \sum_{LS} |u_{LS}(k)|^2~.    \label{phi_DD^2_LS}   
\end{equation}
The probabilities of the different $LS$-components are 
\begin{equation}
    P_{\Delta-\Delta}^{LS}= \int\limits_0^{+\infty}\,dk\, k^2\, |u_{LS}(k)|^2~,~~~\sum_{LS} P_{\Delta-\Delta}^{LS} = P_{\Delta-\Delta}~.    \label{P^LS}
\end{equation}
In the CCF model, the probabilities of the $^3S_1,~^3D_1,~^7D_1$, and $^7G_1$  $\Delta-\Delta$ states are, respectively:
$P_{\Delta-\Delta}^{01}=0.33\%,~P_{\Delta-\Delta}^{21}=0.03\%,~P_{\Delta-\Delta}^{23}=0.97\%$, and $P_{\Delta-\Delta}^{43}=0.02\%$.
The total probability of the $\Delta-\Delta$ states is $P_{\Delta-\Delta}=1.35\%$.

\begin{figure}
\begin{center}
  \includegraphics[scale = 0.52]{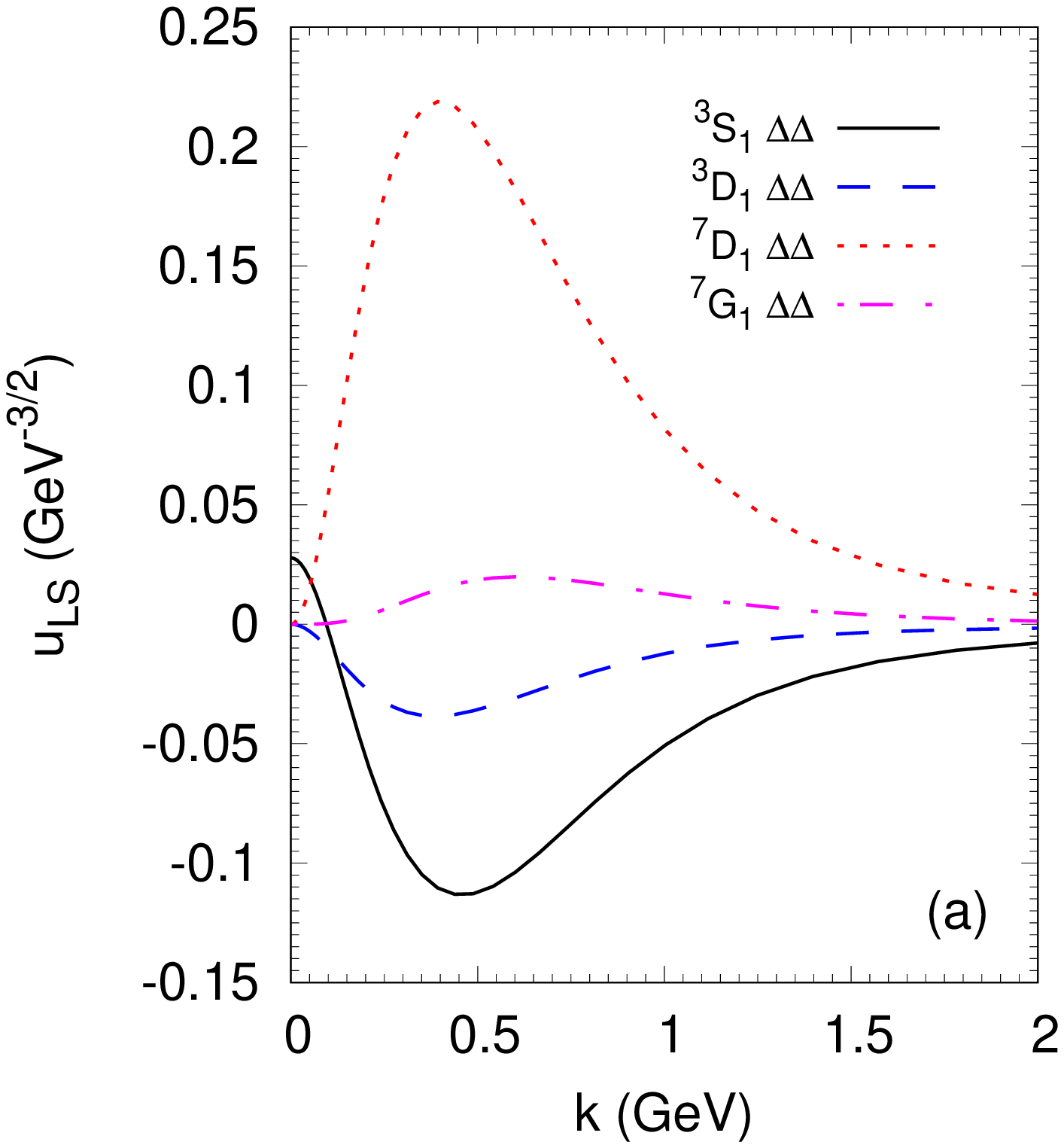}
  \includegraphics[scale = 0.52]{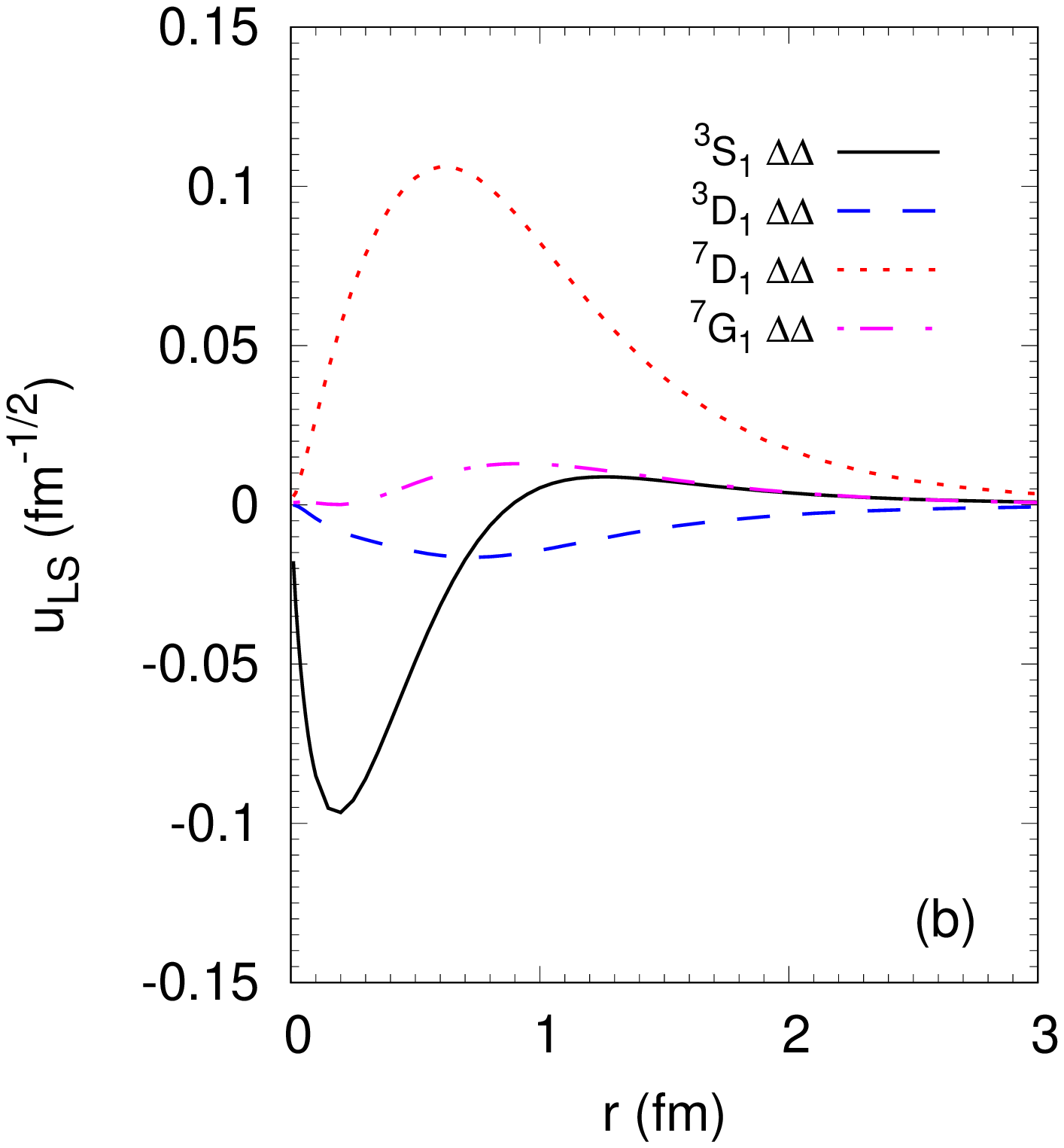}
\end{center}
\caption{\label{fig:DD_WF_CCF} The wave functions of the $\Delta-\Delta$ system in momentum (a) and configuration (b) space.}
\end{figure}

\begin{figure}
\begin{center}
   \includegraphics[scale = 0.60]{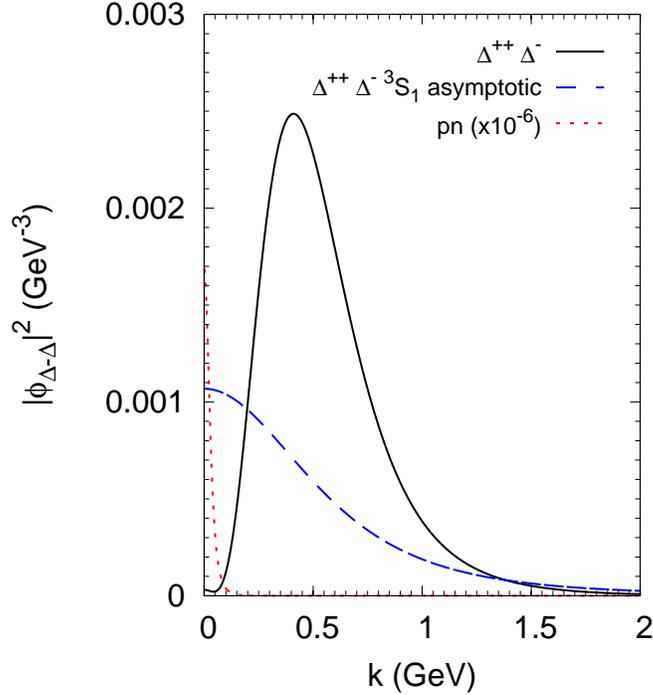}
\end{center}
\caption{\label{fig:phiDD2} Momentum distribution of the struck $\Delta$ in the $\Delta-\Delta$ c.m. frame.
  Solid line: CCF model, Eq.(\ref{phi_DD^2_LS}). Long-dashed line: large-distance asymptotic solution,
  Eq.(\ref{phi_k}), obtained by setting $m_{\Delta}=m_{\Delta_R}=1.232$ GeV. Both lines are multiplied
  by an extra factor of 1/2 which is the isospin fraction of the $\Delta^{++}-\Delta^-$ component. The ordinary
  deuteron c.m. momentum distribution multiplied by a factor of $10^{-6}$ is shown by the short-dashed line.}
\end{figure}

For the purpose of comparison with other potential models we have also calculated
the radial wave functions in configuration space which are obtained by a Fourier-Bessel transformation
\begin{equation}
   \frac{u_{LS}(r)}{r}=\sqrt{\frac{2}{\pi}} \int\limits_0^\infty dk k^2 j_L(kr) u_{LS}(k)~.   \label{u_LS_r}    
\end{equation}
Fig.~\ref{fig:DD_WF_CCF} displays the partial wave functions of the $\Delta-\Delta$ system in momentum and coordinate
representations\footnote{The $k$-space partial waves behave as $u_{LS}(k) \propto k^L$ in the limit $k \to 0$.
The $r$-space partial waves satisfy $u_{LS}(r) \propto r^{L+1}$ in the limit $r \to 0$.}.
All $LS$ partial waves in momentum space are maximal around the absolute value $k \simeq 0.4-0.6$ GeV.
The resulting c.m. momentum distribution is plotted in Fig.~\ref{fig:phiDD2} by the solid line.
The result of the coupled-channel model calculation is by far different from the simple large-distance asymptotic form
\begin{equation}
   \phi_{\Delta-\Delta}(\mathbf{k})=\frac{(\kappa P_{\Delta-\Delta})^{1/2}/\pi}{\kappa^2+\mathbf{k}^2}~,  \label{phi_k}
\end{equation}
where the range parameter is $\kappa=\sqrt{2 \mu E_b}$ with the reduced mass $\mu=m_{\Delta} m_{\Delta_R}/(m_{\Delta}+m_{\Delta_R})$
and the binding energy $E_b=m_{\Delta}+m_{\Delta_R}-m_d$. Note that, owing to the large binding energy, the c.m. momentum distribution
of the $\Delta-\Delta$ system is much harder than that of the ordinary deuteron.

The shapes of the $r$-space wave functions are similar to those of other potential
models with $\Delta$ degrees-of-freedom (cf. Fig.~2 in ref. \cite{Haapakoski:1974qnh},
Fig.~10 in ref. \cite{Niephaus:1979mw}, and Fig.~14 in ref. \cite{Wiringa:1984tg}).
In particular, the wave function of the dominating $^7D_1$ $\Delta-\Delta$ component
is quite close to that of ref. \cite{Niephaus:1979mw}. There are some moderate differences
for other components, e.g., in the CCF model the wave function of the $^3S_1$  $\Delta-\Delta$
component has a node at $r \simeq 0.9$ fm which is a feature of the particular coupled-channel
model realization (see, however, ref. \cite{Dymarz:1986km} where a node in the $^3S_1$
$\Delta-\Delta$ component at $r\simeq 0.5$ fm has been reported too). Our feeling is
that the differences in the momentum distribution Eq.(\ref{phi_DD^2_LS}) will be quite small
between the various models. The main difference between the models and, thus, the major
uncertainty concerns the total probability of the $\Delta-\Delta$ configuration, which varies
between $\sim 0.3\%$ and $\sim 1\%$. In some sense the CCF model applied in this work represents
the upper limit on the $\Delta-\Delta$ admixture in the deuteron.

\section{Background processes}
\label{bgr}

\subsection{Pion charge exchange}
\label{CEX}

\begin{figure}
\begin{center}
   \includegraphics[scale = 0.60]{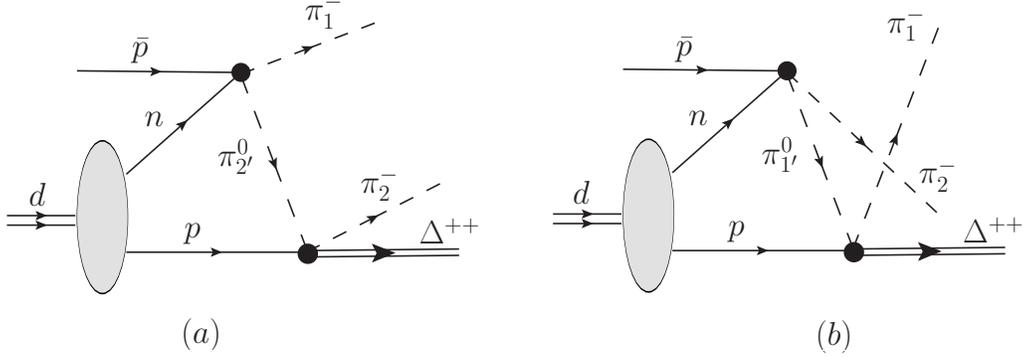}
\end{center}
\caption{\label{fig:bgr_CEX} The background processes due to inelastic CEX of the neutral pion on the proton.}
\end{figure}

The antiproton may annihilate with the neutron producing a $\pi^-\pi^0$ pair. The neutral pion may then experience inelastic
CEX scattering on the proton producing a $\pi^- \Delta^{++}$ pair. This CEX background process is depicted
in Fig.~\ref{fig:bgr_CEX}.
The amplitude of Fig.~\ref{fig:bgr_CEX} can be calculated starting from the $S$-matrix.

However, a more economic way to derive it is to use the vertex function $\Gamma_{d \to np}(p_d,p_p)$
which is defined similar to Eq.(\ref{Gamma_d2DD}):
\begin{equation}
  \frac{i \Gamma_{d \to np}(p_d,p_p)}{p_n^2-m_N^2+i\epsilon} = \left(\frac{2E_pE_d}{E_n}\right)^{1/2} (2\pi)^{3/2} \phi(\mathbf{k})~,
  \label{Gamma_d2np}
\end{equation}
where $\phi(\mathbf{k})$ is the deuteron wave function in momentum space (spin indices are implicit), $\mathbf{k}=-\mathbf{p}_p+\mathbf{p}_d/2$,
$E_p=\sqrt{\mathbf{p}_p^2+m_N^2}$, $E_n=E_d-E_p$. The invariant matrix element of Fig.~\ref{fig:bgr_CEX}a can be written as
\begin{eqnarray}
  iM_{\pi_1\pi_2\Delta_R;\bar p d}^{(a)} &=& \int \frac{d^4p_p}{(2\pi)^4} iM_{\pi p}(p_2,p_\Delta;p_{2^\prime},p_p) \frac{i}{p_p^2-m_N^2+i\epsilon}
                                     \frac{i}{p_{2^\prime}^2-m_\pi^2+i\epsilon} \nonumber \\
    & & \times iM_{\bar p n}(p_1,p_{2^\prime};p_{\bar p},p_n) \frac{i}{p_n^2-m_N^2+i\epsilon} \Gamma_{d \to np}(p_d,p_p)~.      \label{M^a_Feynman}
\end{eqnarray}
The integration contour over $dp_p^0$ can be closed in the lower part of the complex plane where only the pole of the proton propagator
at $p_p^0=E_p-i\epsilon$ contributes, such that
\begin{equation}
  \int \frac{dp_p^0}{2\pi} \frac{i}{(p_p^0)^2-E_p^2+i\epsilon} = \frac{1}{2E_p}~.  \label{contourInt}
\end{equation}
Hence we obtain
\begin{eqnarray}
  iM_{\pi_1\pi_2\Delta_R;\bar p d}^{(a)} &=& \int \frac{d^3p_p}{(2\pi)^{3/2}} \left(\frac{E_d}{2E_pE_n}\right)^{1/2} iM_{\pi p}(p_2,p_\Delta;p_{2^\prime},p_p)
                                     \frac{i}{p_{2^\prime}^2-m_\pi^2+i\epsilon} \nonumber \\
    & & \times iM_{\bar p n}(p_1,p_{2^\prime};p_{\bar p},p_n) \phi(\mathbf{k})~.      \label{M^a}
\end{eqnarray}

A kinematically interesting scenario for the ``signal'' process of antiproton annihilation on the $\Delta-\Delta$ state emerges in the case
that both $t=(p_1-p_{\bar p})^2$ and $u=(p_2-p_{\bar p})^2$ (Fig.~\ref{fig:IA}) are large, i.e. $t \sim u \sim -s/2$, since one has to resolve
a short time interval of the deuteron existing in a $\Delta-\Delta$ state. Thus the $\bar p n \to \pi^- \pi^0$ amplitude is hard and can be
factorized out in Eq.(\ref{M^a}) by neglecting the neutron Fermi motion.  Such regime corresponds to both pions having momenta with large
positive $z$-components. Hence the momentum transfer $q=p_p-p_\Delta$ in the CEX process $\pi^0 p \to \pi^- \Delta^{++}$ is small, $q^2 \ll p_{2t}^2$.
Under these assumptions the inverse propagator of the pion can be simplified:
\begin{eqnarray}
  &&  p_{2^\prime}^2 -m_\pi^2 + i\epsilon = (p_2-q)^2  -m_\pi^2 + i\epsilon
  = 2p_2^z\left(q^z-q^0\frac{E_2}{p_2^z} + \frac{\mathbf{p}_{2t}\mathbf{q}_t}{p_2^z} + \frac{q^2}{2p_2^z} + i\epsilon \right)  \nonumber \\
  && = 2p_2^z ( p_p^z + \Delta^0_\pi + i\epsilon )~,    \label{invPionProp}
\end{eqnarray}
where
\begin{equation}
  \Delta^0_\pi = -p_\Delta^z + (E_\Delta-m_N)\frac{E_2}{p_2^z}
  + \frac{\mathbf{p}_{2t}(\mathbf{p}_{pt}-\mathbf{p}_{\Delta t})}{p_2^z}~.  \label{Delta^0_pi}
\end{equation}
In Eq.(\ref{Delta^0_pi}) we neglected the term $q^2/2p_2^z$ and the Fermi motion of the proton.

In the calculation of the pion CEX amplitude $M_{\pi p}(p_2,p_\Delta;p_{2^\prime},p_p)$ we put the four-momentum $p_{2^\prime}$
of the intermediate pion on mass shell by setting $p_p^z=-\Delta^0_\pi$ for fixed proton transverse momentum $\mathbf{p}_{pt}$.
After this setting the pion CEX amplitude becomes independent of the longitudinal momentum of the proton.
This allows us to separate the integral over $dp_p^z$ in Eq.(\ref{M^a}) with the inverse pion propagator
of Eq.(\ref{invPionProp}):
\begin{equation}
   I_z = \int dp_p^z \frac{\phi(-\mathbf{p}_p)}{p_p^z + \Delta^0_\pi + i\epsilon}~,     \label{I_z_def}
\end{equation}
as given in the deuteron rest frame.
The deuteron wave function in momentum space can be expressed as follows (cf. \cite{Frankfurt:1996uz}):
\begin{equation}
   \phi(\mathbf{k})
   =\frac{1}{\sqrt{4\pi}}\left(u(k)+\frac{w(k)}{\sqrt{8}}S(\mathbf{k})\right) \chi^M   \label{phi}
\end{equation}
with the spin tensor operator
\begin{equation}
   S(\mathbf{k})=\frac{3(\bvec{\sigma}_p\mathbf{k})(\bvec{\sigma}_n\mathbf{k})}{k^2} 
               -\bvec{\sigma}_p\bvec{\sigma}_n~,   \label{Sspin}
\end{equation}
and $\chi^M$ being the eigenfunction of the spin~=~1 state with spin projection $M=0,\pm1$.
We will apply the analytical parameterization of the S- and D-wave components in the spirit of the Paris \cite{Lacombe:1981eg} model,
however, with the values of parameters adjusted according to the CCF model \cite{Haidenbauer:1993pw}
\footnote{The coefficients $c_j$ ($d_j$) in Eq.(\ref{uw}) are those of CCF model times $+(-)\sqrt{2/\pi}$. It means that $w(k) < 0$
  at small $k$. To avoid misunderstanding: the sign conventions for the $D$-wave in Eq.(\ref{LS_repr}) and Eq.(\ref{phi}) are different.}
\begin{equation}
   u(k)=\sum_j \frac{c_j}{k^2+m_j^2}~,~~~w(k)=\sum_j \frac{d_j}{k^2+m_j^2}~,   \label{uw}
\end{equation}
with additional conditions $\sum_j c_j =0$  and $\sum_j d_j = \sum_j d_j/m_j^2 =  \sum_j d_j m_j^2 = 0$.
These conditions guarantee the decrease of both wave functions $\propto 1/k^4$ at large $k$ and $w(k) \propto k^2$ at small $k$.
The latter guarantees the absence of a pole at $k=0$ in the product $w(k)S(\mathbf{k})$. 

The integration contour over $dp_p^z$ in Eq.(\ref{I_z_def}) can be closed in the upper part of the complex plane
where only the poles of the wave function at $p_p^z=im_{jt}, m_{jt}=\sqrt{m_j^2+p_{pt}^2}$ contribute.
This leads to the following expression for the longitudinal momentum integral:
\begin{equation}
  I_z = \frac{1}{\sqrt{4\pi}} \sum_j \frac{\pi}{m_{jt}(\Delta^0_\pi + im_{jt})} \phi_j^M(-\mathbf{p}_p)~,     \label{I_z}
\end{equation}
where
\begin{equation}
  \phi_j^M(-\mathbf{p}_p) =\left(c_j+\frac{d_j}{\sqrt{8}}S(-\mathbf{p}_p)\right) \chi^M          \label{phi_j^M}
\end{equation}
with $\mathbf{p}_p=(\mathbf{p}_{pt},im_{jt})$.
Using Eq.(\ref{I_z}), after some algebra Eq.(\ref{M^a}) can finally be transformed to the following expression:
\begin{eqnarray}
  M_{\pi_1\pi_2\Delta_R;\bar p d}^{(a)} &=&  -\frac{m_d^{1/2} M_{\bar p n}(p_1,p_{\bar p}+p_n-p_1;p_{\bar p},p_n)}{16\pi m_N p_2^z}  \nonumber \\
   && \times \int d^2p_{pt}  M_{\pi p}(p_2,p_\Delta;p_2+p_\Delta-p_p,p_p)
      \sum_j \frac{\phi_j^M(-\mathbf{p}_{pt},-im_{jt})}{m_{jt}(\Delta^0_\pi + im_{jt})}~,      \label{M^a_final}
\end{eqnarray}
where $p_n=(m_N,0)$ and $p_p=(\sqrt{m_N^2+p_{pt}^2+(\Delta^0_\pi)^2},\mathbf{p}_{pt},-\Delta^0_\pi)$ are the neutron and proton
four-momenta in the elementary matrix elements. Note that the summation over spin indices of intermediate proton and neutron
is implicitly assumed in Eq.(\ref{M^a_final}).

\subsection{Three-pion annihilation}
\label{3pi}

\begin{figure}
\begin{center}
   \includegraphics[scale = 0.60]{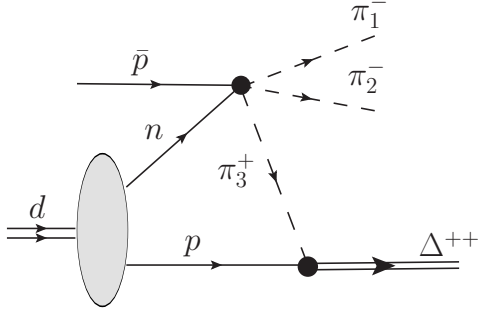}
\end{center}
\caption{\label{fig:bgr_3pi} The background process initiated by antiproton annihilation on the neutron into three pions.}
\end{figure}

Fig.~\ref{fig:bgr_3pi} shows another possible background channel due to the two-step process
$\bar p n \to \pi^- \pi^- \pi^+,~\pi^+ p \to \Delta^{++}$.    
Similar to Eq.(\ref{M^a}), the invariant matrix element of Fig.~\ref{fig:bgr_3pi} can be expressed as
\begin{eqnarray}
  iM_{\pi_1\pi_2\Delta_R;\bar p d}^{(3\pi)} &=& \int \frac{d^3p_p}{(2\pi)^{3/2}} \left(\frac{E_d}{2E_pE_n}\right)^{1/2} iM_{\pi p}(p_\Delta;p_3,p_p)
                                     \frac{i}{p_3^2-m_\pi^2+i\epsilon} \nonumber \\
    & & \times iM_{\bar p n}(p_1,p_2,p_3;p_{\bar p},p_n) \phi(\mathbf{k})~,      \label{M^3pi}
\end{eqnarray}
where the intermediate proton is put on mass shell, i.e. $p_p^0=E_p=\sqrt{\mathbf{p}_p^2+m_N^2}$, $p_3=p_\Delta-p_p$, $E_n=E_d-E_p$,
and $\mathbf{k}=-\mathbf{p}_p+\mathbf{p}_d/2$.

Since all three particles in the $\pi N \Delta$ vertex have small momenta, the simplification of the pion propagator in Eq.(\ref{M^3pi})
by neglecting proton Fermi motion in the spirit of Eq.(\ref{invPionProp}) is generally impossible. Thus, we simplified Eq.(\ref{M^3pi})
by only replacing proton and neutron energies in the denominator by the nucleon mass. The resulting expression in the deuteron rest frame is
\begin{equation}
  M_{\pi_1\pi_2\Delta_R;\bar p d}^{(3\pi)} = - \frac{m_d^{1/2}}{4\pi^{3/2}m_N} \int d^2p_{pt}
            \int\limits_{-\infty}^{+\infty}dp_p^z 
            \frac{M_{\pi p}(p_\Delta;p_3,p_p) M_{\bar p n}(p_1,p_2,p_3;p_{\bar p},p_n) \phi(-\mathbf{p}_p)}{p_3^2-m_\pi^2+i\epsilon}~.
            \label{M^3pi_simplified}
\end{equation}
At fixed proton transverse momentum $\mathbf{p}_{pt}$, the pion propagator may have up to two poles at $p_p^z=\Delta_1$ and $p_p^z=\Delta_2$
with $\Delta_1 \leq \Delta_2$. The poles are given by the zeros of the function
\begin{equation}
  F(\mathbf{p}_p)=p_3^2-m_\pi^2=(p_\Delta-p_p)^2-m_\pi^2=m_\Delta^2+m_N^2-m_\pi^2 - 2 E_\Delta E_p + 2 \mathbf{p}_\Delta \mathbf{p}_p~.   \label{F}
\end{equation}
The calculation of the poles $\Delta_{1,2}$ is described in Appendix \ref{app_piPoles}. In order to avoid numerical problems related to the poles,
we added a small artificial width to the pion. Thus, we replaced in Eq.(\ref{M^3pi_simplified}) $\epsilon \to m_\pi\Gamma_\pi^{\rm art}$ with
$\Gamma_\pi^{\rm art} \sim 10$ MeV. This allows to compute the three-dimensional integral over the proton momentum in the usual way. 
To achieve a smooth dependence of the matrix element on the momentum of the $\Delta$, the numerical integration on the proton transverse momentum
has been performed separately in the subregions with and without pion poles, while the integration over $dp_p^z$ has been performed separately
in the intervals $p_p^z < \Delta_1$, $\Delta_1 < p_p^z < \Delta_2$ and $\Delta_2 <  p_p^z$. The moderate influence of the  choice of the artificial
pion width on the results is displayed in Fig.\ref{fig:sig_alpha_plab15gev_pDt0gev_gampi} below.

One note is in order here. For simplicity, we perform the background calculations using the NR description of the deuteron.
Since the ordinary deuteron wave function in momentum space is quite narrow (cf. Fig.~\ref{fig:phiDD2}) the NR approximation
should be indeed reasonable in evaluating momentum space integrals like in Eqs.(\ref{M^a_final}),(\ref{M^3pi}), provided that
the elementary amplitudes do not strongly grow in certain regions of momentum space. For example, in the case of pion inelastic CEX,
the $\pi^0 p \to \pi^- \Delta^{++}$ amplitude drops quickly with transverse momentum transfer and, thus, the integration over proton
transverse momentum in Eq.(\ref{M^a_final}) is unproblematic. However, the $\bar p n \to \pi^- \pi^- \pi^+$ amplitude $M_{\bar p n}(s^\prime)$
extracted from the fit to the available experimental data (see Appendix \ref{NbarN23pi}) strongly grows with decreasing
$s^\prime=(p_{\bar p}+p_d-p_p)^2$. This makes the integral in Eq.(\ref{M^3pi}) sensitive to the lower limit of $p_p^z$. Hence, in the
spirit of the LC approach, we have restricted the longitudinal proton momentum integral by the condition $(E_p-p_p^z)/m_d < 1$.

\section{Numerical results}
\label{results}

The differential cross section  of the background processes is expressed by Eq.(\ref{dsigma}) where one has to replace
$M^{(0)}$ by $M^{(a)}+M^{(b)}$ for the pion CEX background (see Eq.(\ref{M^a_final}) and the same equation
with interchange $p_1 \leftrightarrow p_2$ for $M^{(b)}$) or by $M^{(3\pi)}$ (see Eq.(\ref{M^3pi_simplified}) for the
three-pion annihilation background. Interference between signal and background processes is neglected.
The calculation of the differential cross section\footnote{In this section we will denote the residual delta as ``$\Delta$''
dropping the lower index ``R'' for brevity. The struck delta will be denoted as ``$\Delta_S$''.}
$E_{\Delta} d^4\sigma/d^3p_{\Delta}dt$ for the background is numerically exhaustive, since it requires integration over pion
azimuthal angle. Thus we have calculated the following quantity:
\begin{equation}
  E_{\Delta} \frac{d^5\sigma}{d^3p_{\Delta} d\Omega_\pi}
  = \frac{\overline{|M_{\pi_1\pi_2\Delta;\bar p d}|^2} p_1^2}{32 (2\pi)^5 p_{\rm lab} m_d \kappa}~, \label{d^5sigma}
\end{equation}
where $\Omega_\pi$ is the solid angle defining the direction of the momentum $\mathbf{p}_1$ in the deuteron rest frame,
$\kappa=|p_1E_2+(p_1- \tilde{\mathbf{p}} \cdot \mathbf{p}_1 /p_1)E_1|$,
$\tilde{\mathbf{p}} = \mathbf{p}_{\bar p} - \mathbf{p}_{\Delta}$.
In Eq.(\ref{d^5sigma}), $\overline{|M_{\pi_1\pi_2\Delta;\bar p d}|^2}$ should be replaced by the corresponding background or signal
expression. For the signal, Eq.(\ref{M2_simpl}) is applied in the case of the NR description, while in the case of
the LC description we have
\begin{equation}
   \overline{|M^{(0)}_{\pi_1\pi_2\Delta;\bar p d}|^2} 
    =  \frac{2(m_\Delta^2+\bm{k}^2)^{1/2}m_d}{(m_d-m_\Delta)(2-\alpha)^2} 
       (2\pi)^3 |\phi_{\Delta-\Delta}(\bm{k})|^2 \overline{|M_{\pi_1\pi_2;\bar p \Delta_S}|^2}~.   \label{M2_LC}
\end{equation}
All signal cross sections shown on the figures below include an extra factor of 1/2 which is the isospin fraction
of the $\Delta^- - \Delta^{++}$ component.

\begin{figure}
  \includegraphics[scale = 0.37]{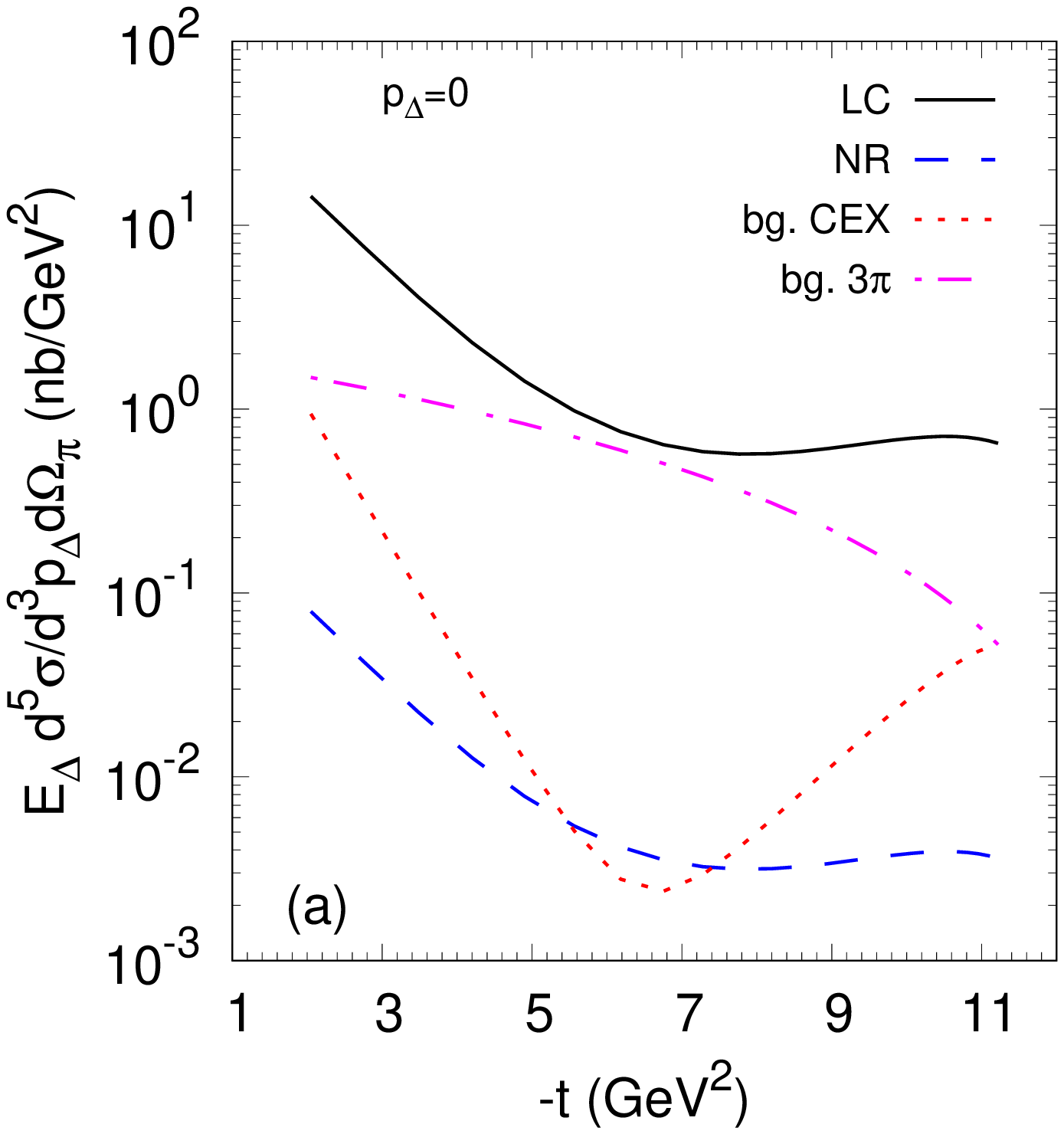}
  \includegraphics[scale = 0.37]{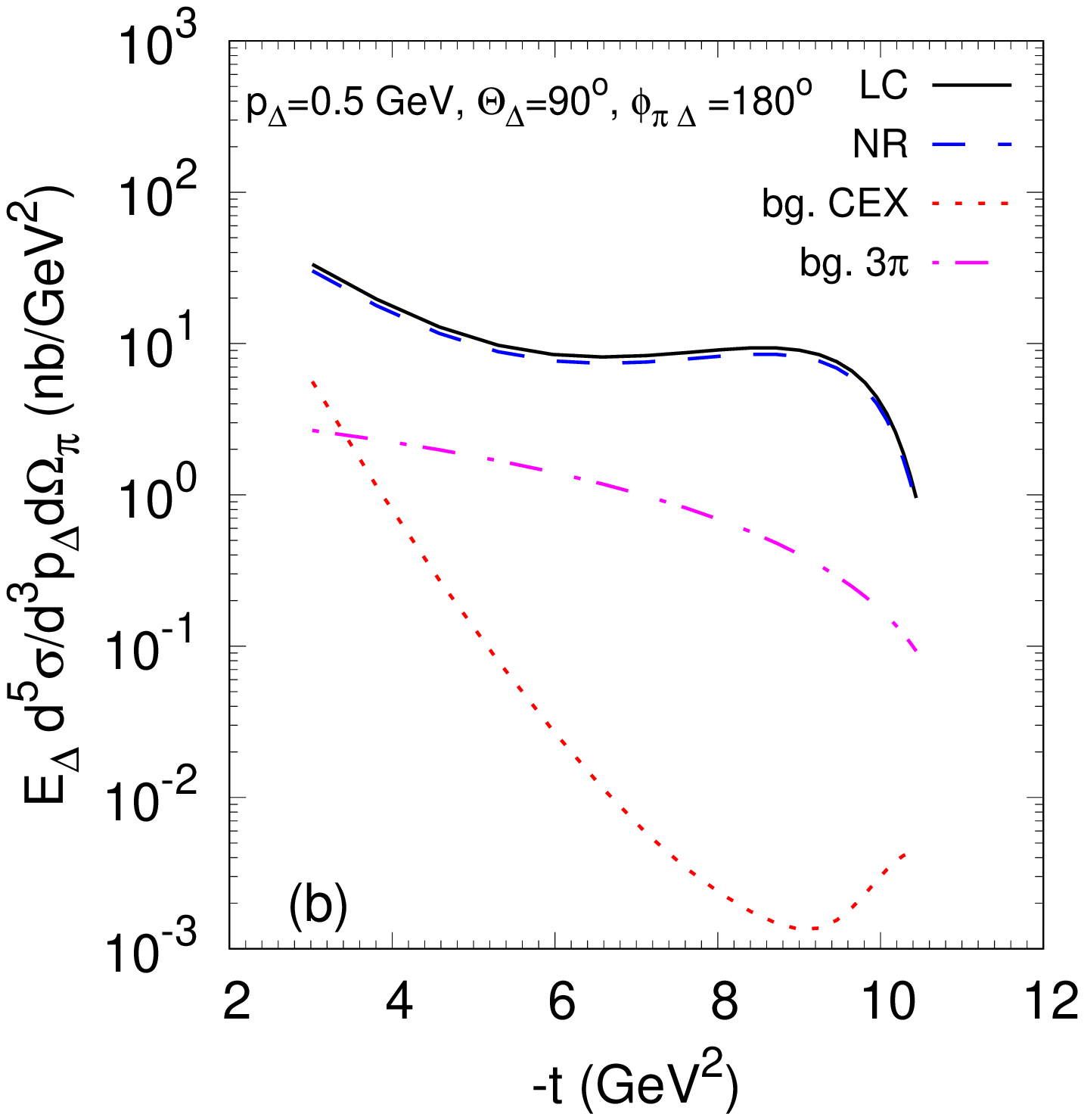}
  \includegraphics[scale = 0.37]{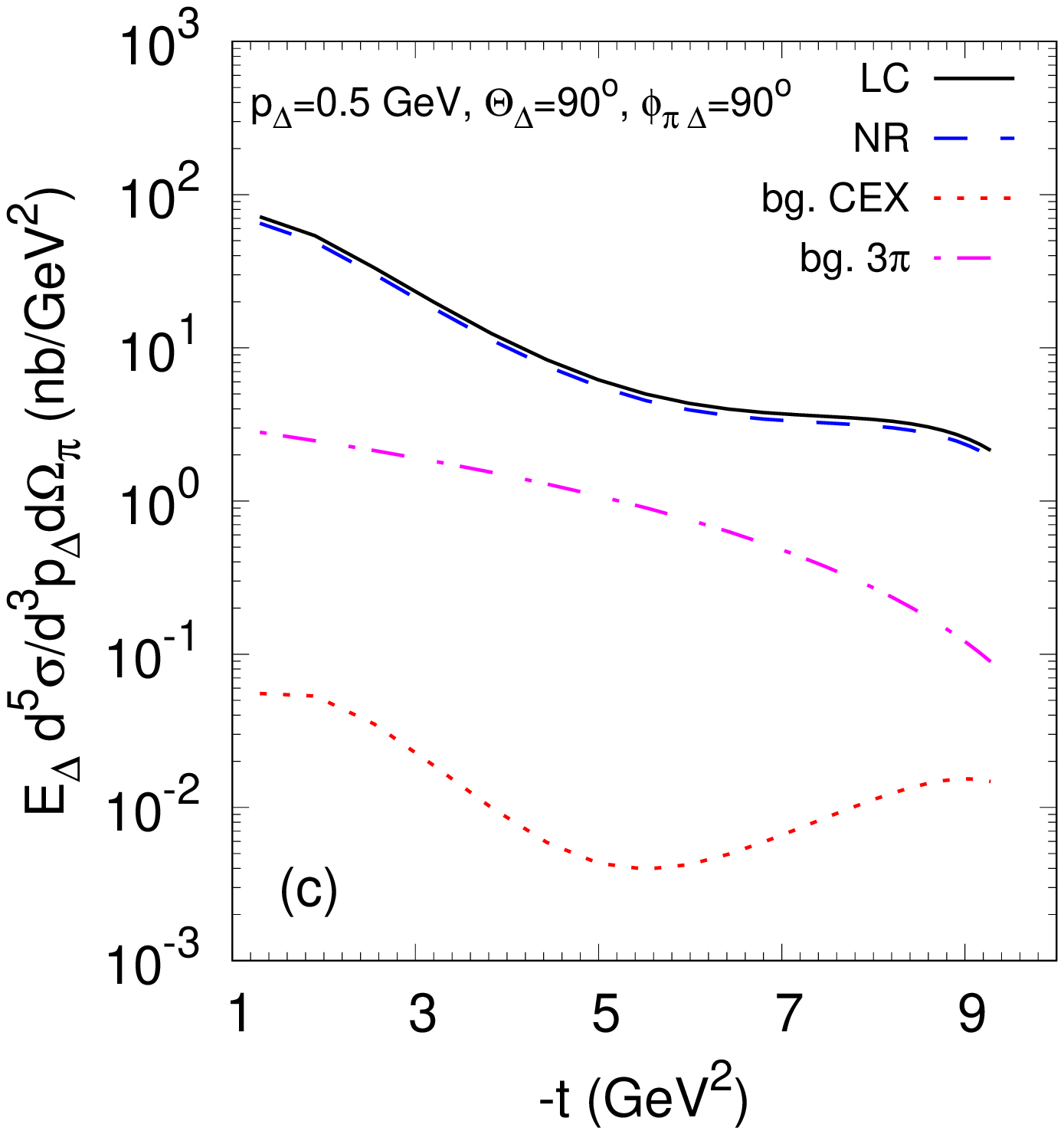}
  \includegraphics[scale = 0.37]{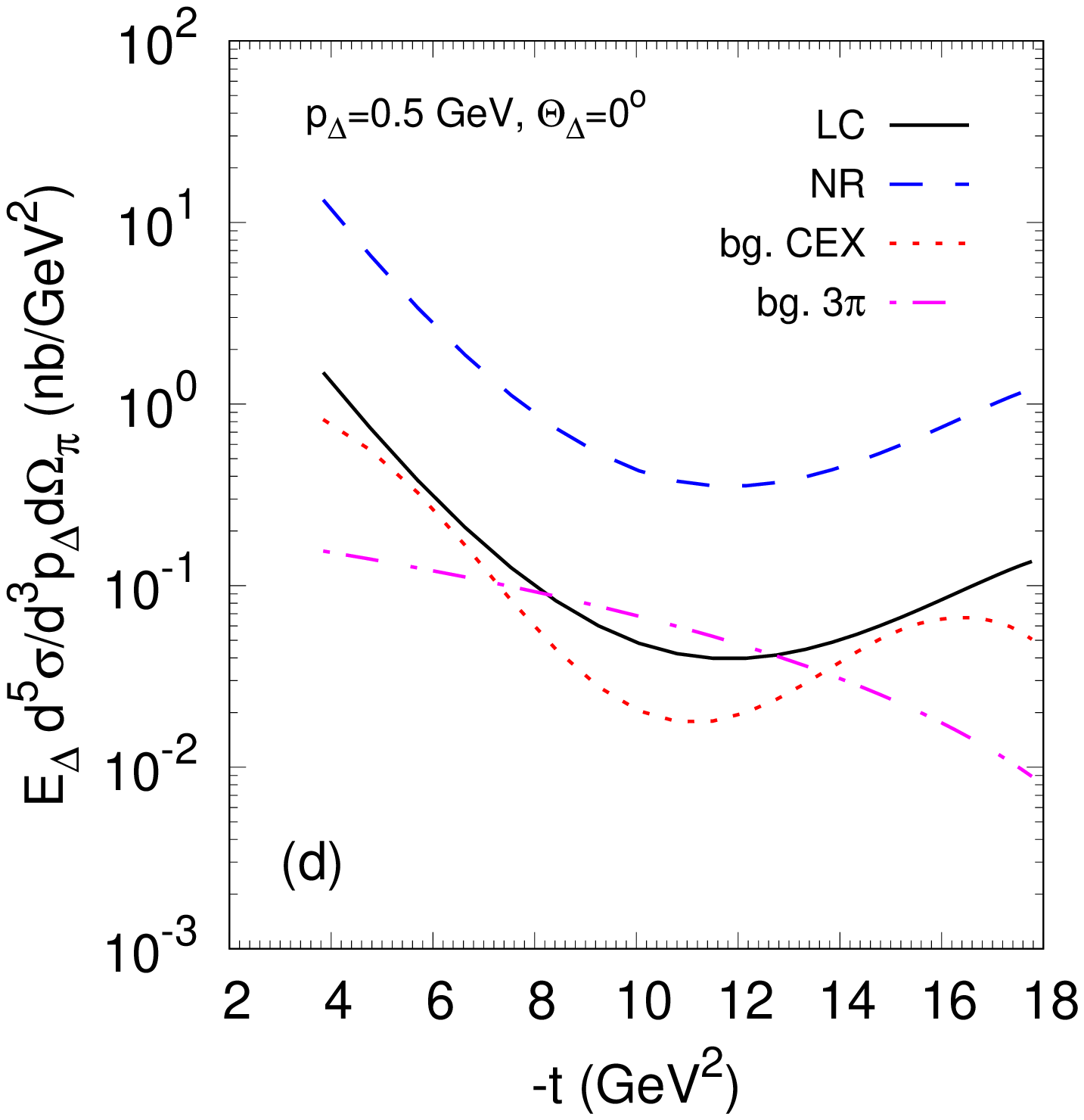}
  \includegraphics[scale = 0.37]{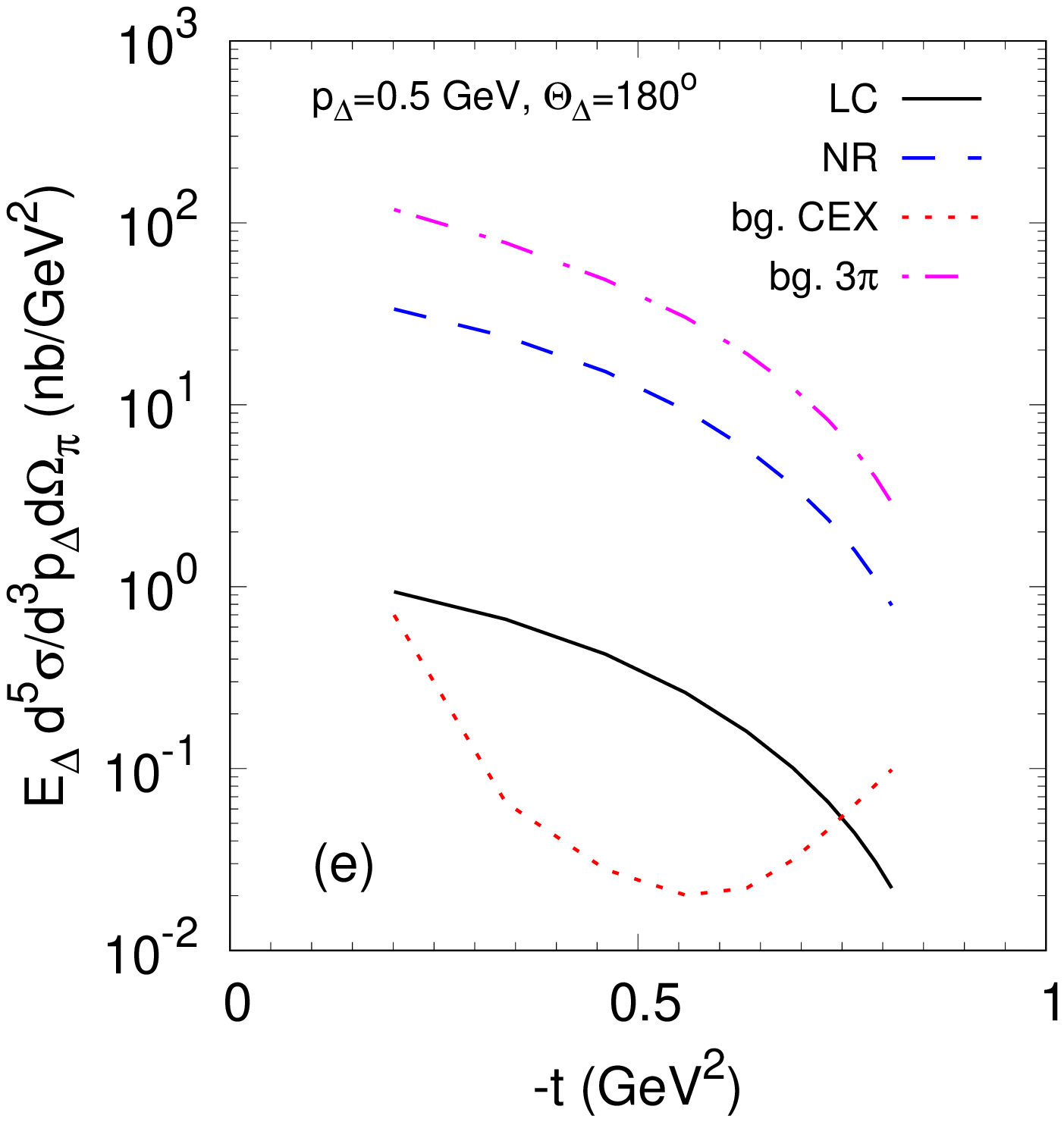}
  \includegraphics[scale = 0.37]{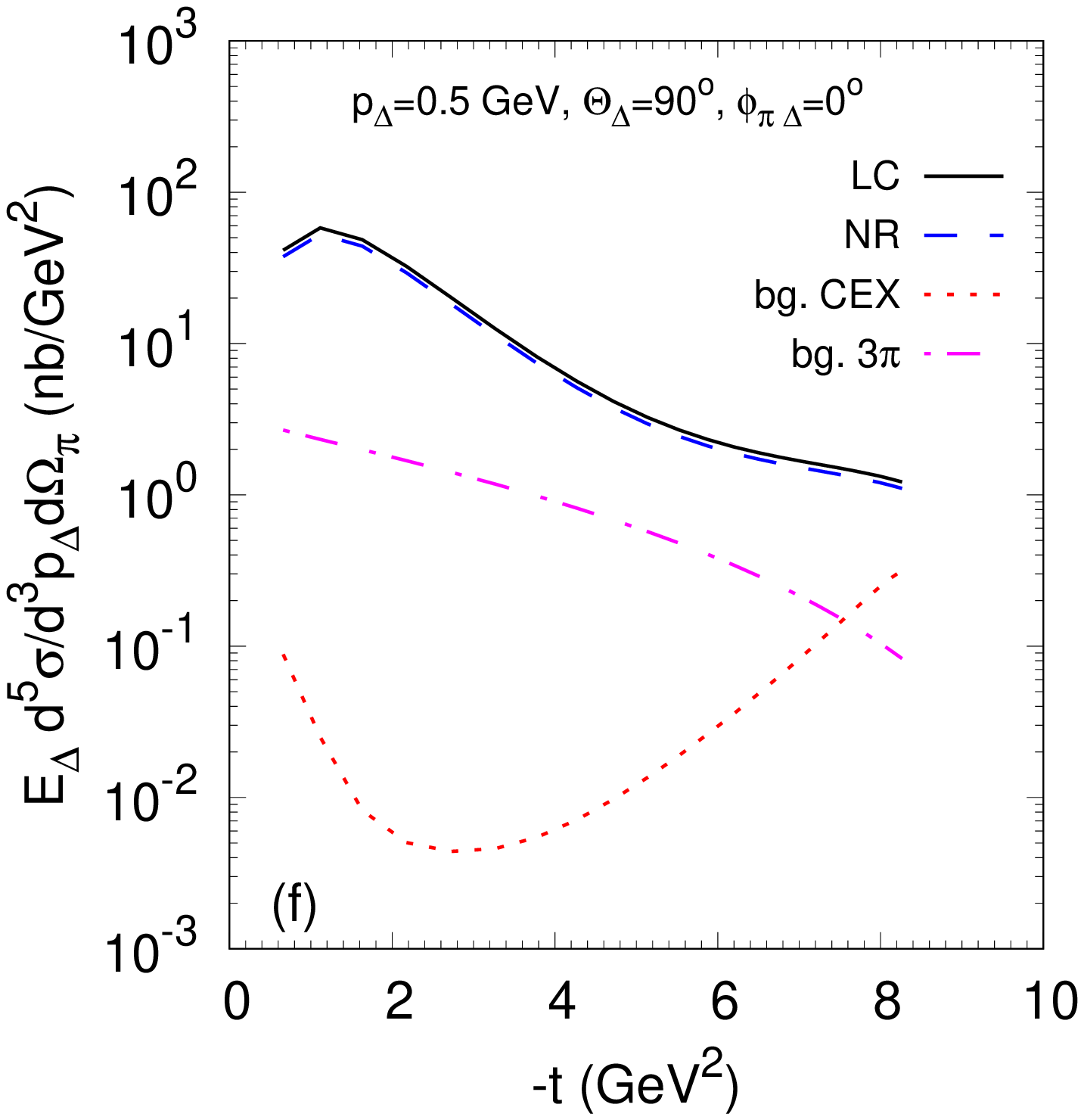}
  \caption{\label{fig:d^5sig} Differential cross section of the reaction $\bar p d \to \pi^- \pi^- \Delta^{++}$
    at $p_{\rm lab}=10$ GeV/c as a function of the Mandelstam $t$, which is defined as the four-momentum transfer squared between
    one of the outgoing pions (1$^{\rm{}st}$ pion) and the antiproton (Eq.(\ref{t})).
    The solid, dashed, dotted, and dot-dashed lines correspond to the LC signal, NR signal, CEX background, and $3\pi$ background,
    respectively.
    Different panels display the results for the different values
    of the momentum $p_\Delta$ and the polar angle $\Theta_\Delta$  of the $\Delta$-isobar and of the relative azimuthal angle
    $\phi_{\pi\Delta}=\phi_\pi-\phi_\Delta$ between 1$^{\rm{}st}$ pion and $\Delta$. All quantities refer to the deuteron rest frame.
    The residual $\Delta^{++}$ is assumed to be on the mass shell.}  
\end{figure}
Fig.~\ref{fig:d^5sig} shows the differential $\bar p d \to \pi^- \pi^- \Delta^{++}$ cross sections calculated at 10 GeV/c beam momentum.
The cross sections are plotted in the $t$-interval where both pions have $z$-components of their momenta larger than 1 GeV/c. On one hand,
this condition is needed in order to ensure the ``softness'' of the pion CEX (otherwise our calculation becomes inapplicable).
On the other hand, the most interesting case of hard $\bar p \Delta^- \to \pi^- \pi^-$ interaction -- i.e. when $\Theta_{c.m.} \simeq 90\degree$
and both pions have momenta with $z$-components close to $p_{\rm lab}/2$ -- is fully covered.
We have considered the representative cases of the $\Delta$ at rest and of the $\Delta$ at 0.5 GeV/c momentum, emitted at polar angles
$\Theta_\Delta=0\degree, 90\degree$ and $180\degree$.
In the case of $\Theta_\Delta=90\degree$ the cross section also depends on the azimuthal angle between $\pi$ and $\Delta$.

At zero momentum of the residual $\Delta$-resonance, the LC calculation produces much larger signal cross section than the NR calculation does.
This can be traced back to the struck $\Delta$ momentum distribution in the $\Delta-\Delta$ c.m. frame (Fig.~\ref{fig:phiDD2}).
In the NR calculation, the momentum $k$ of the struck $\Delta$ is zero, while in the LC calculation $k^z=0.4$ GeV/c corresponding to $\alpha=1.3$
for the residual $\Delta$ at rest. When the residual $\Delta$ moves transversely to the beam direction with momentum 0.5 GeV/c, the difference
between LC and NR calculations is practically invisible. If the residual $\Delta$ moves in positive or negative $z$ direction with 0.5 GeV/c
momentum, then the intrinsic momentum is $k^z=-0.14$ GeV/c or $3.77$ GeV/c, respectively, i.e. in regions where the struck $\Delta$ momentum
distribution is strongly suppressed, and thus the LC calculation predicts much smaller signal cross section as compared to the NR calculation.

The characteristic shape of the $\Delta-\Delta$ momentum distribution (Fig.~\ref{fig:phiDD2}) is certainly of primary interest.
One expects that it should be visible in the $\alpha$-distributions of the residual $\Delta$:
\begin{equation}
  \alpha \beta \frac{d^5\sigma}{d\alpha d\beta d\phi_{\pi\Delta} p_{\Delta t} dp_{\Delta t} dp_\Delta^2}
  = \frac{\overline{|M_{\pi_1\pi_2\Delta;\bar p d}|^2} p_{1t} {\cal A}(p_\Delta^2)}%
         {16(2\pi)^4 p_{\rm lab} m_d \kappa_t}~,                   \label{dsig/dalpha}
\end{equation}
where 
\begin{equation}
  \beta=\frac{2(E_1+p_1^z)}{E_{\bar p}+m_d-E_{\Delta}+p_{\rm lab}-p_{\Delta}^z}        \label{alpha_1}
\end{equation}
is the LC momentum fraction of one of the outgoing pions (``1$^{\rm{}st}$ pion''), and
$\phi_{\pi\Delta} = \phi_\pi-\phi_\Delta$ is the relative azimuthal angle between the 1$^{\rm{}st}$ pion and the $\Delta$.
The quantity
\begin{equation}
  \kappa_t = 2\left|p_{1t}\frac{E_2+p_2^z}{E_1+p_1^z} + p_{1t} + \frac{\mathbf{p}_{\Delta t}\cdot\mathbf{p}_{1t}}{p_{1t}}\right|   \label{kappa_t}
\end{equation}
originates from expressing the phase space volume of the outgoing particles in terms of the LC momentum fractions and transverse momenta. 

To take into account the possible off-shellness of the residual $\Delta$ we have also introduced in Eq.(\ref{dsig/dalpha})
the spectral function of the $\Delta$-resonance
\begin{equation}
   {\cal A}(p_\Delta^2)=\frac{\sqrt{p_\Delta^2} \Gamma_\Delta / \pi}{(p_\Delta^2-m_\Delta^2)^2+m_\Delta^2 \Gamma_\Delta^2}~,   \label{calA}
\end{equation}
normalized as
\begin{equation}
    \int\limits_{(m_\pi+m_N)^2}^{+\infty} {\cal A}(M^2) dM^2 = 1~.         \label{normcalA}    
\end{equation}
The off-shell background matrix elements are obtained in the usual way, i.e. by the replacements $m_\Delta \to \sqrt{p_\Delta^2}$.
The expressions Eqs.(\ref{M2_simpl}),(\ref{M2_LC}) for the moduli squared of the signal matrix elements and the relation Eq.(\ref{k_int})
between the LC momentum fraction $\alpha$ and the internal momentum $\bm{k}$ are not modified due to the $\Delta^{++}$ off-shellness.
In the numerical results below we have set the residual $\Delta^{++}$ on its mass shell.

\begin{figure}[ht]
  \includegraphics[scale = 0.50]{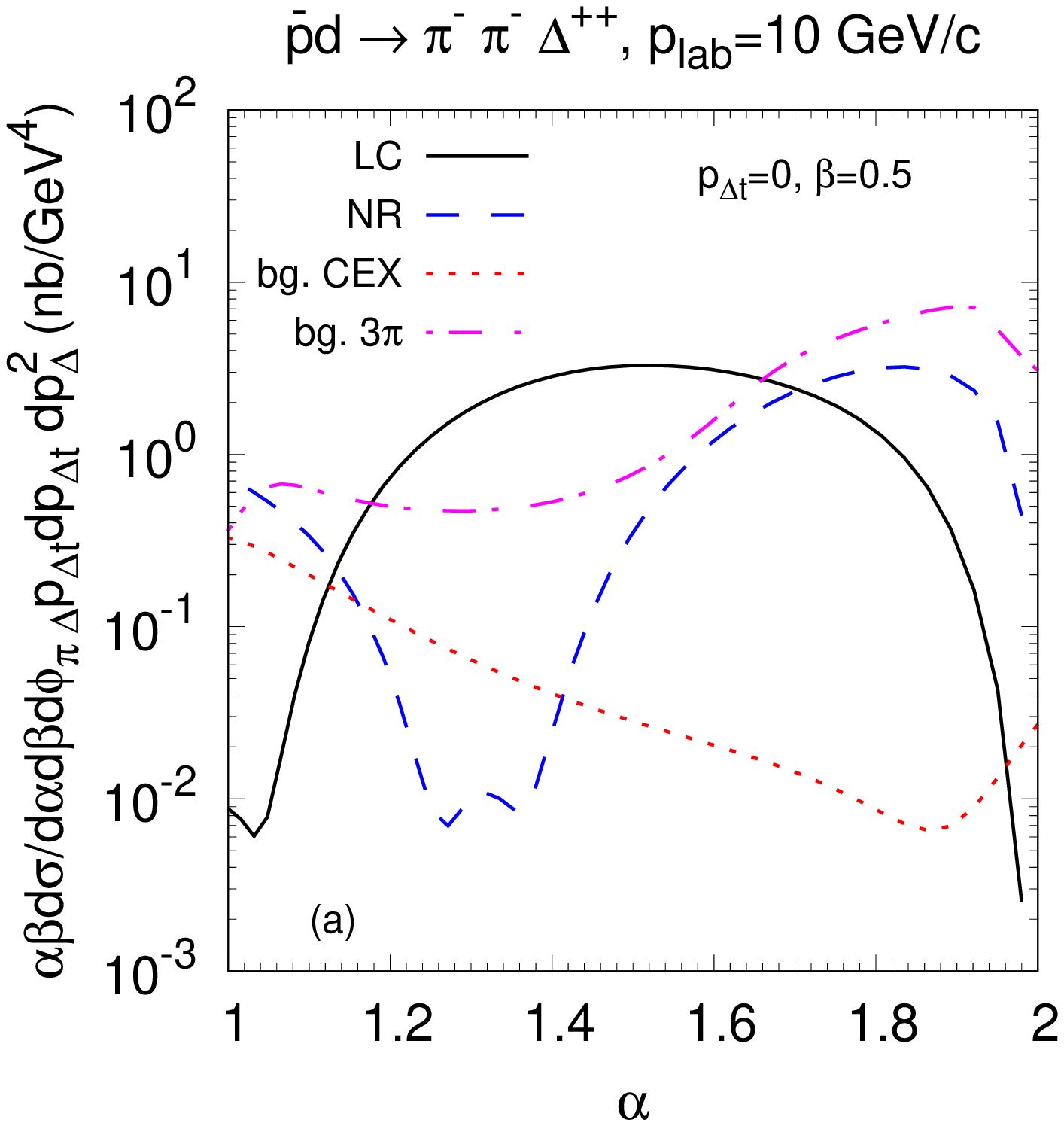}
  \includegraphics[scale = 0.50]{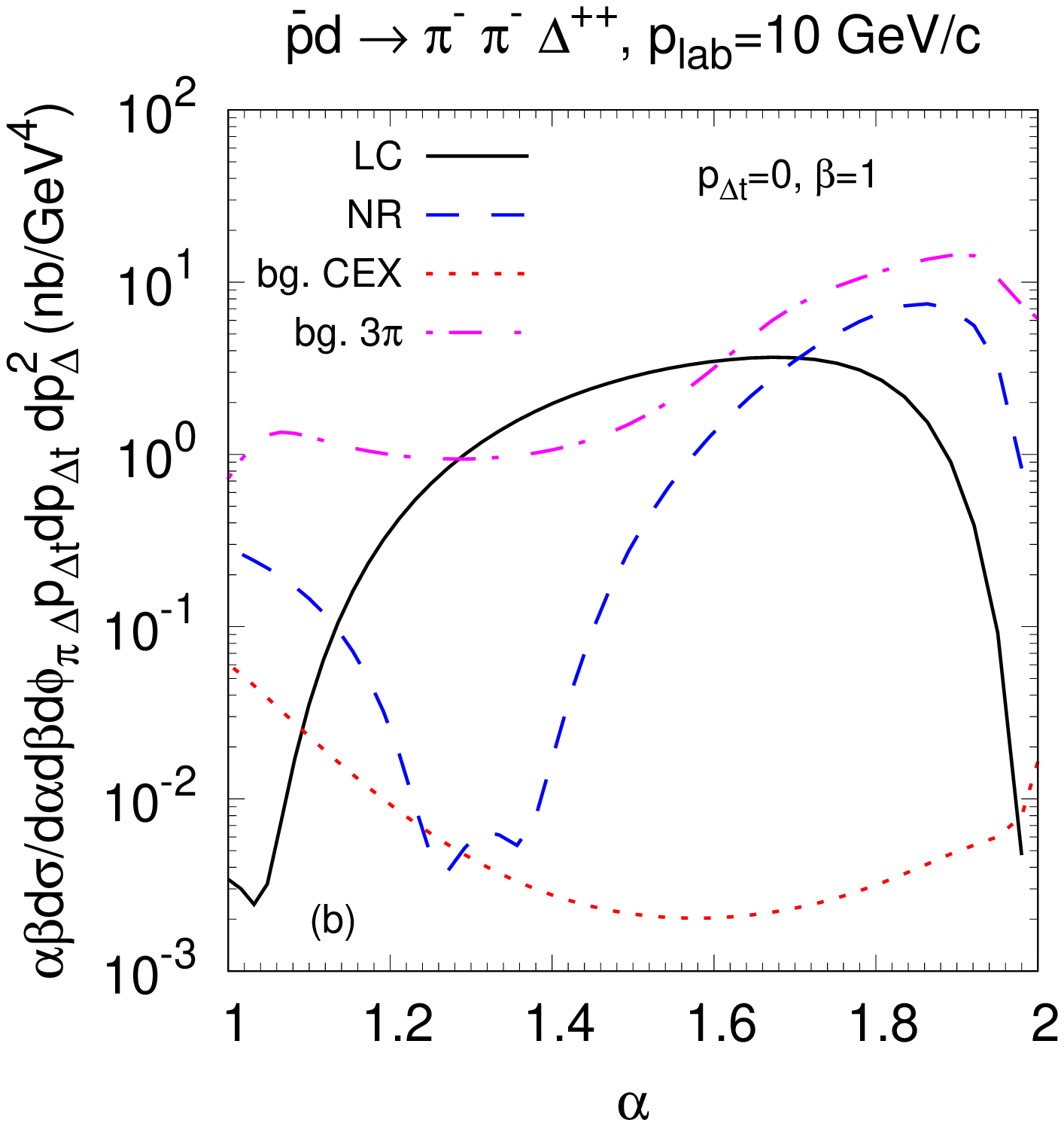}
  \includegraphics[scale = 0.50]{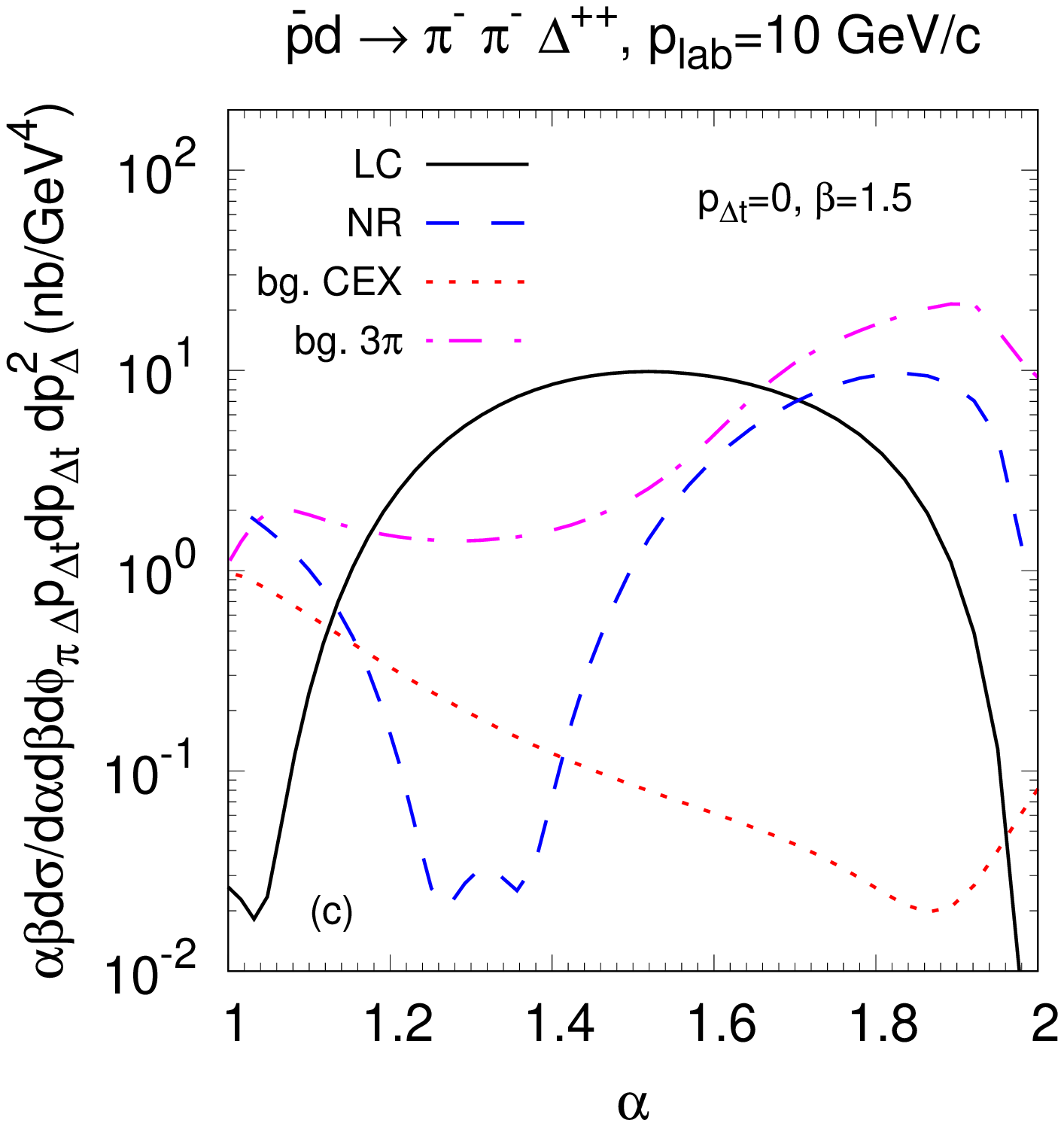}
  \caption{\label{fig:sig_alpha_plab10gev_pDt0gev} Differential cross section
    of the reaction $\bar p d \to \pi^- \pi^- \Delta^{++}$
    at $p_{\rm lab}=10$ GeV/c as a function of the LC momentum fraction
    $\alpha$ of the residual $\Delta$ (Eq.(\ref{alpha})).
    The $\Delta$ transverse momentum is fixed to zero.  
    The LC  momentum fraction of the 1$^{\rm{}st}$ pion
    $\beta=0.5$ (a), $\beta=1$ (b) and $\beta=1.5$ (c).
    Notations for lines are the same as in Fig.~\ref{fig:d^5sig}.}
\end{figure}
Fig.~\ref{fig:sig_alpha_plab10gev_pDt0gev} displays $\alpha$ distributions of the residual $\Delta$ at zero transverse momentum
for $\beta=0.5, 1$ and 1.5 for beam momentum $p_{\rm lab}=10$ GeV/c. Indeed, the shape of the $\alpha$ dependence of the signal
cross section reflects the shape of the momentum dependence of the $\Delta-\Delta$ configuration\footnote{The matrix elements
for $\beta=0.5$ and $\beta=1.5$ are identical due to two identical pions in the final state. Thus, in these two cases the cross
sections differ only due to the factor $\kappa_t$ of Eq.(\ref{kappa_t}).}. The latter has a maximum at $k=0.41$ GeV/c.
In the case of $p_{\Delta t}=0$ this maximum is reached at $\alpha=1.80$ and $0.95$ (NR), or at $\alpha=1.32$ and $0.68$ (LC).
Thus, due to the presence of the internal-momentum-dependent denominator in Eq.(\ref{k_int}) the strength of the $\alpha$-distribution
is shifted to smaller values of $\alpha$ (i.e. larger positive $p_{\Delta}^z$) in the case of LC calculation as compared to the NR one.
Therefore, due to relativistic effects, the signal should be clearly visible at intermediate values of $\alpha$ because
the background quickly decreases towards small $\alpha$. For $\beta=1.5$ the signal is more pronounced. This can be understood
by using the approximate relation $\beta = 1 + \cos(\Theta_{c.m.})$. Thus, $\beta=1$ corresponds to $\Theta_{c.m.}=90\degree$ while
$\beta=1.5$ corresponds to $\Theta_{c.m.}=60\degree$. In the latter case, as shown in Fig.~\ref{fig:pbarD-2pi-pi-_90deg}
of Appendix \ref{NbarD2pipi}, the elementary $\bar p \Delta^- \to \pi^- \pi^-$ differential cross section grows more slowly
with decreasing $s^\prime$ at small $s^\prime$.
(In order to avoid misunderstanding, we would like to point out that $s$ in the abscissa of Fig.~\ref{fig:pbarD-2pi-pi-_90deg}
has the meaning of $s^\prime$ in Eq.(\ref{sprime_to_s}).)
As a result, the distortion of the $\alpha$ dependence of the signal due to the elementary $\bar p \Delta^- \to \pi^- \pi^-$
differential cross section growing towards $\alpha \to 2$ is less pronounced for $\beta=1.5$ than for
$\beta=1$. Hence, we set $\beta=1.5$ as the default case. 

\begin{figure}[ht]
  \includegraphics[scale = 0.60]{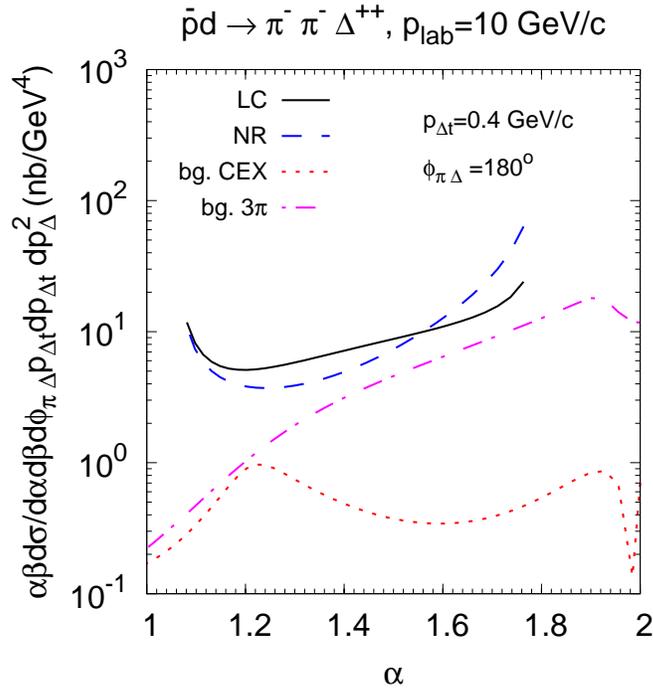}
  \caption{\label{fig:sig_alpha_plab10gev_pDt0.4gev} 
    Same as in Fig.~\ref{fig:sig_alpha_plab10gev_pDt0gev}c but for
    $\Delta$ transverse momentum of 0.4 GeV/c.
    The polar angle between the 1$^{\rm{}st}$ pion and
    the residual $\Delta$ is $\phi_{\pi\Delta}=180\degree$.
    The range of $\alpha$ within which the signal calculations are shown is restricted
    by the kinematic region where the struck $\Delta$ is time-like.}
\end{figure}
At finite transverse momentum of the residual $\Delta$, the range of $\alpha$ where the struck $\Delta$ is still time-like becomes narrower.
This is demonstrated in Fig.~\ref{fig:sig_alpha_plab10gev_pDt0.4gev}. At the limiting values of $\alpha$ the signal cross section diverges
because the density matrix of a spin-3/2 particle (the numerator in Eq.(\ref{iG^munuq})) becomes singular for $m_\Delta \to 0$. In other words,
our calculation becomes unreliable for far-offshell struck $\Delta$. Below we focus on the kinematics with $p_{\Delta t}=0$.

\begin{figure}[ht]
  \includegraphics[scale = 0.50]{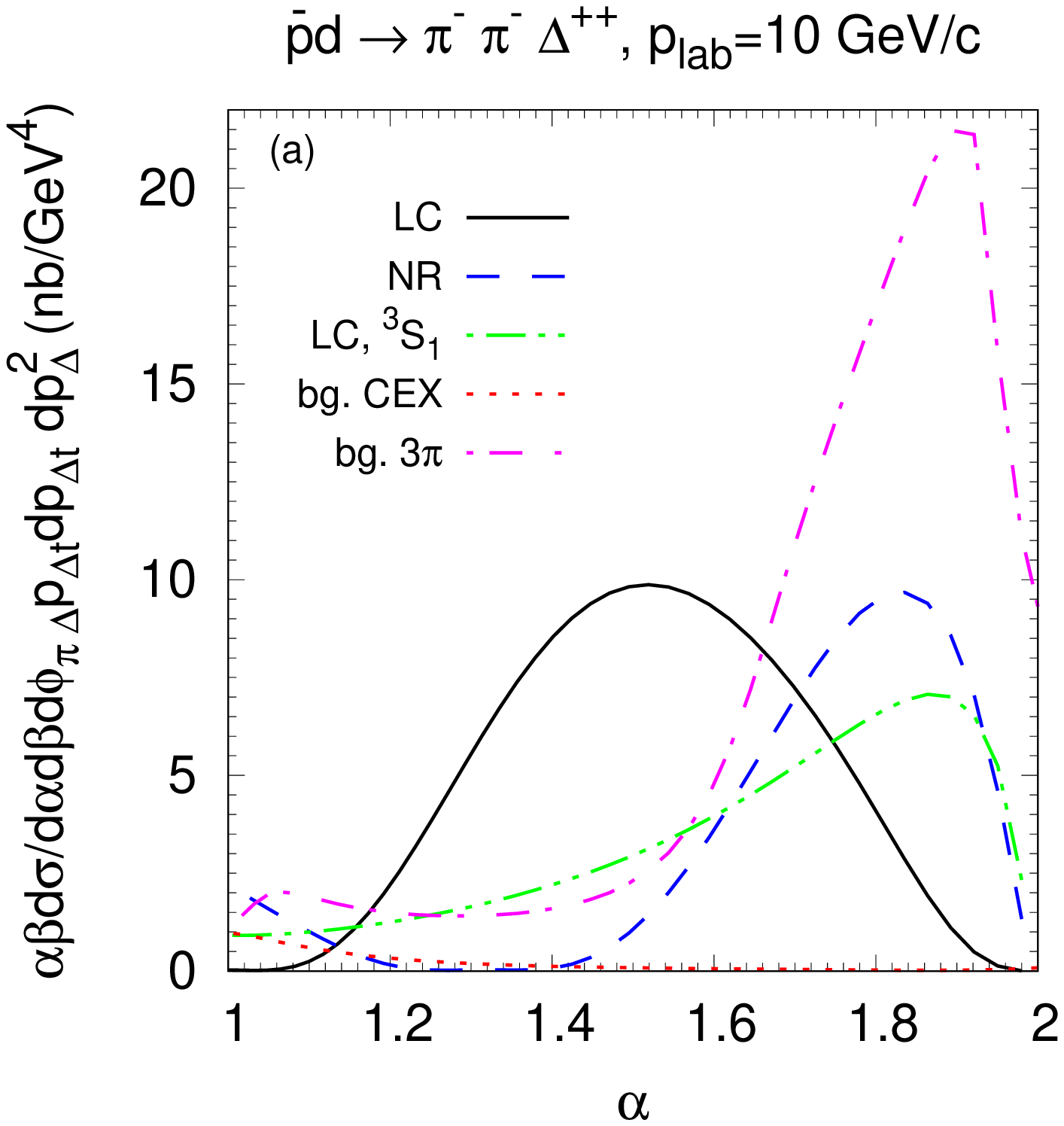}
  \includegraphics[scale = 0.50]{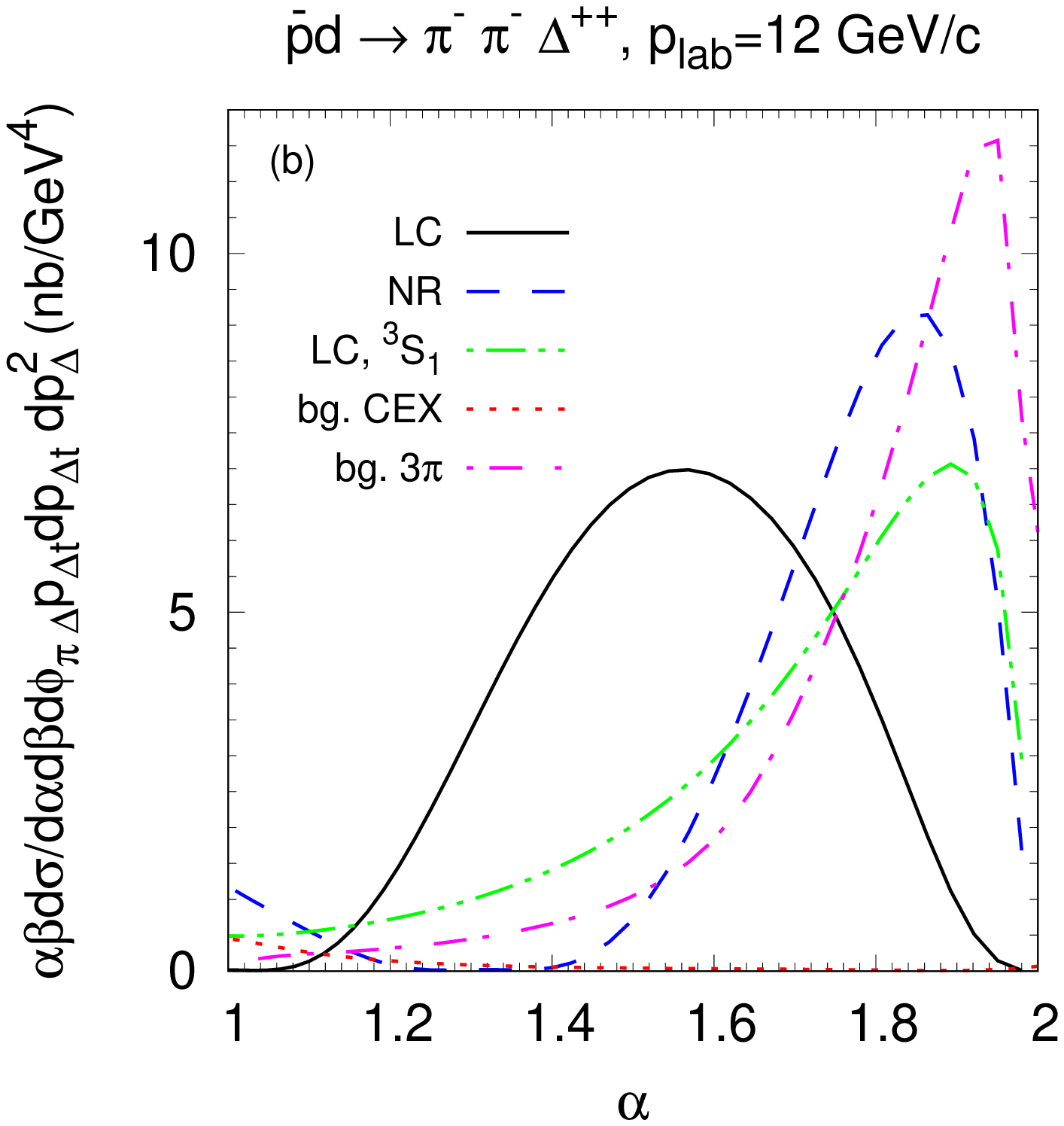}
  \includegraphics[scale = 0.50]{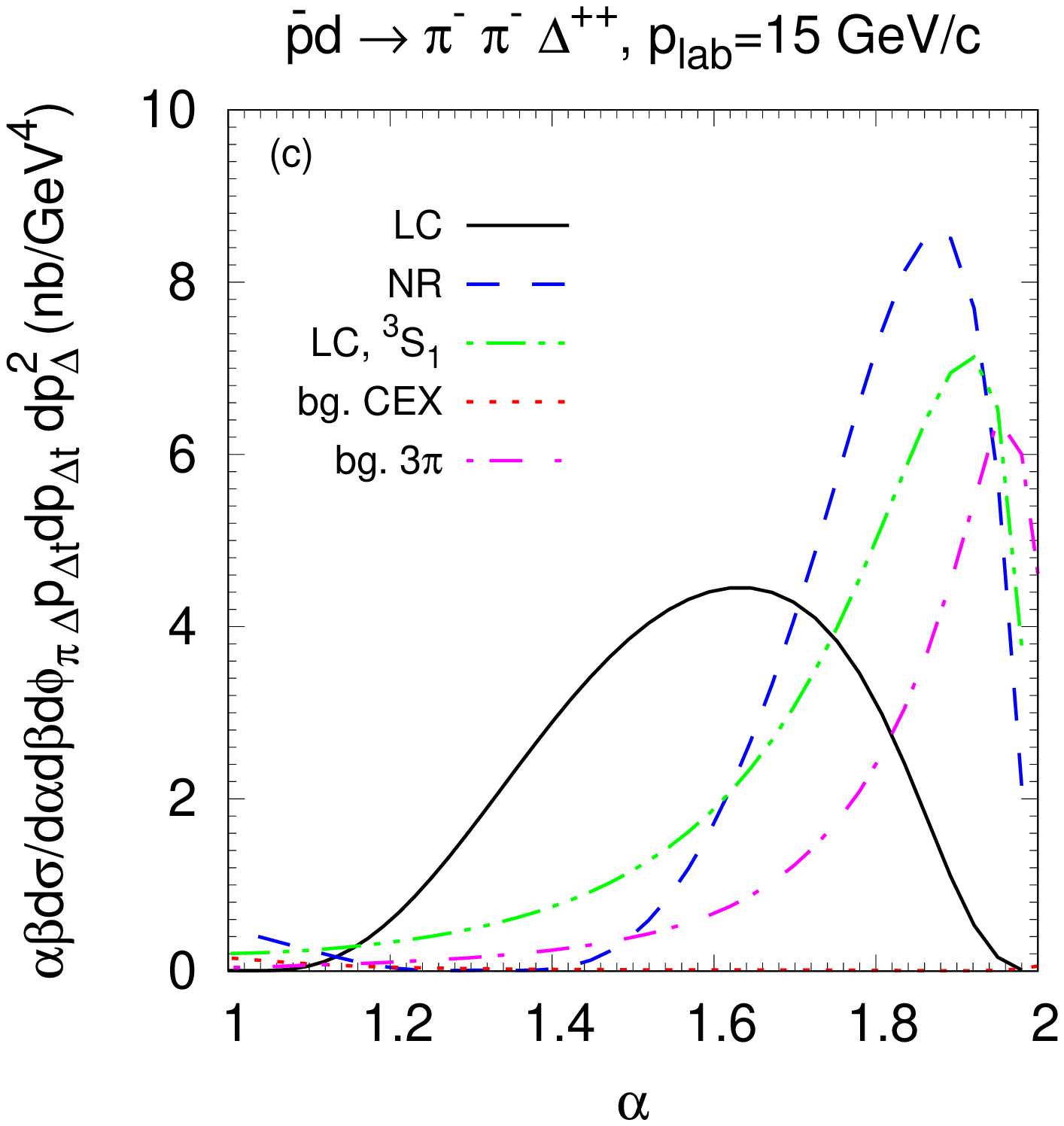}
  \caption{\label{fig:sig_alpha_pDt0gev_nonlog} Differential cross section of
    the reaction $\bar p d \to \pi^- \pi^- \Delta^{++}$ vs $\alpha$ of the residual $\Delta$
    for $p_{\rm lab}=10$ GeV/c (a) 12 GeV/c (b) and 15 GeV/c (c).
    Calculations are done with $p_{\Delta t}=0, \beta=1.5$.
    The dot-dot-dashed (green) line shows the LC signal calculated with the large-distance asymptotic
    form of the $\Delta-\Delta$ wave function, Eq.(\ref{phi_k}).
    The notations for other lines are the same as in Fig.~\ref{fig:d^5sig}.}
\end{figure}
Fig.~\ref{fig:sig_alpha_pDt0gev_nonlog} compares the $\alpha$ distributions
at beam momenta $p_{\rm lab}=10, 12$ and 15 GeV/c on the linear scale.
Both signal and background cross sections slightly decrease with increasing $p_{\rm lab}$.
However, the background decreases faster and becomes smoother at higher beam momenta.
Hence, the peak in the signal cross section becomes more pronounced with increasing $p_{\rm lab}$.
We also observe a strong influence of the underlying model for the
$\Delta-\Delta$ wave function on the results: the LC calculation with the
$^3S_1$ wave function produces an $\alpha$ distribution enhanced at larger $\alpha$ values.
This is related to the larger high-momentum tail of the $^3S_1$
wave function, as seen from Fig.~\ref{fig:phiDD2}. 
The NR $\Delta-\Delta$ wave function -- as compared to the LC one -- leads to
the signal $\alpha$ distribution shifted to larger $\alpha$ values, but its shape is still clearly
distinguishable from the background.

\begin{figure}[ht]
  \includegraphics[scale = 0.60]{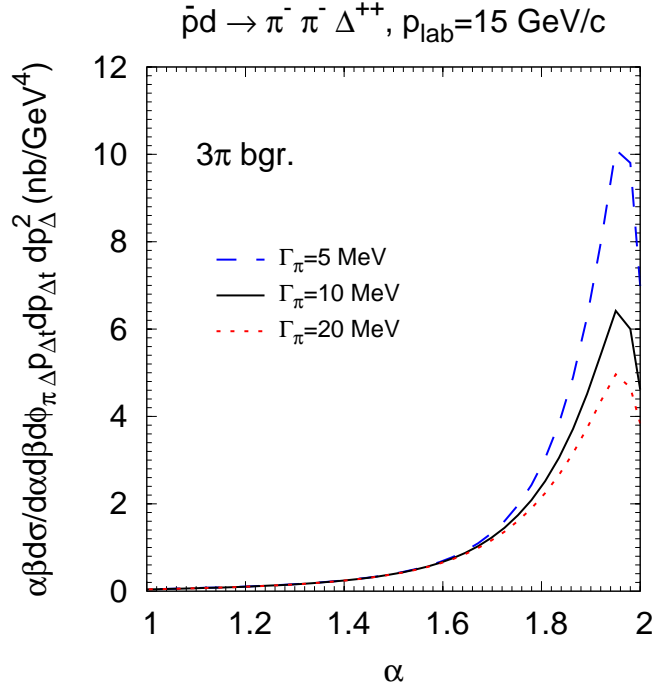}
  \caption{\label{fig:sig_alpha_plab15gev_pDt0gev_gampi} The three-pion
    annihilation background cross section as a function of $\alpha$ of the
    residual $\Delta$ for $p_{\rm lab}=15$ GeV/c at the
      kinematic condition $p_{\Delta t}=0$, $\beta=1.5$.
    The curves show the calculations with different choice of the artificial pion width as indicated.}
\end{figure}
As discussed in sec. \ref{3pi}, in the calculation of the three-pion
background we had to introduce a finite value for the width of the intermediate $\pi^+$. 
Fig.~\ref{fig:sig_alpha_plab15gev_pDt0gev_gampi} displays the influence of the choice of the pion width in our calculations.
The height of the peak close to $\alpha=2$ depends on the choice of the pion width. However, the background in the range $\alpha < 1.7$
is stable against variation of $\Gamma_\pi$.

We have performed a Monte-Carlo sampling of events in the
three-body phase space of outgoing $\pi_1,\pi_2,\Delta$ according to the probability
\begin{equation}
  dP \propto \overline{|M_{\pi_1\pi_2\Delta;\bar p d}|^2} d\Phi_3(p_{\bar p}+p_d;p_1,p_2,p_{\Delta}) {\cal A}(p_\Delta^2) dp_\Delta^2~,   \label{dP}
\end{equation}
where
\begin{equation}
  d\Phi_3(p_{\bar p}+p_d;p_1,p_2,p_{\Delta}) = \delta^{(4)}(p_1+p_2+p_{\Delta}-p_{\bar p}-p_d) \frac{d^3p_{\Delta}}{(2\pi)^32E_{\Delta}}
                \frac{d^3p_1}{(2\pi)^32E_1} \frac{d^3p_2}{(2\pi)^32E_2}    \label{dPhi_3}
\end{equation}
is the three-body phase space volume element.
\begin{figure}[ht]
  \includegraphics[scale = 0.60]{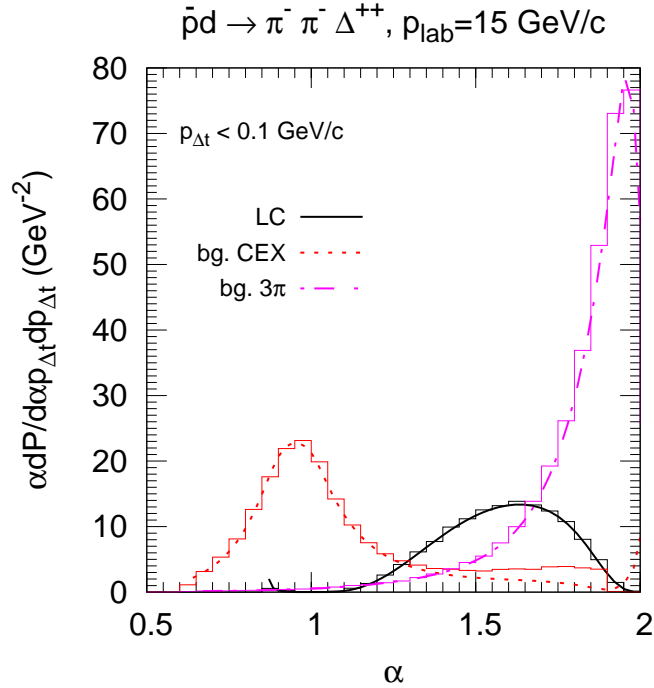}
  \caption{\label{fig:dP_dalpha_plab15gev_pDt0gev_phi_piD_int}
    Histograms: probability distributions of the residual $\Delta^{++}$ in $\alpha$
    at $p_{\rm lab}=15$ GeV/c in the kinematics with
    $p_{\Delta}^2=m_\Delta^2$; $p_{\Delta t} < 0.1$ GeV/c; $\beta=1.4\div1.6$; $p_1^z, p_2^z > 1$ GeV/c.
    Smooth lines: the analytical results of Fig.~\ref{fig:sig_alpha_pDt0gev_nonlog}c multiplied by constant factors
    for appropriate normalization.}
\end{figure}
Fig.~\ref{fig:dP_dalpha_plab15gev_pDt0gev_phi_piD_int} shows the
$\alpha$ distributions of simulated $\bar p d \to \pi^- \pi^- \Delta^{++}$ signal and background events
for small transverse momentum of the residual $\Delta^{++}$.
As we see, the sampled and analytical distributions practically coincide.
Some deviation of the CEX background from the analytical result is due to its
strong sensitivity to the transverse momentum of $\Delta$ at $\alpha \simeq 2$.
(In the simulations we included the cut $p_{\Delta t} < 1.9$ GeV/c for the CEX background).
Note that the absolute values of the differential cross section are not
accessible from Fig.~\ref{fig:dP_dalpha_plab15gev_pDt0gev_phi_piD_int} since
the sampled distributions are normalized to unity after integration over $\alpha$ and $p_{\Delta t}$.

\section{Summary and conclusions}
\label{summary}

We have theoretically studied the effect of the $\Delta^--\Delta^{++}$ configuration of the deuteron on the
differential cross sections of the exclusive reaction $\bar p d \to \pi^- \pi^- \Delta^{++}$. For the analysis
we used the ordinary deuteron wave functions and the wave functions of the $\Delta-\Delta$ configuration according
to the CCF model of ref. \cite{Haidenbauer:1993pw}.
The signal cross section is proportional to the wave function squared of the $\Delta-\Delta$
configuration in momentum space and the matrix element squared of the $\bar p \Delta^- \to \pi^- \pi^-$ process.
The latter has been calculated within the $N,\Delta$ exchange model. Two types of possible background sources due to
the following two-step processes have been considered:
(i) $\bar p n \to \pi^- \pi^0$, $\pi^0 p \to \pi^- \Delta^{++}$ and
(ii) $\bar p n \to \pi^- \pi^- \pi^+$, $\pi^+ p \to \Delta^{++}$.
The discussion was focused on kinematics with large momentum transfer between $\bar p$ and both $\pi^-$ mesons.

We have found that the pion CEX background (i) is important for forward
production of $\Delta^{++}$ (cf. Fig.~\ref{fig:d^5sig}d). In this case the
$\Delta^{++}$ may experience large longitudinal momentum transfer from the
scattered pion.
In other situations the CEX background is strongly suppressed relative to the
three-pion annihilation background (ii). 
The latter background grows strongly for backward $\Delta^{++}$, because in
this case the c.m. energy of the colliding $\bar p n$ system
is small which leads to a large $\bar p n \to \pi^- \pi^- \pi^+$ amplitude.

Owing to the large binding of the $\Delta-\Delta$ configuration in the
deuteron, the momentum distribution becomes significantly harder than in the
ordinary deuteron.
This leads to important relativistic corrections which have
been taken into account in this work within the LC theory.
Moreover, the coupled channel models of the deuteron with strong tensor
interaction predict the dominance of the $^7D_1$ $\Delta-\Delta$ state which
produces a pronounced maximum at about 0.5 GeV/c c.m. momentum.
We have demonstrated that the combination of LC and coupled channel effects leads to a specific shape
of the $\alpha$ distribution of the residual $\Delta^{++}$ peaking at $\alpha \simeq 1.5\div1.6$
for zero transverse momentum, which manifests the maximum in the $\Delta-\Delta$ c.m. momentum distribution.
This behaviour of the signal cross section is clearly distinguishable from the
three-pion annihilation background smoothly increasing with $\alpha$.
We have also found that there is a broad kinematic range of residual $\Delta^{++}$
($\alpha = 1.2\div1.7,~p_{\Delta t} < 0.4$ GeV/c), where the one-step signal process dominates over
the two-step background processes.
Even if the $\Delta-\Delta$ probability would be reduced by a factor of five
down to $\sim 0.3\%$ -- the $\alpha$-distribution of the $\Delta$ at low transverse momentum
(Fig.~\ref{fig:sig_alpha_pDt0gev_nonlog}c) would still allow to see the contribution of the $\bar p$ annihilation
on the $\Delta-\Delta$ component.
These findings can be used not only to test the presence of the $\Delta-\Delta$ configuration in the deuteron,
but also to explore its c.m. momentum distribution.

On the basis of our model we have developed a Monte-Carlo event generator which can be applied
for detailed feasibility studies with the PANDA detector system.
The results of these studies will be published elsewhere.
Note that a complementary test of the $^7D_1$ $\Delta-\Delta$ state dominance would
be possible with a polarized deuteron target at PANDA.

Finally, we note that the previous experimental analyses quoted in sec. \ref{intro} do not take
into account the LC wave function, and thus their conclusions on limits to the probability of
a $\Delta-\Delta$ configuration need to be taken with caution.

\begin{acknowledgments}
  Stimulating discussions with J.~Ritman are gratefully acknowledged.
  The research of M.S. was supported by the U.S. Department of Energy,
  Office of Science, Office of Nuclear Physics, under Award No. DE-FG02-93ER40771.
  The computational resources of the Frankfurt Center for Scientific Computing (FUCHS-CSC) have been used
  in this work.
\end{acknowledgments}

\bibliography{pbarDelDel}

\begin{thebibliography}{42}%
\makeatletter
\providecommand \@ifxundefined [1]{%
 \@ifx{#1\undefined}
}%
\providecommand \@ifnum [1]{%
 \ifnum #1\expandafter \@firstoftwo
 \else \expandafter \@secondoftwo
 \fi
}%
\providecommand \@ifx [1]{%
 \ifx #1\expandafter \@firstoftwo
 \else \expandafter \@secondoftwo
 \fi
}%
\providecommand \natexlab [1]{#1}%
\providecommand \enquote  [1]{``#1''}%
\providecommand \bibnamefont  [1]{#1}%
\providecommand \bibfnamefont [1]{#1}%
\providecommand \citenamefont [1]{#1}%
\providecommand \href@noop [0]{\@secondoftwo}%
\providecommand \href [0]{\begingroup \@sanitize@url \@href}%
\providecommand \@href[1]{\@@startlink{#1}\@@href}%
\providecommand \@@href[1]{\endgroup#1\@@endlink}%
\providecommand \@sanitize@url [0]{\catcode `\\12\catcode `\$12\catcode
  `\&12\catcode `\#12\catcode `\^12\catcode `\_12\catcode `\%12\relax}%
\providecommand \@@startlink[1]{}%
\providecommand \@@endlink[0]{}%
\providecommand \url  [0]{\begingroup\@sanitize@url \@url }%
\providecommand \@url [1]{\endgroup\@href {#1}{\urlprefix }}%
\providecommand \urlprefix  [0]{URL }%
\providecommand \Eprint [0]{\href }%
\providecommand \doibase [0]{http://dx.doi.org/}%
\providecommand \selectlanguage [0]{\@gobble}%
\providecommand \bibinfo  [0]{\@secondoftwo}%
\providecommand \bibfield  [0]{\@secondoftwo}%
\providecommand \translation [1]{[#1]}%
\providecommand \BibitemOpen [0]{}%
\providecommand \bibitemStop [0]{}%
\providecommand \bibitemNoStop [0]{.\EOS\space}%
\providecommand \EOS [0]{\spacefactor3000\relax}%
\providecommand \BibitemShut  [1]{\csname bibitem#1\endcsname}%
\let\auto@bib@innerbib\@empty
%</preamble>
\bibitem [{\citenamefont {Haapakoski}\ and\ \citenamefont
  {Saarela}(1974)}]{Haapakoski:1974qnh}%
  \BibitemOpen
  \bibfield  {author} {\bibinfo {author} {\bibfnamefont {P.}~\bibnamefont
  {Haapakoski}}\ and\ \bibinfo {author} {\bibfnamefont {M.}~\bibnamefont
  {Saarela}},\ }\href {\doibase 10.1016/0370-2693(74)90395-5} {\bibfield
  {journal} {\bibinfo  {journal} {Phys. Lett.}\ }\textbf {\bibinfo {volume}
  {53B}},\ \bibinfo {pages} {333} (\bibinfo {year} {1974})}\BibitemShut
  {NoStop}%
%%CITATION = PHLTA,53B,333;%%
\bibitem [{\citenamefont {Arenh\"ovel}(1975)}]{Arenhovel:1975dx}%
  \BibitemOpen
  \bibfield  {author} {\bibinfo {author} {\bibfnamefont {H.}~\bibnamefont
  {Arenh\"ovel}},\ }\href {\doibase 10.1007/BF01409596} {\bibfield  {journal}
  {\bibinfo  {journal} {Z. Phys.}\ }\textbf {\bibinfo {volume} {A275}},\
  \bibinfo {pages} {189} (\bibinfo {year} {1975})}\BibitemShut {NoStop}%
%%CITATION = ZEPYA,A275,189;%%
\bibitem [{\citenamefont {Dymarz}\ and\ \citenamefont
  {Khanna}(1990)}]{Dymarz:1990ac}%
  \BibitemOpen
  \bibfield  {author} {\bibinfo {author} {\bibfnamefont {R.}~\bibnamefont
  {Dymarz}}\ and\ \bibinfo {author} {\bibfnamefont {F.~C.}\ \bibnamefont
  {Khanna}},\ }\href {\doibase 10.1016/0375-9474(90)90127-8} {\bibfield
  {journal} {\bibinfo  {journal} {Nucl. Phys.}\ }\textbf {\bibinfo {volume}
  {A516}},\ \bibinfo {pages} {549} (\bibinfo {year} {1990})}\BibitemShut
  {NoStop}%
%%CITATION = NUPHA,A516,549;%%
\bibitem [{\citenamefont {Haidenbauer}\ \emph {et~al.}(1993)\citenamefont
  {Haidenbauer}, \citenamefont {Holinde},\ and\ \citenamefont
  {Johnson}}]{Haidenbauer:1993pw}%
  \BibitemOpen
  \bibfield  {author} {\bibinfo {author} {\bibfnamefont {J.}~\bibnamefont
  {Haidenbauer}}, \bibinfo {author} {\bibfnamefont {K.}~\bibnamefont
  {Holinde}}, \ and\ \bibinfo {author} {\bibfnamefont {M.~B.}\ \bibnamefont
  {Johnson}},\ }\href {\doibase 10.1103/PhysRevC.48.2190} {\bibfield  {journal}
  {\bibinfo  {journal} {Phys. Rev.}\ }\textbf {\bibinfo {volume} {C48}},\
  \bibinfo {pages} {2190} (\bibinfo {year} {1993})}\BibitemShut {NoStop}%
%%CITATION = PHRVA,C48,2190;%%
\bibitem [{\citenamefont {Glozman}\ \emph {et~al.}(1993)\citenamefont
  {Glozman}, \citenamefont {Neudatchin},\ and\ \citenamefont
  {Obukhovsky}}]{Glozman:1993pm}%
  \BibitemOpen
  \bibfield  {author} {\bibinfo {author} {\bibfnamefont {L.~{\relax Ya}.}\
  \bibnamefont {Glozman}}, \bibinfo {author} {\bibfnamefont {V.~G.}\
  \bibnamefont {Neudatchin}}, \ and\ \bibinfo {author} {\bibfnamefont {I.~T.}\
  \bibnamefont {Obukhovsky}},\ }\href {\doibase 10.1103/PhysRevC.48.389}
  {\bibfield  {journal} {\bibinfo  {journal} {Phys. Rev.}\ }\textbf {\bibinfo
  {volume} {C48}},\ \bibinfo {pages} {389} (\bibinfo {year}
  {1993})}\BibitemShut {NoStop}%
%%CITATION = PHRVA,C48,389;%%
\bibitem [{\citenamefont {Glozman}\ and\ \citenamefont
  {Kuchina}(1994)}]{Glozman:1994xe}%
  \BibitemOpen
  \bibfield  {author} {\bibinfo {author} {\bibfnamefont {L.~{\relax Ya}.}\
  \bibnamefont {Glozman}}\ and\ \bibinfo {author} {\bibfnamefont {E.~I.}\
  \bibnamefont {Kuchina}},\ }\href {\doibase 10.1103/PhysRevC.49.1149,
  10.1103/PhysRevC.53.2560} {\bibfield  {journal} {\bibinfo  {journal} {Phys.
  Rev.}\ }\textbf {\bibinfo {volume} {C49}},\ \bibinfo {pages} {1149} (\bibinfo
  {year} {1994})},\ \bibinfo {note} {[Erratum: Phys.
  Rev.C53,2560(1996)]}\BibitemShut {NoStop}%
%%CITATION = PHRVA,C49,1149;%%
\bibitem [{\citenamefont {Benz}\ and\ \citenamefont
  {Soding}(1974)}]{Benz:1974au}%
  \BibitemOpen
  \bibfield  {author} {\bibinfo {author} {\bibfnamefont {P.}~\bibnamefont
  {Benz}}\ and\ \bibinfo {author} {\bibfnamefont {P.}~\bibnamefont {Soding}},\
  }\href {\doibase 10.1016/0370-2693(74)90064-1} {\bibfield  {journal}
  {\bibinfo  {journal} {Phys. Lett.}\ }\textbf {\bibinfo {volume} {52B}},\
  \bibinfo {pages} {367} (\bibinfo {year} {1974})}\BibitemShut {NoStop}%
%%CITATION = PHLTA,52B,367;%%
\bibitem [{\citenamefont {Braun}\ \emph {et~al.}(1974)\citenamefont {Braun},
  \citenamefont {Brick}, \citenamefont {Fridman}, \citenamefont {Gerber},
  \citenamefont {Juillot}, \citenamefont {Maurer}, \citenamefont {Michalon},
  \citenamefont {Michalon-Mentzer}, \citenamefont {Strub},\ and\ \citenamefont
  {Voltolini}}]{Braun:1974fy}%
  \BibitemOpen
  \bibfield  {author} {\bibinfo {author} {\bibfnamefont {H.}~\bibnamefont
  {Braun}}, \bibinfo {author} {\bibfnamefont {D.}~\bibnamefont {Brick}},
  \bibinfo {author} {\bibfnamefont {A.}~\bibnamefont {Fridman}}, \bibinfo
  {author} {\bibfnamefont {J.-P.}\ \bibnamefont {Gerber}}, \bibinfo {author}
  {\bibfnamefont {P.}~\bibnamefont {Juillot}}, \bibinfo {author} {\bibfnamefont
  {G.}~\bibnamefont {Maurer}}, \bibinfo {author} {\bibfnamefont
  {A.}~\bibnamefont {Michalon}}, \bibinfo {author} {\bibfnamefont {M.-E.}\
  \bibnamefont {Michalon-Mentzer}}, \bibinfo {author} {\bibfnamefont
  {R.}~\bibnamefont {Strub}}, \ and\ \bibinfo {author} {\bibfnamefont
  {C.}~\bibnamefont {Voltolini}},\ }\href {\doibase 10.1103/PhysRevLett.33.312}
  {\bibfield  {journal} {\bibinfo  {journal} {Phys. Rev. Lett.}\ }\textbf
  {\bibinfo {volume} {33}},\ \bibinfo {pages} {312} (\bibinfo {year}
  {1974})}\BibitemShut {NoStop}%
%%CITATION = PRLTA,33,312;%%
\bibitem [{\citenamefont {Allasia}\ \emph {et~al.}(1986)\citenamefont {Allasia}
  \emph {et~al.}}]{Allasia:1986kg}%
  \BibitemOpen
  \bibfield  {author} {\bibinfo {author} {\bibfnamefont {D.}~\bibnamefont
  {Allasia}} \emph {et~al.},\ }\href {\doibase 10.1016/0370-2693(86)91035-X}
  {\bibfield  {journal} {\bibinfo  {journal} {Phys. Lett.}\ }\textbf {\bibinfo
  {volume} {B174}},\ \bibinfo {pages} {450} (\bibinfo {year}
  {1986})}\BibitemShut {NoStop}%
%%CITATION = PHLTA,B174,450;%%
\bibitem [{\citenamefont {Bertin}\ \emph {et~al.}(1997)\citenamefont {Bertin}
  \emph {et~al.}}]{Bertin:1997dw}%
  \BibitemOpen
  \bibfield  {author} {\bibinfo {author} {\bibfnamefont {A.}~\bibnamefont
  {Bertin}} \emph {et~al.} (\bibinfo {collaboration} {OBELIX}),\ }\href
  {\doibase 10.1016/S0370-2693(97)00490-5} {\bibfield  {journal} {\bibinfo
  {journal} {Phys. Lett.}\ }\textbf {\bibinfo {volume} {B403}},\ \bibinfo
  {pages} {177} (\bibinfo {year} {1997})}\BibitemShut {NoStop}%
%%CITATION = PHLTA,B403,177;%%
\bibitem [{\citenamefont {Denisov}\ \emph {et~al.}(1999)\citenamefont
  {Denisov}, \citenamefont {Kuksa},\ and\ \citenamefont
  {Lykasov}}]{Denisov:1998sq}%
  \BibitemOpen
  \bibfield  {author} {\bibinfo {author} {\bibfnamefont {O.~{\relax Yu}.}\
  \bibnamefont {Denisov}}, \bibinfo {author} {\bibfnamefont {S.~D.}\
  \bibnamefont {Kuksa}}, \ and\ \bibinfo {author} {\bibfnamefont {G.~I.}\
  \bibnamefont {Lykasov}},\ }\href {\doibase 10.1016/S0370-2693(99)00562-6}
  {\bibfield  {journal} {\bibinfo  {journal} {Phys. Lett.}\ }\textbf {\bibinfo
  {volume} {B458}},\ \bibinfo {pages} {415} (\bibinfo {year}
  {1999})}\BibitemShut {NoStop}%
\bibitem [{\citenamefont {Frankfurt}\ and\ \citenamefont
  {Strikman}(1979)}]{Frankfurt:1977vc}%
  \BibitemOpen
  \bibfield  {author} {\bibinfo {author} {\bibfnamefont {L.~L.}\ \bibnamefont
  {Frankfurt}}\ and\ \bibinfo {author} {\bibfnamefont {M.~I.}\ \bibnamefont
  {Strikman}},\ }\href {\doibase 10.1016/0550-3213(79)90018-X} {\bibfield
  {journal} {\bibinfo  {journal} {Nucl. Phys.}\ }\textbf {\bibinfo {volume}
  {B148}},\ \bibinfo {pages} {107} (\bibinfo {year} {1979})}\BibitemShut
  {NoStop}%
%%CITATION = NUPHA,B148,107;%%
\bibitem [{\citenamefont {Frankfurt}\ and\ \citenamefont
  {Strikman}(1981)}]{Frankfurt:1981mk}%
  \BibitemOpen
  \bibfield  {author} {\bibinfo {author} {\bibfnamefont {L.}~\bibnamefont
  {Frankfurt}}\ and\ \bibinfo {author} {\bibfnamefont {M.}~\bibnamefont
  {Strikman}},\ }\href@noop {} {\bibfield  {journal} {\bibinfo  {journal}
  {Phys. Rep.}\ }\textbf {\bibinfo {volume} {76}},\ \bibinfo {pages} {215}
  (\bibinfo {year} {1981})}\BibitemShut {NoStop}%
%%CITATION = PRPLC,76,215;%%
\bibitem [{\citenamefont {Berestetskii}\ \emph {et~al.}(1971)\citenamefont
  {Berestetskii}, \citenamefont {Lifshitz},\ and\ \citenamefont
  {Pitaevskii}}]{BLP}%
  \BibitemOpen
  \bibfield  {author} {\bibinfo {author} {\bibfnamefont {V.~B.}\ \bibnamefont
  {Berestetskii}}, \bibinfo {author} {\bibfnamefont {E.~M.}\ \bibnamefont
  {Lifshitz}}, \ and\ \bibinfo {author} {\bibfnamefont {L.~P.}\ \bibnamefont
  {Pitaevskii}},\ }\href@noop {} {\emph {\bibinfo {title} {Relativistic Quantum
  Theory}}}\ (\bibinfo  {publisher} {Pergamon Press},\ \bibinfo {year}
  {1971})\BibitemShut {NoStop}%
\bibitem [{\citenamefont {Frankfurt}\ \emph {et~al.}(1997)\citenamefont
  {Frankfurt}, \citenamefont {Piasetzky}, \citenamefont {Sargsian},\ and\
  \citenamefont {Strikman}}]{Frankfurt:1996uz}%
  \BibitemOpen
  \bibfield  {author} {\bibinfo {author} {\bibfnamefont {L.~L.}\ \bibnamefont
  {Frankfurt}}, \bibinfo {author} {\bibfnamefont {E.}~\bibnamefont
  {Piasetzky}}, \bibinfo {author} {\bibfnamefont {M.~M.}\ \bibnamefont
  {Sargsian}}, \ and\ \bibinfo {author} {\bibfnamefont {M.~I.}\ \bibnamefont
  {Strikman}},\ }\href {\doibase 10.1103/PhysRevC.56.2752} {\bibfield
  {journal} {\bibinfo  {journal} {Phys. Rev.}\ }\textbf {\bibinfo {volume}
  {C56}},\ \bibinfo {pages} {2752} (\bibinfo {year} {1997})}\BibitemShut
  {NoStop}%
\bibitem [{\citenamefont {Machleidt}\ \emph {et~al.}(1987)\citenamefont
  {Machleidt}, \citenamefont {Holinde},\ and\ \citenamefont
  {Elster}}]{Machleidt:1987hj}%
  \BibitemOpen
  \bibfield  {author} {\bibinfo {author} {\bibfnamefont {R.}~\bibnamefont
  {Machleidt}}, \bibinfo {author} {\bibfnamefont {K.}~\bibnamefont {Holinde}},
  \ and\ \bibinfo {author} {\bibfnamefont {C.}~\bibnamefont {Elster}},\ }\href
  {\doibase 10.1016/S0370-1573(87)80002-9} {\bibfield  {journal} {\bibinfo
  {journal} {Phys. Rept.}\ }\textbf {\bibinfo {volume} {149}},\ \bibinfo
  {pages} {1} (\bibinfo {year} {1987})}\BibitemShut {NoStop}%
%%CITATION = PRPLC,149,1;%%
\bibitem [{\citenamefont {Varshalovich}\ \emph {et~al.}(1988)\citenamefont
  {Varshalovich}, \citenamefont {Moskalev},\ and\ \citenamefont
  {Khersonskii}}]{Varshalovich}%
  \BibitemOpen
  \bibfield  {author} {\bibinfo {author} {\bibfnamefont {D.~A.}\ \bibnamefont
  {Varshalovich}}, \bibinfo {author} {\bibfnamefont {A.~N.}\ \bibnamefont
  {Moskalev}}, \ and\ \bibinfo {author} {\bibfnamefont {V.~K.}\ \bibnamefont
  {Khersonskii}},\ }\href@noop {} {\emph {\bibinfo {title} {Quantum Theory of
  Angular Momentum}}}\ (\bibinfo  {publisher} {World Scientific},\ \bibinfo
  {address} {Singapore},\ \bibinfo {year} {1988})\BibitemShut {NoStop}%
\bibitem [{\citenamefont {Niephaus}\ \emph {et~al.}(1979)\citenamefont
  {Niephaus}, \citenamefont {Gari},\ and\ \citenamefont
  {Sommer}}]{Niephaus:1979mw}%
  \BibitemOpen
  \bibfield  {author} {\bibinfo {author} {\bibfnamefont {G.~H.}\ \bibnamefont
  {Niephaus}}, \bibinfo {author} {\bibfnamefont {M.}~\bibnamefont {Gari}}, \
  and\ \bibinfo {author} {\bibfnamefont {B.}~\bibnamefont {Sommer}},\ }\href
  {\doibase 10.1103/PhysRevC.20.1096} {\bibfield  {journal} {\bibinfo
  {journal} {Phys. Rev.}\ }\textbf {\bibinfo {volume} {C20}},\ \bibinfo {pages}
  {1096} (\bibinfo {year} {1979})}\BibitemShut {NoStop}%
%%CITATION = PHRVA,C20,1096;%%
\bibitem [{\citenamefont {Wiringa}\ \emph {et~al.}(1984)\citenamefont
  {Wiringa}, \citenamefont {Smith},\ and\ \citenamefont
  {Ainsworth}}]{Wiringa:1984tg}%
  \BibitemOpen
  \bibfield  {author} {\bibinfo {author} {\bibfnamefont {R.~B.}\ \bibnamefont
  {Wiringa}}, \bibinfo {author} {\bibfnamefont {R.~A.}\ \bibnamefont {Smith}},
  \ and\ \bibinfo {author} {\bibfnamefont {T.~L.}\ \bibnamefont {Ainsworth}},\
  }\href {\doibase 10.1103/PhysRevC.29.1207} {\bibfield  {journal} {\bibinfo
  {journal} {Phys. Rev.}\ }\textbf {\bibinfo {volume} {C29}},\ \bibinfo {pages}
  {1207} (\bibinfo {year} {1984})}\BibitemShut {NoStop}%
%%CITATION = PHRVA,C29,1207;%%
\bibitem [{\citenamefont {Dymarz}\ and\ \citenamefont
  {Khanna}(1986)}]{Dymarz:1986km}%
  \BibitemOpen
  \bibfield  {author} {\bibinfo {author} {\bibfnamefont {R.}~\bibnamefont
  {Dymarz}}\ and\ \bibinfo {author} {\bibfnamefont {F.~C.}\ \bibnamefont
  {Khanna}},\ }\href {\doibase 10.1103/PhysRevLett.56.1448} {\bibfield
  {journal} {\bibinfo  {journal} {Phys. Rev. Lett.}\ }\textbf {\bibinfo
  {volume} {56}},\ \bibinfo {pages} {1448} (\bibinfo {year}
  {1986})}\BibitemShut {NoStop}%
%%CITATION = PRLTA,56,1448;%%
\bibitem [{\citenamefont {Lacombe}\ \emph {et~al.}(1981)\citenamefont
  {Lacombe}, \citenamefont {Loiseau}, \citenamefont {Vinh~Mau}, \citenamefont
  {Cote}, \citenamefont {Pires},\ and\ \citenamefont
  {de~Tourreil}}]{Lacombe:1981eg}%
  \BibitemOpen
  \bibfield  {author} {\bibinfo {author} {\bibfnamefont {M.}~\bibnamefont
  {Lacombe}}, \bibinfo {author} {\bibfnamefont {B.}~\bibnamefont {Loiseau}},
  \bibinfo {author} {\bibfnamefont {R.}~\bibnamefont {Vinh~Mau}}, \bibinfo
  {author} {\bibfnamefont {J.}~\bibnamefont {Cote}}, \bibinfo {author}
  {\bibfnamefont {P.}~\bibnamefont {Pires}}, \ and\ \bibinfo {author}
  {\bibfnamefont {R.}~\bibnamefont {de~Tourreil}},\ }\href {\doibase
  10.1016/0370-2693(81)90659-6} {\bibfield  {journal} {\bibinfo  {journal}
  {Phys. Lett.}\ }\textbf {\bibinfo {volume} {101B}},\ \bibinfo {pages} {139}
  (\bibinfo {year} {1981})}\BibitemShut {NoStop}%
%%CITATION = PHLTA,101B,139;%%
\bibitem [{\citenamefont {Weinberg}(1966)}]{Weinberg:1966jm}%
  \BibitemOpen
  \bibfield  {author} {\bibinfo {author} {\bibfnamefont {S.}~\bibnamefont
  {Weinberg}},\ }\href {\doibase 10.1103/PhysRev.150.1313} {\bibfield
  {journal} {\bibinfo  {journal} {Phys. Rev.}\ }\textbf {\bibinfo {volume}
  {150}},\ \bibinfo {pages} {1313} (\bibinfo {year} {1966})}\BibitemShut
  {NoStop}%
%%CITATION = PHRVA,150,1313;%%
\bibitem [{\citenamefont {Dmitriev}\ \emph {et~al.}(1986)\citenamefont
  {Dmitriev}, \citenamefont {Sushkov},\ and\ \citenamefont
  {Gaarde}}]{Dmitriev:1986st}%
  \BibitemOpen
  \bibfield  {author} {\bibinfo {author} {\bibfnamefont {V.}~\bibnamefont
  {Dmitriev}}, \bibinfo {author} {\bibfnamefont {O.}~\bibnamefont {Sushkov}}, \
  and\ \bibinfo {author} {\bibfnamefont {C.}~\bibnamefont {Gaarde}},\ }\href
  {\doibase 10.1016/0375-9474(86)90158-2} {\bibfield  {journal} {\bibinfo
  {journal} {Nucl. Phys.}\ }\textbf {\bibinfo {volume} {A459}},\ \bibinfo
  {pages} {503} (\bibinfo {year} {1986})}\BibitemShut {NoStop}%
%%CITATION = NUPHA,A459,503;%%
\bibitem [{\citenamefont {Brown}\ and\ \citenamefont
  {Weise}(1975)}]{Brown:1975di}%
  \BibitemOpen
  \bibfield  {author} {\bibinfo {author} {\bibfnamefont {G.~E.}\ \bibnamefont
  {Brown}}\ and\ \bibinfo {author} {\bibfnamefont {W.}~\bibnamefont {Weise}},\
  }\href {\doibase 10.1016/0370-1573(75)90026-5} {\bibfield  {journal}
  {\bibinfo  {journal} {Phys. Rept.}\ }\textbf {\bibinfo {volume} {22}},\
  \bibinfo {pages} {279} (\bibinfo {year} {1975})}\BibitemShut {NoStop}%
%%CITATION = PRPLC,22,279;%%
\bibitem [{\citenamefont {Tabakin}\ and\ \citenamefont
  {Eisenstein}(1985)}]{Tabakin:1985yv}%
  \BibitemOpen
  \bibfield  {author} {\bibinfo {author} {\bibfnamefont {F.}~\bibnamefont
  {Tabakin}}\ and\ \bibinfo {author} {\bibfnamefont {R.~A.}\ \bibnamefont
  {Eisenstein}},\ }\href {\doibase 10.1103/PhysRevC.31.1857} {\bibfield
  {journal} {\bibinfo  {journal} {Phys. Rev.}\ }\textbf {\bibinfo {volume}
  {C31}},\ \bibinfo {pages} {1857} (\bibinfo {year} {1985})}\BibitemShut
  {NoStop}%
%%CITATION = PHRVA,C31,1857;%%
\bibitem [{\citenamefont {Brodsky}\ and\ \citenamefont
  {Farrar}(1973)}]{Brodsky:1973kr}%
  \BibitemOpen
  \bibfield  {author} {\bibinfo {author} {\bibfnamefont {S.~J.}\ \bibnamefont
  {Brodsky}}\ and\ \bibinfo {author} {\bibfnamefont {G.~R.}\ \bibnamefont
  {Farrar}},\ }\href {\doibase 10.1103/PhysRevLett.31.1153} {\bibfield
  {journal} {\bibinfo  {journal} {Phys. Rev. Lett.}\ }\textbf {\bibinfo
  {volume} {31}},\ \bibinfo {pages} {1153} (\bibinfo {year}
  {1973})}\BibitemShut {NoStop}%
%%CITATION = PRLTA,31,1153;%%
\bibitem [{\citenamefont {Sopkovich}(1962)}]{Sopkovich}%
  \BibitemOpen
  \bibfield  {author} {\bibinfo {author} {\bibfnamefont {N.~J.}\ \bibnamefont
  {Sopkovich}},\ }\href {\doibase 10.1007/BF02754354} {\bibfield  {journal}
  {\bibinfo  {journal} {Nuovo Cimento}\ }\textbf {\bibinfo {volume} {26}},\
  \bibinfo {pages} {186} (\bibinfo {year} {1962})}\BibitemShut {NoStop}%
\bibitem [{\citenamefont {Larionov}\ and\ \citenamefont
  {Lenske}(2017)}]{Larionov:2017hcm}%
  \BibitemOpen
  \bibfield  {author} {\bibinfo {author} {\bibfnamefont {A.~B.}\ \bibnamefont
  {Larionov}}\ and\ \bibinfo {author} {\bibfnamefont {H.}~\bibnamefont
  {Lenske}},\ }\href {\doibase 10.1016/j.physletb.2017.09.007} {\bibfield
  {journal} {\bibinfo  {journal} {Phys. Lett.}\ }\textbf {\bibinfo {volume}
  {B773}},\ \bibinfo {pages} {470} (\bibinfo {year} {2017})}\BibitemShut
  {NoStop}%
\bibitem [{\citenamefont {Eide}\ \emph {et~al.}(1973)\citenamefont {Eide} \emph
  {et~al.}}]{Eide:1973tb}%
  \BibitemOpen
  \bibfield  {author} {\bibinfo {author} {\bibfnamefont {A.}~\bibnamefont
  {Eide}} \emph {et~al.},\ }\href {\doibase 10.1016/0550-3213(73)90176-4}
  {\bibfield  {journal} {\bibinfo  {journal} {Nucl. Phys.}\ }\textbf {\bibinfo
  {volume} {B60}},\ \bibinfo {pages} {173} (\bibinfo {year}
  {1973})}\BibitemShut {NoStop}%
%%CITATION = NUPHA,B60,173;%%
\bibitem [{\citenamefont {Van~de Wiele}\ and\ \citenamefont
  {Ong}(2010)}]{VandeWiele:2010kz}%
  \BibitemOpen
  \bibfield  {author} {\bibinfo {author} {\bibfnamefont {J.}~\bibnamefont
  {Van~de Wiele}}\ and\ \bibinfo {author} {\bibfnamefont {S.}~\bibnamefont
  {Ong}},\ }\href {\doibase 10.1140/epja/i2010-11044-7} {\bibfield  {journal}
  {\bibinfo  {journal} {Eur. Phys. J.}\ }\textbf {\bibinfo {volume} {A46}},\
  \bibinfo {pages} {291} (\bibinfo {year} {2010})}\BibitemShut {NoStop}%
\bibitem [{\citenamefont {Wang}\ \emph {et~al.}(2017)\citenamefont {Wang},
  \citenamefont {Bystritskiy},\ and\ \citenamefont
  {Tomasi-Gustafsson}}]{Wang:2015ybw}%
  \BibitemOpen
  \bibfield  {author} {\bibinfo {author} {\bibfnamefont {Y.}~\bibnamefont
  {Wang}}, \bibinfo {author} {\bibfnamefont {Y.~M.}\ \bibnamefont
  {Bystritskiy}}, \ and\ \bibinfo {author} {\bibfnamefont {E.}~\bibnamefont
  {Tomasi-Gustafsson}},\ }\href {\doibase 10.1103/PhysRevC.95.045202}
  {\bibfield  {journal} {\bibinfo  {journal} {Phys. Rev.}\ }\textbf {\bibinfo
  {volume} {C95}},\ \bibinfo {pages} {045202} (\bibinfo {year}
  {2017})}\BibitemShut {NoStop}%
\bibitem [{\citenamefont {Matsuyama}\ \emph {et~al.}(2007)\citenamefont
  {Matsuyama}, \citenamefont {Sato},\ and\ \citenamefont
  {Lee}}]{Matsuyama:2006rp}%
  \BibitemOpen
  \bibfield  {author} {\bibinfo {author} {\bibfnamefont {A.}~\bibnamefont
  {Matsuyama}}, \bibinfo {author} {\bibfnamefont {T.}~\bibnamefont {Sato}}, \
  and\ \bibinfo {author} {\bibfnamefont {T.~S.~H.}\ \bibnamefont {Lee}},\
  }\href {\doibase 10.1016/j.physrep.2006.12.003} {\bibfield  {journal}
  {\bibinfo  {journal} {Phys. Rept.}\ }\textbf {\bibinfo {volume} {439}},\
  \bibinfo {pages} {193} (\bibinfo {year} {2007})}\BibitemShut {NoStop}%
\bibitem [{\citenamefont {Serot}\ and\ \citenamefont
  {Walecka}(1986)}]{Serot:1984ey}%
  \BibitemOpen
  \bibfield  {author} {\bibinfo {author} {\bibfnamefont {B.~D.}\ \bibnamefont
  {Serot}}\ and\ \bibinfo {author} {\bibfnamefont {J.~D.}\ \bibnamefont
  {Walecka}},\ }\href@noop {} {\bibfield  {journal} {\bibinfo  {journal} {Adv.
  Nucl. Phys.}\ }\textbf {\bibinfo {volume} {16}},\ \bibinfo {pages} {1}
  (\bibinfo {year} {1986})}\BibitemShut {NoStop}%
%%CITATION = ANUPB,16,1;%%
\bibitem [{\citenamefont {Shyam}\ and\ \citenamefont
  {Mosel}(2003)}]{Shyam:2003cn}%
  \BibitemOpen
  \bibfield  {author} {\bibinfo {author} {\bibfnamefont {R.}~\bibnamefont
  {Shyam}}\ and\ \bibinfo {author} {\bibfnamefont {U.}~\bibnamefont {Mosel}},\
  }\href {\doibase 10.1103/PhysRevC.67.065202} {\bibfield  {journal} {\bibinfo
  {journal} {Phys. Rev.}\ }\textbf {\bibinfo {volume} {C67}},\ \bibinfo {pages}
  {065202} (\bibinfo {year} {2003})}\BibitemShut {NoStop}%
\bibitem [{\citenamefont {Oset}\ \emph {et~al.}(1982)\citenamefont {Oset},
  \citenamefont {Toki},\ and\ \citenamefont {Weise}}]{Oset:1981ih}%
  \BibitemOpen
  \bibfield  {author} {\bibinfo {author} {\bibfnamefont {E.}~\bibnamefont
  {Oset}}, \bibinfo {author} {\bibfnamefont {H.}~\bibnamefont {Toki}}, \ and\
  \bibinfo {author} {\bibfnamefont {W.}~\bibnamefont {Weise}},\ }\href
  {\doibase 10.1016/0370-1573(82)90123-5} {\bibfield  {journal} {\bibinfo
  {journal} {Phys. Rept.}\ }\textbf {\bibinfo {volume} {83}},\ \bibinfo {pages}
  {281} (\bibinfo {year} {1982})}\BibitemShut {NoStop}%
%%CITATION = PRPLC,83,281;%%
\bibitem [{\citenamefont {Collins}(1977)}]{Collins}%
  \BibitemOpen
  \bibfield  {author} {\bibinfo {author} {\bibfnamefont {P.~D.~B.}\
  \bibnamefont {Collins}},\ }\href@noop {} {\emph {\bibinfo {title} {An
  introduction to Regge theory and high energy physics}}}\ (\bibinfo
  {publisher} {Cambridge University Press},\ \bibinfo {address} {Cambridge --
  London -- New York -- Melbourne},\ \bibinfo {year} {1977})\BibitemShut
  {NoStop}%
\bibitem [{\citenamefont {Guidal}\ \emph {et~al.}(1997)\citenamefont {Guidal},
  \citenamefont {Laget},\ and\ \citenamefont {Vanderhaeghen}}]{Guidal:1997hy}%
  \BibitemOpen
  \bibfield  {author} {\bibinfo {author} {\bibfnamefont {M.}~\bibnamefont
  {Guidal}}, \bibinfo {author} {\bibfnamefont {J.~M.}\ \bibnamefont {Laget}}, \
  and\ \bibinfo {author} {\bibfnamefont {M.}~\bibnamefont {Vanderhaeghen}},\
  }\href {\doibase 10.1016/S0375-9474(97)00612-X} {\bibfield  {journal}
  {\bibinfo  {journal} {Nucl. Phys.}\ }\textbf {\bibinfo {volume} {A627}},\
  \bibinfo {pages} {645} (\bibinfo {year} {1997})}\BibitemShut {NoStop}%
%%CITATION = NUPHA,A627,645;%%
\bibitem [{\citenamefont {Macnaughton}\ \emph {et~al.}(1977)\citenamefont
  {Macnaughton}, \citenamefont {Butler}, \citenamefont {Coyne},\ and\
  \citenamefont {Fu}}]{Macnaughton:1977cy}%
  \BibitemOpen
  \bibfield  {author} {\bibinfo {author} {\bibfnamefont {J.}~\bibnamefont
  {Macnaughton}}, \bibinfo {author} {\bibfnamefont {W.~R.}\ \bibnamefont
  {Butler}}, \bibinfo {author} {\bibfnamefont {D.~G.}\ \bibnamefont {Coyne}}, \
  and\ \bibinfo {author} {\bibfnamefont {C.}~\bibnamefont {Fu}},\ }\href
  {\doibase 10.1103/PhysRevD.15.1832} {\bibfield  {journal} {\bibinfo
  {journal} {Phys. Rev.}\ }\textbf {\bibinfo {volume} {D15}},\ \bibinfo {pages}
  {1832} (\bibinfo {year} {1977})}\BibitemShut {NoStop}%
%%CITATION = PHRVA,D15,1832;%%
\bibitem [{\citenamefont {Bloodworth}\ \emph {et~al.}(1974)\citenamefont
  {Bloodworth}, \citenamefont {Jackson}, \citenamefont {Merdjanian},
  \citenamefont {Prentice},\ and\ \citenamefont {Yoon}}]{Bloodworth:1974ji}%
  \BibitemOpen
  \bibfield  {author} {\bibinfo {author} {\bibfnamefont {I.~J.}\ \bibnamefont
  {Bloodworth}}, \bibinfo {author} {\bibfnamefont {W.~C.}\ \bibnamefont
  {Jackson}}, \bibinfo {author} {\bibfnamefont {H.}~\bibnamefont {Merdjanian}},
  \bibinfo {author} {\bibfnamefont {J.~D.}\ \bibnamefont {Prentice}}, \ and\
  \bibinfo {author} {\bibfnamefont {T.~S.}\ \bibnamefont {Yoon}},\ }\href
  {\doibase 10.1016/0550-3213(74)90165-5} {\bibfield  {journal} {\bibinfo
  {journal} {Nucl. Phys.}\ }\textbf {\bibinfo {volume} {B81}},\ \bibinfo
  {pages} {231} (\bibinfo {year} {1974})}\BibitemShut {NoStop}%
%%CITATION = NUPHA,B81,231;%%
\bibitem [{\citenamefont {Honecker}\ \emph {et~al.}(1977)\citenamefont
  {Honecker} \emph {et~al.}}]{Honecker:1977me}%
  \BibitemOpen
  \bibfield  {author} {\bibinfo {author} {\bibfnamefont {R.}~\bibnamefont
  {Honecker}} \emph {et~al.},\ }\href {\doibase 10.1016/0550-3213(77)90369-8}
  {\bibfield  {journal} {\bibinfo  {journal} {Nucl. Phys.}\ }\textbf {\bibinfo
  {volume} {B131}},\ \bibinfo {pages} {189} (\bibinfo {year}
  {1977})}\BibitemShut {NoStop}%
%%CITATION = NUPHA,B131,189;%%
\bibitem [{\citenamefont {Eastman}\ \emph {et~al.}(1973)\citenamefont
  {Eastman}, \citenamefont {Ming}, \citenamefont {Oh}, \citenamefont {Parker},
  \citenamefont {Smith},\ and\ \citenamefont {Sprafka}}]{Eastman:1973va}%
  \BibitemOpen
  \bibfield  {author} {\bibinfo {author} {\bibfnamefont {P.~S.}\ \bibnamefont
  {Eastman}}, \bibinfo {author} {\bibfnamefont {M.~Z.}\ \bibnamefont {Ming}},
  \bibinfo {author} {\bibfnamefont {B.~Y.}\ \bibnamefont {Oh}}, \bibinfo
  {author} {\bibfnamefont {D.~L.}\ \bibnamefont {Parker}}, \bibinfo {author}
  {\bibfnamefont {G.~A.}\ \bibnamefont {Smith}}, \ and\ \bibinfo {author}
  {\bibfnamefont {R.~J.}\ \bibnamefont {Sprafka}},\ }\href {\doibase
  10.1016/0550-3213(73)90499-9} {\bibfield  {journal} {\bibinfo  {journal}
  {Nucl. Phys.}\ }\textbf {\bibinfo {volume} {B51}},\ \bibinfo {pages} {29}
  (\bibinfo {year} {1973})}\BibitemShut {NoStop}%
%%CITATION = NUPHA,B51,29;%%
\bibitem [{\citenamefont {Patrignani}\ \emph {et~al.}(2016)\citenamefont
  {Patrignani} \emph {et~al.}}]{Patrignani:2016xqp}%
  \BibitemOpen
  \bibfield  {author} {\bibinfo {author} {\bibfnamefont {C.}~\bibnamefont
  {Patrignani}} \emph {et~al.} (\bibinfo {collaboration} {Particle Data
  Group}),\ }\href {\doibase 10.1088/1674-1137/40/10/100001} {\bibfield
  {journal} {\bibinfo  {journal} {Chin. Phys.}\ }\textbf {\bibinfo {volume}
  {C40}},\ \bibinfo {pages} {100001} (\bibinfo {year} {2016})}\BibitemShut
  {NoStop}%
%%CITATION = CHPHD,C40,100001;%%
\end{thebibliography}%

\newpage

\appendix

\begin{appendices}

\section{Relation between the LC and NR wave functions of the $\Delta-\Delta$ system}
\label{app_LC_vs_NR}

\begin{figure}[ht]
\begin{center}
   \includegraphics[scale = 0.60]{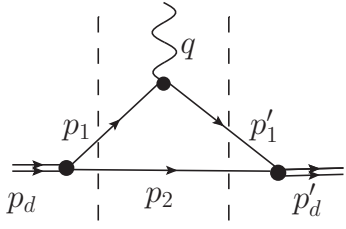}
\end{center}
\caption{\label{fig:emFF} Lowest order contribution to the absorption amplitude
  of a photon on the deuteron. $q$, $p_d$ and $p_d^\prime$ are the four-momenta of the photon, initial and final deuteron,
  respectively. $p_1$, $p_2$ and $p_1^\prime$ are the four-momenta of the intermediate $\Delta$'s. Time axis is from left
to right.} 
\end{figure}

Consider the electromagnetic formfactor of the deuteron viewed as a $\Delta-\Delta$ state, Fig.~\ref{fig:emFF}.
In the kinematics of  high-energy $ed$ scattering in the c.m. frame of colliding particles the four-momentum transfer
from electron to deuteron can be written as follows (see equation on p. 225 of ref. \cite{Frankfurt:1981mk},
note opposite direction of $z$-axis):
\begin{equation}
  q=\left(\frac{2\nu+q^2}{4P},\bm{q}_t,\frac{2\nu-q^2}{4P}\right)~,     \label{q}
\end{equation}
where $P \to +\infty$ is the electron momentum directed along the $z$-axis (correspondingly, $p_d^z=-P$) and $\nu=p_dq$.
At very large $P$ the four-momentum transfer becomes purely transverse.
This allows to consider only the graph of Fig.~\ref{fig:emFF}, since other graphs contain pair production and disappear for $q^z=0$.
The matrix element of Fig.~\ref{fig:emFF} can be calculated within the non-covariant perturbation theory rules \cite{Weinberg:1966jm}
which give the following expression:
\begin{equation}
  M(\bm{q}_t) = - \Gamma_{d \to \Delta\Delta}^2 \int \frac{d^3p_2}{(2\pi)^3} \frac{M_{\gamma^*}(q;p_1)}{2E_{\bm{p}_2}2E_{\bm{p}_1}2E_{\bm{p}_1^\prime}}
  \frac{1}{p_d^0-E_{\bm{p}_2}-E_{\bm{p}_1}+i\epsilon} \frac{1}{p_d^{0\prime}-E_{\bm{p}_2}-E_{\bm{p}_1^\prime}+i\epsilon}~,   \label{M}
\end{equation}
where $E_{\bm{p}_i}=\sqrt{m_\Delta^2+\bm{p}_i^2}, i=1,2,1^\prime$ are the energies of the intermediate $\Delta$'s with
three-momenta $\bm{p}_1=\bm{p}_d-\bm{p}_2$, $\bm{p}_1^\prime=\bm{q}+\bm{p}_d-\bm{p}_2$,
and $M_{\gamma^*}(q;p_1)$ is the invariant matrix element of the electromagnetic transition, $\gamma \Delta \to \Delta$
(the spin indices are implicit).
For simplicity, a constant vertex factor $\Gamma_{d \to \Delta\Delta}$ is assumed.
Introducing the ratios
\begin{equation}
  \alpha_i = \frac{-2p_i^z}{P},~~~i=1,2,1^\prime~,    \label{alpha_i}
\end{equation}
the particle energies can be expressed as
\begin{equation}
  E_{\bm{p}_i}=\frac{|\alpha_i|P}{2} + \frac{m_i^2+\bm{p}_{it}^2}{|\alpha_i|P} + O(1/P^3)~.   \label{E_i}
\end{equation}
Using the relations $\alpha_1+\alpha_2=\alpha_{1^\prime}+\alpha_2=2$, $\bm{p}_{1t}=\bm{p}_{1^{\prime}t}-\bm{q}_t=-\bm{p}_{2t}$, which
follow from three-momentum conservation at the vertices, and the relation
\begin{equation}
  \frac{d^3p_2}{E_{\bm{p}_2}} = \frac{d\alpha_2 d^2p_{2t}}{\alpha_2}~,     \label{Jacobian}
\end{equation}
the matrix element Eq.(\ref{M}) can be expressed as follows:
\begin{equation}
  M(\bm{q}_t) = - \Gamma_{d \to \Delta\Delta}^2 \int\limits_0^2 \frac{d\alpha_2}{2\alpha_2(2-\alpha_2)^2}
  \int \frac{d^2p_{2t}}{(2\pi)^3} \frac{2}{m_d^2-M_{1,2}^2} \frac{2M_{\gamma^*}(\bm{q}_t;\alpha_1,\bm{p}_{1t})}{m_d^2-M_{1^\prime,2}^2}~, \label{M_1}
\end{equation}
where
\begin{eqnarray}
  M_{1,2}^2 &=& \frac{4(m_\Delta^2+p_{2t}^2)}{\alpha_2(2-\alpha_2)},   \label{M_12^2}\\
  M_{1^\prime,2}^2 &=&  \frac{2(m_\Delta^2+p_{2t}^2)}{\alpha_2} + \frac{2(m_\Delta^2+(\bm{q}_t-\bm{p}_{2t})^2)}{2-\alpha_2} -q_t^2~.   \label{M_1prime2^2}
\end{eqnarray}
Using the internal three-momenta $\bm{k}$ and $\bm{k}^\prime$ defined as
\begin{eqnarray}
  \alpha_2 &=& 1 + \frac{k^z}{\sqrt{m_\Delta^2+\bm{k}^2}}~,~~~\bm{k}_t=-\bm{p}_{2t}~,  \label{k}\\
  \alpha_2 &=& 1 + \frac{k^{\prime z}}{\sqrt{m_\Delta^2+\bm{k}^{\prime2}}}~,~~~\bm{k}^\prime_t=\bm{k}_t+\frac{1}{2}\alpha_2\bm{q}_t  ~, \label{k^prime}
\end{eqnarray}
and the relation
\begin{equation}
    \frac{d\alpha_2 d^2p_{2t}}{\alpha_2(2-\alpha_2)}=\frac{d^3k}{\sqrt{m_\Delta^2+\bm{k}^2}}   \label{Jacobian2}
\end{equation}
one can express Eq.(\ref{M_1}) in the form
\begin{equation}
  M(\bm{q}_t) = - \Gamma_{d \to \Delta\Delta}^2 \int \frac{d^3k}{(2\pi)^3\sqrt{m_\Delta^2+\bm{k}^2}(2-\alpha_2)}
   \frac{1}{m_d^2-M_{1,2}^2} \frac{2M_{\gamma^*}(\bm{q}_t;\alpha_1,\bm{k}_t)}{m_d^2-M_{1^\prime,2}^2}~, \label{M_2}
\end{equation}
with $M_{1,2}^2=4(m_\Delta^2+\bm{k}^2)$ and $M_{1^\prime,2}^2=4(m_\Delta^2+\bm{k}^{\prime2})$.
We  can now introduce the LC wave function of the $\Delta-\Delta$ state defined according to ref. \cite{Frankfurt:1981mk}
(see sec. 2.3.1 of ref. \cite{Frankfurt:1981mk}, the vertex function $\chi$ is replaced by $i\Gamma_{d \to \Delta\Delta}$
in our notation):
\begin{equation}
  \psi_{\Delta-\Delta}(k)=\frac{i\Gamma_{d \to \Delta\Delta}}{M_{1,2}^2-m_d^2}~.    \label{psi_DD}
\end{equation}
Then Eq.(\ref{M_2}) can be rewritten in the form
\begin{equation}
  M(\bm{q}_t) = \int \frac{d^3k}{(2\pi)^3\sqrt{m_\Delta^2+\bm{k}^2}(2-\alpha_2)}
   \psi_{\Delta-\Delta}(k) \psi_{\Delta-\Delta}(k^\prime) 2M_{\gamma^*}(\bm{q}_t;2-\alpha_2,\bm{k}_t)~.         \label{M_3}
\end{equation}
Note that in the chosen frame the matrix element (\ref{M_3}) is the only contribution to the Lorentz-invariant matrix element
calculated within the Feynman rules because the graphs with pair production disappear in this frame. 

On the other hand, we can calculate the photo-absorption amplitude in the NR approximation.
In this case we choose the frame, where both the initial and the final deuteron move slowly,
$|\bm{p}_d|,|\bm{p}_d^\prime| \ll m_d$, but the electron is fast.
We start from the $S$-matrix element corresponding to Fig.~\ref{fig:emFF}
(spin indices are suppressed for brevity):
\begin{eqnarray}
  S_{\rm NR} &=& \int d^3r_1 \int d^3r_2  \int d^3r_1^\prime \int d^3r_2^\prime \phi_{\Delta-\Delta,f}^*(\bm{r}_1^\prime,\bm{r}_2^\prime) 
  \phi_{\Delta-\Delta,i}(\bm{r}_1,\bm{r}_2)  \nonumber \\
        && \times \frac{1}{V^2} \int \frac{Vd^3p_1}{(2\pi)^3} \int \frac{Vd^3p_2}{(2\pi)^3} \int \frac{Vd^3p_1^\prime}{(2\pi)^3}
            e^{-i\bm{p}_1\bm{r}_1-i\bm{p}_2\bm{r}_2+i\bm{p}_1^\prime\bm{r}_1^\prime+i\bm{p}_2\bm{r}_2^\prime} S_{\gamma^*}(q;p_1)~,    \label{S_NR}
\end{eqnarray}
where
\begin{equation}
    S_{\gamma^*}(q;p_1) = \frac{(2\pi)^4\delta^{(4)}(p_1^\prime-p_1-q)}{(2E_1V 2E_1^{\prime}V 2q^0V)^{1/2}}\,
     i M_{\gamma^*}(q;p_1)~.    \label{S_gamma}
\end{equation}
Here $E_1=p_d^0-E_{\bm{p}_2}$ and $E_1^{\prime}=p_d^{\prime 0}-E_{\bm{p}_2}$
are the energies of the 1$^{\rm{}st}$ $\Delta$ before and after
$\gamma^*$-absorption (the 2$^{\rm{}nd}$ $\Delta$ is put on the mass shell).

By integrating out the c.m. motion (similar to Eq.(\ref{S^(0)_expanded}) of sec. \ref{signal})
we obtain the usual transition $S$-matrix in a factorized form:
\begin{equation}
   S_{\rm NR} = \frac{(2\pi)^4\delta^{(4)}(p_d^\prime-p_d-q)}{(2p_d^0V 2p_d^{\prime 0}V 2q^0V)^{1/2}}\,
                   i M_{\rm NR}~,     \label{S_NR_1} 
\end{equation}
where the invariant matrix element is
\begin{equation}
   M_{\rm NR}= \int d^3p_2 \left(\frac{p_d^0 p_d^{\prime 0}}{E_1 E_1^{\prime}}\right)^{1/2} 
   \phi_{\Delta-\Delta}^*\left(\frac{\bm{p}_d^\prime}{2}-\bm{p}_2\right) 
   \phi_{\Delta-\Delta}\left(\frac{\bm{p}_d}{2}-\bm{p}_2\right)
   M_{\gamma^*}(q;p_d-p_2)~.                              \label{M_NR}
\end{equation}
Note that one can obtain Eq.(\ref{M_NR}) also by treating the graph Fig.~\ref{fig:emFF} as a Feynman diagram
and then using the relation (\ref{Gamma_d2DD}).

It follows from Eq.(2.22) of ref. \cite{Frankfurt:1981mk} (where one should replace the nucleon
mass by the $\Delta$ mass)
that the function $\psi_{\Delta-\Delta}(k)/(m_\Delta^2+\bm{k}^2)^{1/4}$ satisfies the
non-relativistic Schr\"odinger equation for the $\Delta-\Delta$ bound state
with binding energy $2m_\Delta-m_d$ and the potential corresponding to the
kernel of the Bethe-Salpeter-type equation.
Thus the function $\psi_{\Delta-\Delta}(k)/(m_\Delta^2+\bm{k}^2)^{1/4}$ should
be proportional to the NR wave function $\phi_{\Delta-\Delta}(k)$.
The proportionality factor can be obtained by taking the limit $q=0$ (forward
$ed$ scattering) and assuming a very narrow wave function
$\phi_{\Delta-\Delta}(k)$ peaking at $k=0$.
In this case the LC and NR expressions, i.e. Eqs.(\ref{M_3}) and (\ref{M_NR})
with $\bm{p}_d=\bm{p}_d^\prime=0$, should coincide.
This leads to the following relation
\begin{equation}
    \frac{2 \psi_{\Delta-\Delta}^2(k)}{(2\pi)^3 (m_\Delta^2+\bm{k}^2)^{1/2}} = \frac{m_d}{m_d-m_\Delta} \phi_{\Delta-\Delta}^2(k)~.    \label{LC_vs_NR}
\end{equation}

\section{Poles of the pion propagator}
\label{app_piPoles}

To determine the poles of the pion propagator in the three-pion annihilation background (see Fig.~\ref{fig:bgr_3pi})
for fixed values of the proton transverse momentum, let us consider the $\pi^+ p \to \Delta^{++}$ transition in the frame
where the $\Delta$ has a momentum with a large negative $z$-component.
In that frame, the four-momenta of the $\Delta$, pion
and proton are, respectively,
\begin{eqnarray}
  p_\Delta &=& (P+\frac{m_{\Delta t}^2}{2P},\mathbf{p}_{\Delta t}, -P)~,  \\
  p_\pi &=& (P\alpha_\pi+\frac{m_{\pi t}^2}{2P\alpha_\pi},\mathbf{p}_{\pi t}, -P\alpha_\pi)~, \\
  p_p &=& (P\alpha_p+\frac{m_{p t}^2}{2P\alpha_p}, \mathbf{p}_{p t}, -P\alpha_p)~,
\end{eqnarray}
where the transverse masses are $m_{\Delta t}=\sqrt{\mathbf{p}_{\Delta t}^2+m_\Delta^2}$, $m_{\pi t}=\sqrt{\mathbf{p}_{\pi t}^2+m_\pi^2}$
with $\mathbf{p}_{\pi t}=\mathbf{p}_{\Delta t}-\mathbf{p}_{p t}$,
and $m_{p t}=\sqrt{\mathbf{p}_{p t}^2+m_N^2}$, and $P \to +\infty$. $\alpha_\pi$ and $\alpha_p$ are the longitudinal-boost-invariant
momentum fractions of the $\Delta$ carried by the pion and proton, respectively. They satisfy the condition $\alpha_\pi+\alpha_p=1$.
In the lab. frame the fractions can be expressed as
\begin{equation}
  \alpha_\pi=\frac{E_\pi-p_\pi^z}{E_{\Delta}-p_{\Delta}^z}~,~~~\alpha_p=\frac{E_p-p_p^z}{E_{\Delta}-p_{\Delta}^z}~.
\end{equation}
Energy conservation can be expressed as
\begin{equation}
  \frac{m_{\pi t}^2}{\alpha_\pi} + \frac{m_{p t}^2}{\alpha_p} = m_{\Delta t}^2~.   \label{encons}
\end{equation}
This equation can be easily solved with respect to $\alpha_p$:
\begin{equation}
  \alpha_p = (A \pm \sqrt{A^2-B})/2~,     \label{alphap}
\end{equation}
where
\begin{equation}
  A=\frac{m_{p t}^2+m_{\Delta t}^2-m_{\pi t}^2}{m_{\Delta t}^2}~,~~~B=\frac{4m_{p t}^2}{m_{\Delta t}^2}~.
\end{equation}
The fraction of the deuteron momentum carried by the proton is then
\begin{equation}
  \tilde\alpha_p = \tilde\alpha_\Delta \alpha_p~,    \label{tildealphap}
\end{equation}
where in the deuteron rest frame
\begin{equation}
  \tilde \alpha_p = \frac{E_p-p_p^z}{m_d/2}~,~~~\tilde \alpha_\Delta = \frac{E_\Delta-p_\Delta^z}{m_d/2}~.
\end{equation}
Thus the two poles of the pion propagator are given by
\begin{equation}
  p_p^z=\Delta_{1,2}=\frac{m_{p t}^2}{m_d\tilde \alpha_p} - \frac{m_d\tilde \alpha_p}{4}~.   \label{poles}
\end{equation}
The $\Delta_1$ ($\Delta_2$) is obtained by choosing $+$ ($-$) sign in Eq.(\ref{alphap}).

\section{Elementary amplitudes}
\label{elem}

\begin{subappendices}

\renewcommand{\thesubsection}{C\arabic{subsection}}
  
\subsection{$\bar N N \to \pi \pi$}
\label{NbarN2pipi}

\begin{figure}
\includegraphics[scale = 0.6]{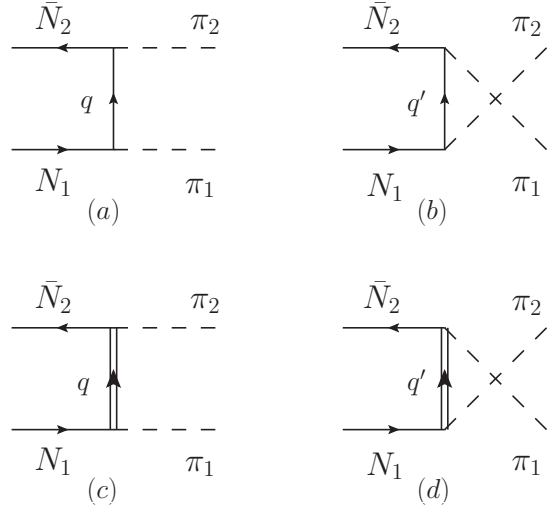}
\caption{\label{fig:NbarN2pipi} Feynman graphs included in the calculation of
  the $\bar N_2 N_1 \to \pi_2 \pi_1$ amplitude.
  Graphs (a) and (b) contain the exchange of a nucleon. Graphs (c)
  and (d) contain the exchange of a $\Delta$-isobar.
  The four-momenta of the exchange particle are denoted as $q$ and $q^\prime$,
  respectively, in the $t$-channel (a,c) and $u$-channel (b,d) graphs, where
  $t=q^2$, $u=q^{\prime 2}$.}
\end{figure}
The amplitude of antinucleon-nucleon annihilation into two pions
is described within the nucleon and $\Delta$ exchange model as displayed in
Fig.~\ref{fig:NbarN2pipi}.
For the $\pi NN$ and $\pi N\Delta$ interactions we apply the following
Lagrangians:
\begin{eqnarray}
  {\cal L}_{\pi NN} &=& \frac{f_{\pi NN}}{m_\pi} \bar\psi \gamma^\mu \gamma^5 \bvec{\tau} \psi \partial_\mu \bvec{\pi}~,  \label{L_piNN}\\
  {\cal L}_{\pi N \Delta} &=& \frac{f_{\pi N \Delta}}{m_\pi} \bar\psi^\mu \bvec{T} \psi \partial_\mu \bvec{\pi} + h.c.~,    \label{L_piND}
\end{eqnarray}
where $f_{\pi NN}=1.008$, $f_{\pi N \Delta}=2.202$ \cite{Dmitriev:1986st}. Here, $\bvec{T}$ is the isospin transition $1/2 \to 3/2$
operator (cf. ref. \cite{Brown:1975di}):
\begin{equation}
  \bvec{T}_{\tau_\Delta \tau_N} = \sum_{l=0,\pm1} <\frac{3}{2} \tau_\Delta|\frac{1}{2}\tau_N;1l> \bvec{t}^{(l)*}~, \label{T}
\end{equation}
with $\bvec{t}^{(0)}=(0,0,1), \bvec{t}^{(\pm1)}=\mp\frac{1}{\sqrt{2}}(1,\pm i,0)$ being the eigenvectors of $\hat I^2$ and $\hat I_3$
operators for $I=1$ in Cartesian basis.
The invariant matrix elements of the graphs (a) and (c) are
\begin{eqnarray}
  M_{\bar NN}^{(a)} &=& \left(\frac{f_{\pi NN} F_{\pi NN}(t)}{m_\pi}\right)^2 \frac{{\cal I}_{\bar NN}^{(a)} \sqrt{\Omega}}{t-m_N^2+i\epsilon}
                    \bar u(-p_2,-\lambda_2) \slashed{k}_2 (\slashed{q}-m_N) \slashed{k}_1 u(p_1,\lambda_1)~,  \label{calM_barNN^a}\\
  M_{\bar NN}^{(c)} &=& -\left(\frac{f_{\pi N\Delta} F_{\pi N\Delta}(t,m_\Delta^2)}{m_\pi}\right)^2 
                      \frac{{\cal I}_{\bar NN}^{(c)} \sqrt{\Omega}}{t-m_{\Delta}^2+i\epsilon} \nonumber \\
                      && \times \bar u(-p_2,-\lambda_2)  (\slashed{q}+m_\Delta) k_{2\mu} k_{1\nu} {\cal P}^{\mu\nu}(q) u(p_1,\lambda_1)~,
                                                                                                             \label{calM_barNN^c}
\end{eqnarray}
where $p_1,\lambda_1$ and $p_2,\lambda_2$ are the four-momenta and spin
projections of the nucleon and anti\-nucleon, respectively, and
$k_1, k_2$ are the four-momenta of the pions.
The Dirac spinors are normalized as $\bar u_{p_1,\lambda_1}u_{p_1,\lambda_1}=-\bar u_{-p_2,-\lambda_2} u_{-p_2,-\lambda_2} = 2m_N$.
In obtaining Eqs.(\ref{calM_barNN^a}),(\ref{calM_barNN^c}) we used the Dirac
propagator of the nucleon
\begin{equation}
  iG(q)=\frac{i(\slashed{q}+m_N)}{q^2-m_N^2+i\epsilon}~,   \label{iGq}
\end{equation}
and the Rarita-Schwinger propagator of the $\Delta$-isobar
\begin{equation}
  iG^{\mu\nu}(q) = \frac{-i(\slashed{q}+m_\Delta)}{q^2-m_{\Delta}^2+i\epsilon} {\cal P}^{\mu\nu}(q)~,   \label{iG^munuq}
\end{equation}
where
\begin{equation}
  {\cal P}^{\mu\nu}(q) = g^{\mu\nu} - \frac{\gamma^\mu\gamma^\nu}{3}
  - \frac{2 q^\mu q^\nu}{3m_\Delta^2} + \frac{q^\mu\gamma^\nu-q^\nu\gamma^\mu}{3m_\Delta}~.   \label{calP^munu}
\end{equation}
The isospin factors are expressed as 
\begin{eqnarray}
  {\cal I}_{\bar NN}^{(a)} &=& (-1)^{1/2+\tau_2}  \sum_{\tau_N=\pm1/2} (\bvec{t}^{(l_2)*} \cdot \bvec{\tau})_{-\tau_2,\tau_N}
  (\bvec{t}^{(l_1)*} \cdot \bvec{\tau})_{\tau_N,\tau_1}~,     \label{I_barNN^a}\\
  {\cal I}_{\bar NN}^{(c)} &=& (-1)^{1/2+\tau_2} \sum_{\tau_\Delta=\pm1/2,\pm3/2} (\bvec{t}^{(l_2)*} \cdot \bvec{T}^\dag)_{-\tau_2,\tau_\Delta}
  (\bvec{t}^{(l_1)*} \cdot \bvec{T})_{\tau_\Delta,\tau_1}~,    \label{I_barNN^c}
\end{eqnarray}
where $l_1,l_2=0,\pm1$ are the isospin projections of the pions,
and $\tau_1,\tau_2=\pm1/2$ are the isospin projections
of nucleon and antinucleon, respectively.
The common factor $(-1)^{1/2+\tau_2}$ originates from the definition of the
physical antineutron state as $|\bar n>=-|\bar N,\tau=+1/2>$ as follows from the relation $\hat G|\bar N,\tau> = - |N,\tau>$,
where $\hat G=\exp(i \pi \hat I_2) \hat C$ is the $G$-parity transformation operator \cite{Tabakin:1985yv}.  
For the $u$-channel graphs (b,d), the matrix elements are obtained by
replacing $k_1 \leftrightarrow k_2$, $q \to q^\prime$ and $t \to u$ in Eqs.(\ref{calM_barNN^a}),(\ref{calM_barNN^c})
and the isospin factors -- by replacing $l_1 \leftrightarrow l_2$ in Eqs.(\ref{I_barNN^a}),(\ref{I_barNN^c}).
For the channels with incoming antiproton the values of isospin factors are listed in Table~\ref{tab:Iiso1}.
\begin{table}[htb]
  \caption{\label{tab:Iiso1} Isospin factors in the nucleon and $\Delta$ exchange amplitudes of Fig.~\ref{fig:NbarN2pipi}.}
  \begin{center}
    \begin{tabular}{|c|c|c|c|c|}
    \hline
    Channel                     & ${\cal I}_{\bar NN}^{(a)}$ & ${\cal I}_{\bar NN}^{(b)}$ & ${\cal I}_{\bar NN}^{(c)}$ & ${\cal I}_{\bar NN}^{(d)}$\\
    \hline
    $\bar p p \to \pi^- \pi^+$  &  -2              &  0               &  -1/3            & -1              \\   
    $\bar p p \to \pi^0 \pi^0$  &   1              &  1               &   2/3            &  2/3            \\
    $\bar p n \to \pi^- \pi^0$  &   $-\sqrt{2}$    &  $\sqrt{2}$      &   $\sqrt{2}/3$   &  $-\sqrt{2}/3$  \\
    \hline
    \end{tabular}
  \end{center}
\end{table}

To describe the finite size of the hadrons, we included formfactors in Eqs.(\ref{calM_barNN^a}),(\ref{calM_barNN^c}). Their choice is defined
by the asymptotic scaling law \cite{Brodsky:1973kr} at $s \to \infty, t/s={\rm const}$: 
\begin{equation}
   \frac{d\sigma}{dt} = \frac{f(t/s)}{s^n}~,~~~n=\sum n_i - 2~,     \label{counting}
\end{equation}
where $n_i$ is the number of the constituents in the incoming and outgoing  particles ($n_B=3, n_M=2$). Hence, $n=8$ for $\bar N N \to \pi \pi$.
By counting the powers of $s$ (assuming $t \sim u \sim s$) one can deduce from the expression
\begin{equation}
  \frac{d\sigma}{dt} 
   = \frac{\overline{|M_{\bar NN}^{(a)}+M_{\bar NN}^{(b)}+M_{\bar NN}^{(c)}+M_{\bar NN}^{(d)}|^2}}{64\pi(s/4-m_N^2)s}~,   \label{dsigmadt}
\end{equation}
the powers of the vertex formfactors:
\begin{eqnarray}
   F_{\pi NN}(t)     &=& \left(\frac{\Lambda_{\pi NN}^2-m_N^2}{\Lambda_{\pi NN}^2-t}\right)^2~,             \label{F_piNN}\\
   F_{\pi N\Delta}(t,M^2) &=& \left(\frac{\Lambda_{\pi N\Delta}^2-M^2}{\Lambda_{\pi N\Delta}^2-t}\right)^{5/2}~.  \label{F_piND}
\end{eqnarray}

Finally, following ref. \cite{Sopkovich} the attenuation factor
$\sqrt{\Omega}$ is introduced in Eqs.(\ref{calM_barNN^a}),(\ref{calM_barNN^c})
to describe the initial-state interaction in the $\bar N N$ channel.
For simplicity, we assume this factor to be energy- and
angular-momentum-independent \cite{Larionov:2017hcm}.

\begin{figure}
\includegraphics[scale = 0.6]{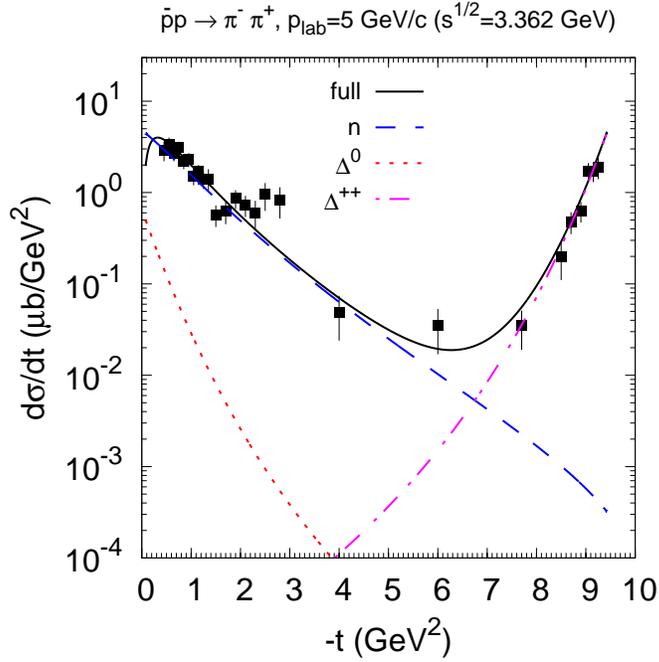}
\caption{\label{fig:pbarp2pi-pi+_5gev} Differential cross section of the $\bar p p \to \pi^- \pi^+$ process at $p_{\rm lab}=5$ GeV/c.
  Solid line: full calculation. Dashed, dotted and dash-dotted lines display the separate contributions of neutron, $\Delta^0$ and
  $\Delta^{++}$ exchange, respectively. Experimental data are from ref. \cite{Eide:1973tb}.}
\end{figure}
The values of the cutoff parameters $\Lambda_{\pi NN}=2.0$ GeV and
$\Lambda_{\pi N\Delta}=1.8$ GeV have been adjusted to describe the shape of
the $t$-dependence of the differential cross section
$d\sigma_{\bar p p \to \pi^- \pi^+}/dt$ at $p_{\rm lab}=5$ GeV/c. 
After this, the parameter $\Omega=0.008$ has been chosen to describe the absolute values of $d\sigma_{\bar p p \to \pi^- \pi^+}/dt$
close to $\Theta_{c.m.}=90\degree$ ($-t=4.7$ Gev$^2$). This value of $\Omega$ is within the range of values from ref. \cite{Larionov:2017hcm},
where meson-exchange models have been applied for the calculation of the $\bar p p \to \bar\Lambda \Lambda$ cross section. 

Fig.~\ref{fig:pbarp2pi-pi+_5gev} shows the comparison with experimental data for the fitted values of the parameters.
At small $-t$ (forward c.m. angles) the main contribution is given by neutron exchange, while at large $-t$ (backward c.m. angles)
the cross section is almost entirely due to $\Delta^{++}$ exchange.
These features are in line with other calculations \cite{VandeWiele:2010kz,Wang:2015ybw}.

\begin{figure}
\includegraphics[scale = 0.6]{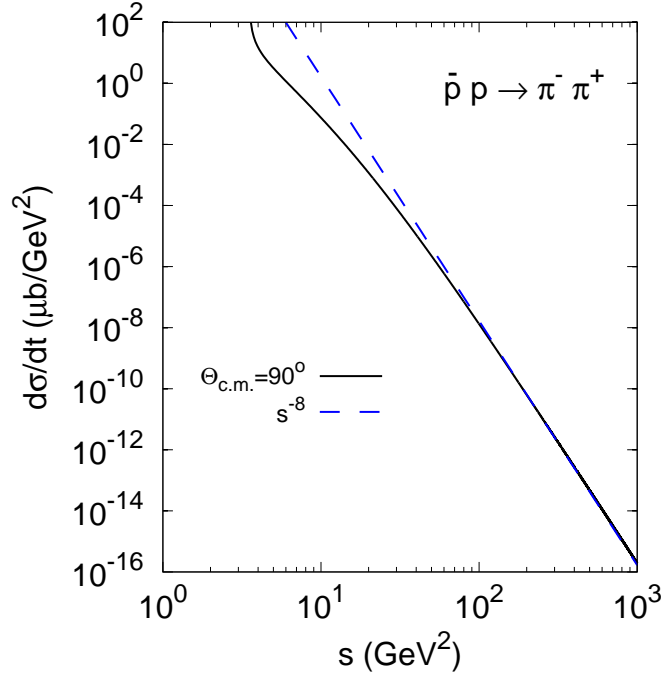}
\caption{\label{fig:pbarp2pi-pi+_90deg} Solid line:
  differential cross section of the $\bar p p \to \pi^- \pi^+$ process
  calculated from Eq.(\ref{dsigmadt}) at $\Theta_{c.m.}=90\degree$ 
  as a function of invariant $s$. Dashed line: large $s$
  asymptote $\propto s^{-8}$.} 
\end{figure}
In Fig.~\ref{fig:pbarp2pi-pi+_90deg} we display the $s$ dependence
of $d\sigma_{\bar p p \to \pi^- \pi^+}/dt$ at $-t=s/2-m_N^2-m_\pi^2$.
The quark counting rule at large $s$ is reproduced exactly.

\subsection{$\bar N \Delta \to \pi \pi$}
\label{NbarD2pipi}

\begin{figure}
\includegraphics[scale = 0.6]{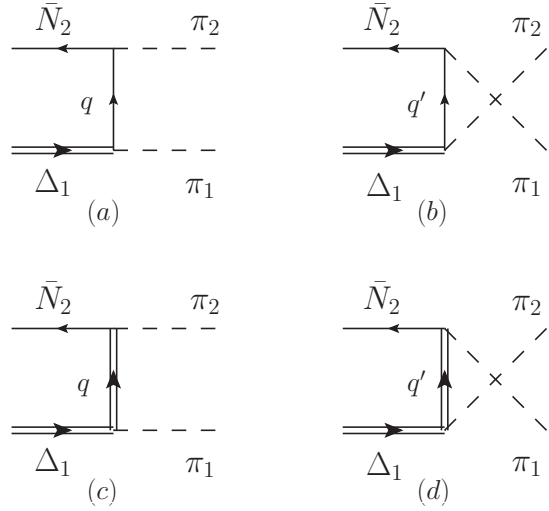}
\caption{\label{fig:NbarD2pipi} Same as Fig.~\ref{fig:NbarN2pipi}, but for the $\bar N_2 \Delta_1 \to \pi_2 \pi_1$ amplitude.}
\end{figure}
The Feynman graphs included in the nucleon-delta annihilation amplitude into two pions are shown in Fig.~\ref{fig:NbarD2pipi}.
The $\pi NN$ and $\pi N\Delta$ coupling Lagrangians were already explained in the previous subsection
(see Eqs.(\ref{L_piNN}),(\ref{L_piND})). The $\pi \Delta\Delta$ coupling Lagrangian can be defined as follows (cf. ref. \cite{Matsuyama:2006rp}):
\begin{equation}
  {\cal L}_{\pi\Delta\Delta} = \frac{f_{\pi\Delta\Delta}}{m_\pi} \bar\psi^\nu \gamma^5 \gamma^\mu  \bvec{T}_\Delta \psi_\nu
                            \partial_\mu \bvec{\pi}~,  \label{L_piDD}
\end{equation}
where
\begin{equation}
  (\bvec{T}_\Delta)_{\tau_{\Delta_2} \tau_{\Delta_1}}
  = \frac{\sqrt{15}}{2} \sum_{l=0,\pm1} <\frac{3}{2} \tau_{\Delta_2} |\frac{3}{2} \tau_{\Delta_1};1l> \bvec{t}^{(l)*}~, \label{T_D}
\end{equation}
is the isospin operator for $I=3/2$.
Within the SU(6) chiral constituent quark model the following relation holds \cite{Matsuyama:2006rp}:
\begin{equation}
    f_{\pi\Delta\Delta}=\frac{6}{5} f_{\pi NN}~.          \label{f_piDD}
\end{equation}
The invariant matrix elements of the $t$-channel graphs (a) and (c) of Fig.~\ref{fig:NbarD2pipi} are
\begin{eqnarray}
  M_{\bar N \Delta}^{(a)} &=& \frac{f_{\pi NN} F_{\pi NN}(t) f_{\pi N\Delta} F_{\pi N\Delta}(t,m_N^2)}{m_\pi^2} 
                    \frac{{\cal I}_{\bar N\Delta}^{(a)} \sqrt{\Omega}}{t-m_N^2+i\epsilon} \nonumber \\
       && \times  \bar u(-p_2,-\lambda_2) \slashed{k}_2 \gamma^5 (\slashed{q}+m_N) k_{1\mu} u^\mu(p_1,\lambda_1)~,  \label{calM_barND^a}\\
  M_{\bar N \Delta}^{(c)} &=& -\frac{f_{\pi N\Delta} F_{\pi N\Delta}(t,m_\Delta^2) f_{\pi\Delta\Delta}  F_{\pi \Delta\Delta}(t)}{m_\pi^2} 
                     \frac{{\cal I}_{\bar N\Delta}^{(c)} \sqrt{\Omega}}{t-m_{\Delta}^2+i\epsilon} \nonumber \\
                     && \times \bar u(-p_2,-\lambda_2) (\slashed{q}+m_\Delta) k_2^\mu {\cal P}_{\mu\nu}(q) \gamma^5 \slashed{k}_1 u^\nu(p_1,\lambda_1)~,
                                                        \label{calM_barND^c}                     
\end{eqnarray}
where the isospin factors are
\begin{eqnarray}
  {\cal I}_{\bar N\Delta}^{(a)} &=& (-1)^{1/2+\tau_2}  \sum_{\tau_N=\pm1/2} (\bvec{t}^{(l_2)*} \cdot \bvec{\tau})_{-\tau_2,\tau_N}
  (\bvec{t}^{(l_1)*} \cdot \bvec{T}^\dag)_{\tau_N,\tau_{\Delta_1}}~,     \label{I_barND^a}\\
  {\cal I}_{\bar N\Delta}^{(c)} &=& (-1)^{1/2+\tau_2} \sum_{\tau_\Delta=\pm1/2,\pm3/2} (\bvec{t}^{(l_2)*} \cdot \bvec{T}^\dag)_{-\tau_2,\tau_\Delta}
  (\bvec{t}^{(l_1)*} \cdot \bvec{T}_\Delta)_{\tau_\Delta,\tau_{\Delta_1}}~.    \label{I_barND^c}
\end{eqnarray}
The $u$-channel matrix elements of the graphs (b) and (d) of Fig.~\ref{fig:NbarD2pipi} are obtained from
Eqs.(\ref{calM_barND^a}),(\ref{calM_barND^c}) by the replacements $k_1 \leftrightarrow k_2$, $q \to q^\prime$ and $t \to u$,
and the corresponding isospin factors -- by replacement $l_1 \leftrightarrow l_2$ in Eqs.(\ref{I_barND^a}),(\ref{I_barND^c}).
After some algebra we get the following values for the channel $\bar p \Delta^- \to \pi^- \pi^-$:
${\cal I}_{\bar N\Delta}^{(a)}={\cal I}_{\bar N\Delta}^{(b)}=\sqrt{2}$,
${\cal I}_{\bar N\Delta}^{(c)}={\cal I}_{\bar N\Delta}^{(d)}=1/\sqrt{2}$. 

To get the high energy asymptotic behavior of Eq.(\ref{counting})
with $n=8$, the $\pi \Delta\Delta$ vertex formfactor should be taken in the form
\begin{equation}
   F_{\pi \Delta\Delta}(t) = \left(\frac{\Lambda_{\pi \Delta\Delta}^2-m_\Delta^2}{\Lambda_{\pi \Delta\Delta}^2-t}\right)^3~.  \label{F_piDD}
\end{equation}
The value of the cutoff $\Lambda_{\pi \Delta\Delta}$ is quite uncertain. However, we expect that it should not
strongly deviate from $\Lambda_{\pi N\Delta}$ in the hard regime $-t \sim -u \sim s/2, s \to +\infty$.
This is supported by the result of the previous subsection, that the cutoffs $\Lambda_{\pi N\Delta}$
and $\Lambda_{\pi NN}$ are also quite similar. Thus, to reduce the number of free parameters,
we set $\Lambda_{\pi \Delta\Delta}=\Lambda_{\pi N\Delta}$.

\begin{figure}[ht]
\includegraphics[scale = 0.6]{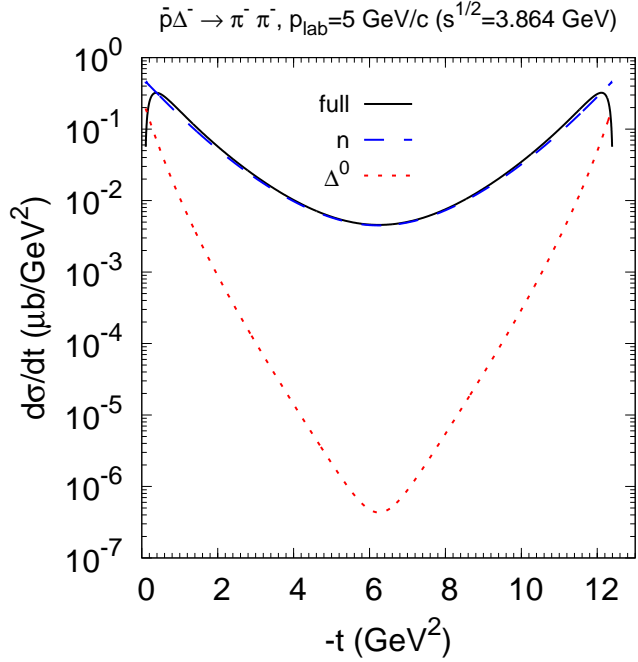}
\caption{\label{fig:pbarD-2pi-pi-_5gev} Differential cross section of the $\bar p \Delta^- \to \pi^- \pi^-$ process at $p_{\rm lab}=5$ GeV/c.
  Solid line: full calculation. Dashed and dotted lines display the separate contributions of neutron and $\Delta^0$ exchange, respectively.}
\end{figure}
Fig.~\ref{fig:pbarD-2pi-pi-_5gev} shows the $t$ dependence of the $\bar p \Delta^- \to \pi^- \pi^-$ differential cross section at
5 GeV/c beam momentum. (The cross section is symmetric with respect to replacement $t \leftrightarrow u$.)
We see that $\Delta^0$ exchange is important at small $-t$, but becomes almost negligible at $\Theta_{c.m.}=90\degree$
(i.e. at $-t=(s-m_N^2-m_\Delta^2)/2-m_\pi^2$).
\begin{figure}[ht]
\includegraphics[scale = 0.6]{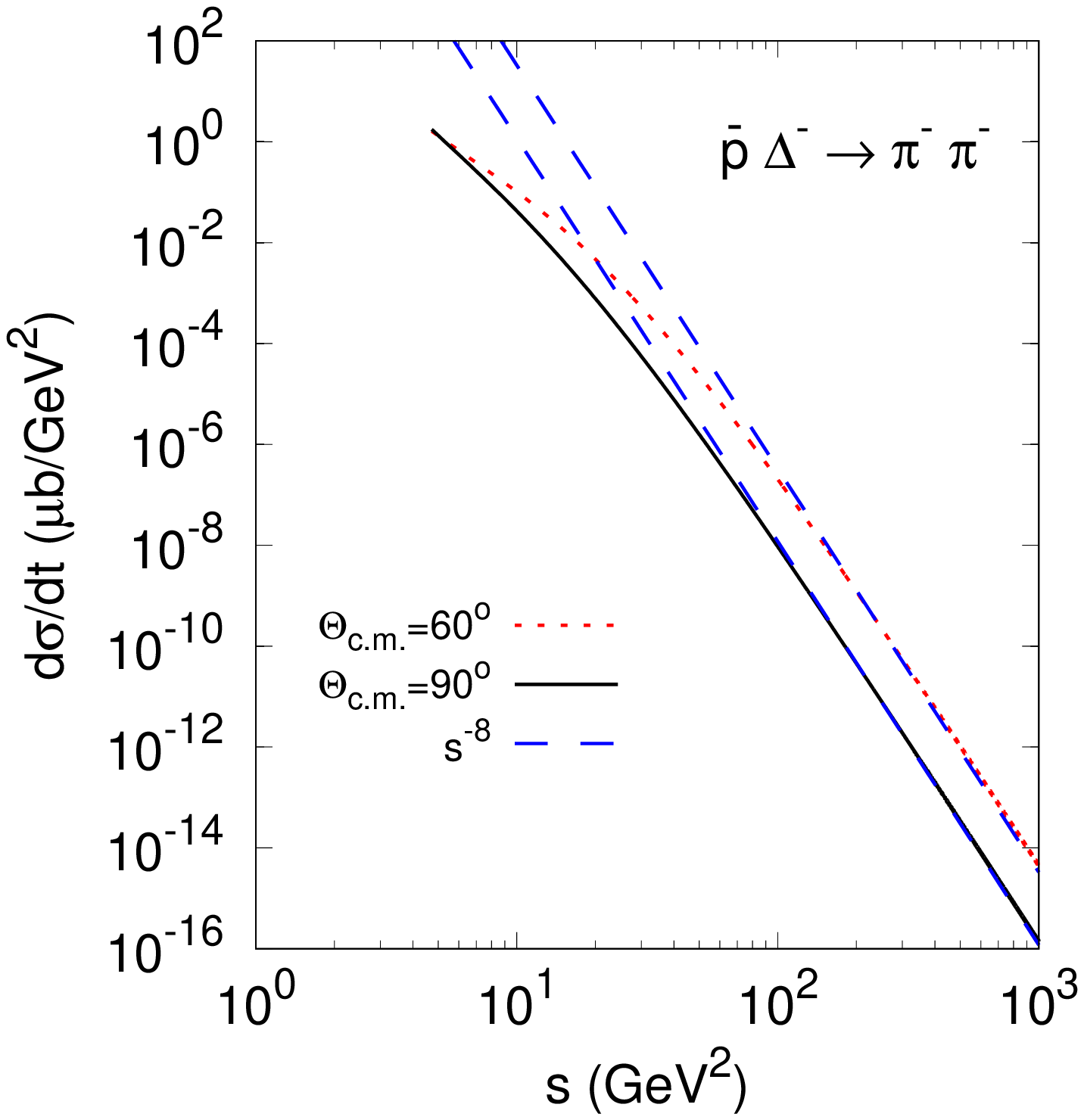}
\caption{\label{fig:pbarD-2pi-pi-_90deg} Differential cross section of
  the $\bar p \Delta^- \to \pi^- \pi^-$ process at $\Theta_{c.m.}=90\degree$
  (solid line) and $60\degree$ (dotted line). The large-$s$ asymptotic behavior $\propto s^{-8}$
  is shown for both cases by the dashed lines.}
\end{figure}
Fig.~\ref{fig:pbarD-2pi-pi-_90deg} displays the $s$ dependence of $d\sigma/dt$ at $\Theta_{c.m.}=90\degree$ and $60\degree$
for the $\bar p \Delta^- \to \pi^- \pi^-$ process. At large $s$, the quark counting rule is exactly respected.

\subsection{$\pi N \to \pi^\prime \Delta$ charge exchange}

\label{piN2piD}

\begin{figure}
\includegraphics[scale = 0.6]{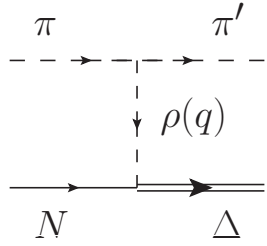}
\caption{\label{fig:piCEX} Feynman graph of the inelastic pion CEX amplitude $\pi N \to \pi^\prime \Delta$
  due to $t$-channel $\rho$-meson exchange.}
\end{figure}

The amplitude of Fig.~\ref{fig:piCEX} has been evaluated with the $\rho\pi\pi$ interaction Lagrangian \cite{Serot:1984ey}
\begin{equation}
    {\cal L}_{\rho\pi\pi}=g_{\rho\pi\pi} [\partial^\mu \bvec{\pi} \times  \bvec{\pi}] \cdot  \bvec{\rho}_\mu~,   \label{L_rhopipi} 
\end{equation}
where $g^2_{\rho\pi\pi}/4\pi=2.88$ (cf. ref. \cite{VandeWiele:2010kz}) such that the  $\rho$ decay width
\begin{equation}
  \Gamma_{\rho \to \pi\pi} = \frac{g^2_{\rho\pi\pi}}{4\pi} \frac{m_\rho}{12} \left(1-\frac{4m_\pi^2}{m_\rho^2}\right)^{3/2} \label{Gamma_rho2pipi}
\end{equation}
is equal to the phenomenological value $0.150$ GeV at the pole mass $m_\rho=0.771$ GeV. 
The $\rho N \Delta$ interaction Lagrangian has been taken in the form with derivative coupling \cite{Shyam:2003cn,Matsuyama:2006rp}:
\begin{equation}
    {\cal L}_{\rho N \Delta} = \frac{i f_{\rho N \Delta}}{m_{\rho}} 
     (\bar\psi^\mu \gamma^\nu \gamma^5 \bvec{T} \psi - \bar \psi \gamma^\nu \gamma^5  \bvec{T}^\dag  \psi^\mu)
     (\partial_\nu \bvec{\rho}_\mu - \partial_\mu \bvec{\rho}_\nu)~.    \label{L_DeltaNrho}
\end{equation}
We will use the value of the coupling constant $f_{\rho N \Delta}=14.0$ which is about two times larger than
in refs. \cite{Shyam:2003cn,Matsuyama:2006rp} but agrees with estimations in ref. \cite{Oset:1981ih}.

The invariant amplitude corresponding to Fig.~\ref{fig:piCEX} can be expressed as
\begin{eqnarray}
  M_{\pi N} &=& -i \frac{g_{\rho\pi\pi} f_{\rho N \Delta}}{m_\rho} \frac{{\cal I}_{\pi N}}{t-m_\rho^2+i\epsilon}
  (k+k^\prime)^\mu \left(-g_{\mu\nu}+\frac{q_\mu q_\nu}{m_\rho^2}\right)      \nonumber \\
  && \times [-\bar u^\nu(p_\Delta,\lambda_\Delta) \slashed{q} \gamma^5 u(p,\lambda)
             +\bar u^\alpha(p_\Delta,\lambda_\Delta) q_\alpha \gamma^\nu \gamma^5 u(p,\lambda)]~,    \label{calM_piN}
\end{eqnarray}
where $k$ and $k^\prime$ are the four-momenta of the incoming and outgoing pion, respectively, and $t=q^2$.
The Rarita-Schwinger vector-spinors of the $\Delta$-resonance are normalized as
$\bar u^\mu(p_\Delta,\lambda_\Delta) u_\mu(p_\Delta,\lambda_\Delta)=-2m_\Delta$.
The isospin factor is
\begin{equation}
    {\cal I}_{\pi N} = \bvec{T}_{\tau_\Delta \tau} \cdot [\bvec{t}^{(l)} \times \bvec{t}^{(l^\prime) *}]~,      \label{I_piN}
\end{equation}
where $\tau=\pm1/2$ and $\tau_\Delta=\pm1/2,\pm3/2$ are the isospin projections of nucleon and $\Delta$, respectively,
and $l,l^\prime=0,\pm1$ are the isospin projections of the incoming and outgoing pion, respectively.
For the relevant channel $\pi^0 p \to \pi^- \Delta^{++}$ (and also for the
channel $\pi^+ p \to \pi^0 \Delta^{++}$) we obtain ${\cal I}_{\pi N}=i$. 

Small $-t$ scattering at high energies is well described within Regge theory, which approximates the exchange of a set
of particles with the same internal quantum numbers (such as $B, I, S$ etc.) by the exchange of a
Regge trajectory \cite{Collins,Guidal:1997hy}. In particular, the $\rho$ meson trajectory includes the
$a_2(1320)$, $\rho_3(1690)$, and $a_4(2040)$ states. The reggeization of the amplitude, Eq.(\ref{calM_piN}),
is reached by replacing 
\begin{equation}
      \frac{1}{t-m_\rho^2+i\epsilon}  \to  
      {\cal P}^\rho_{\rm Regge} = \left(\frac{s}{s_0}\right)^{\alpha_\rho(t)-1} 
      \frac{\pi \alpha_\rho^\prime}{\sin(\pi\alpha_\rho(t))}
      \frac{-1+\exp(-i\pi\alpha_\rho(t))}{2} \frac{1}{\Gamma(\alpha_\rho(t))}~,           \label{rho_reggeization}
\end{equation}
where $s_0=1$ GeV$^2$,~$\alpha_\rho(t)=\alpha_{\rho 0} + \alpha_\rho^\prime t$ with an intercept $\alpha_{\rho 0}$
and a slope $\alpha_\rho^\prime$ determined from the data on exclusive reactions assuming linearity of the $\rho$ meson
trajectory and imposing the condition that $\alpha_\rho(m_\rho^2)=1$.

\begin{figure}
\begin{center}
\includegraphics[scale = 0.5]{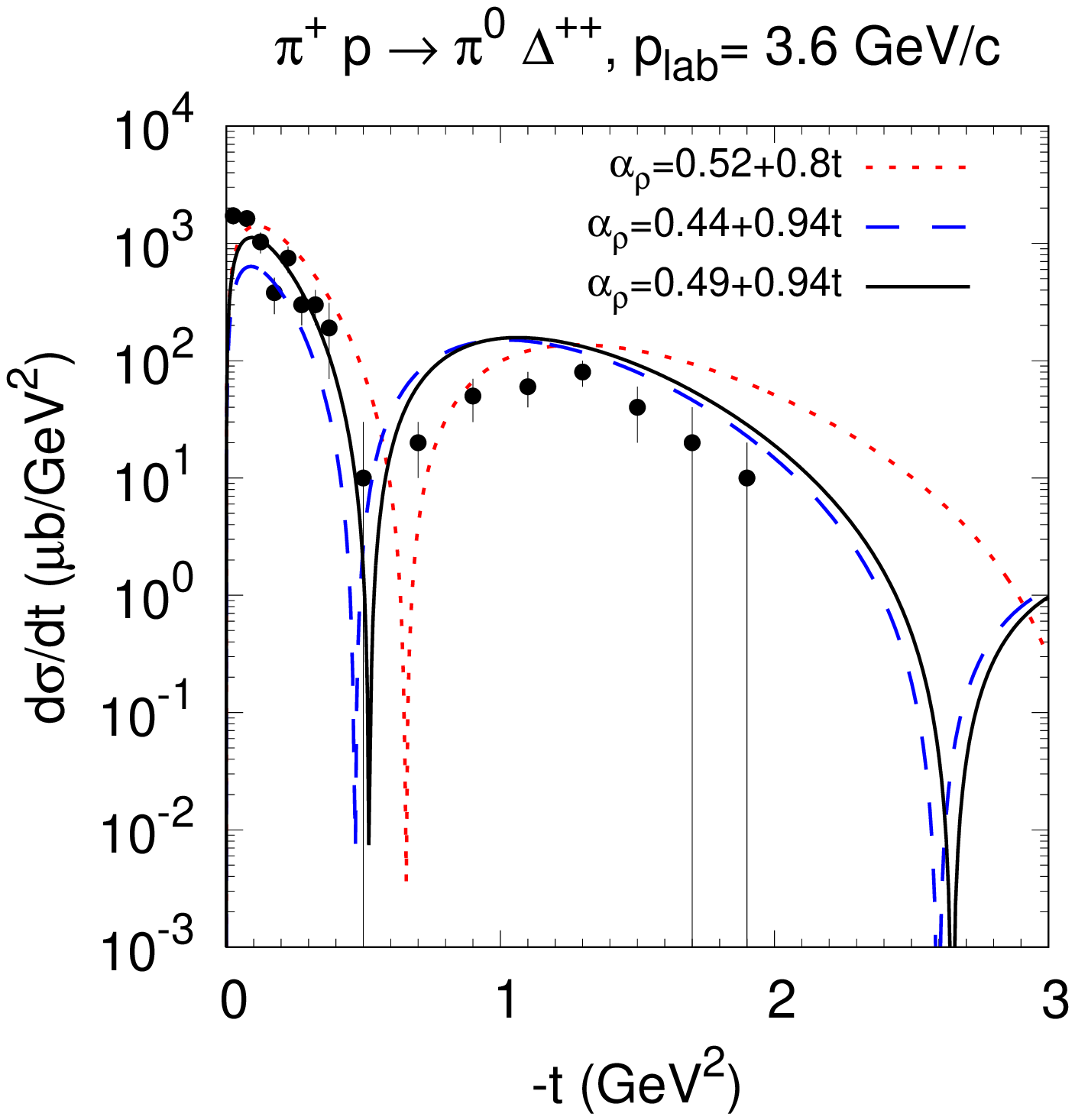}
\includegraphics[scale = 0.5]{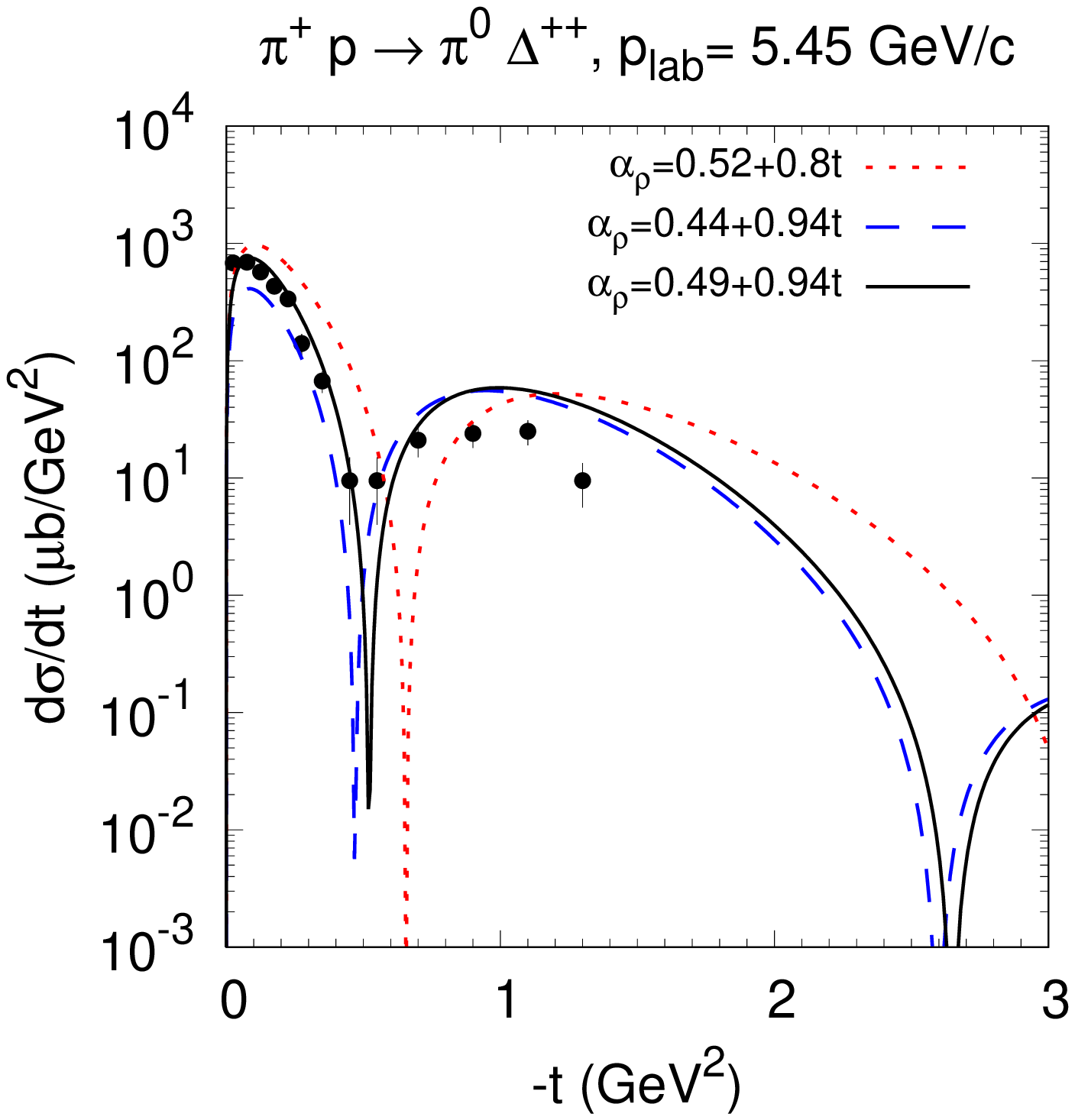}
\includegraphics[scale = 0.5]{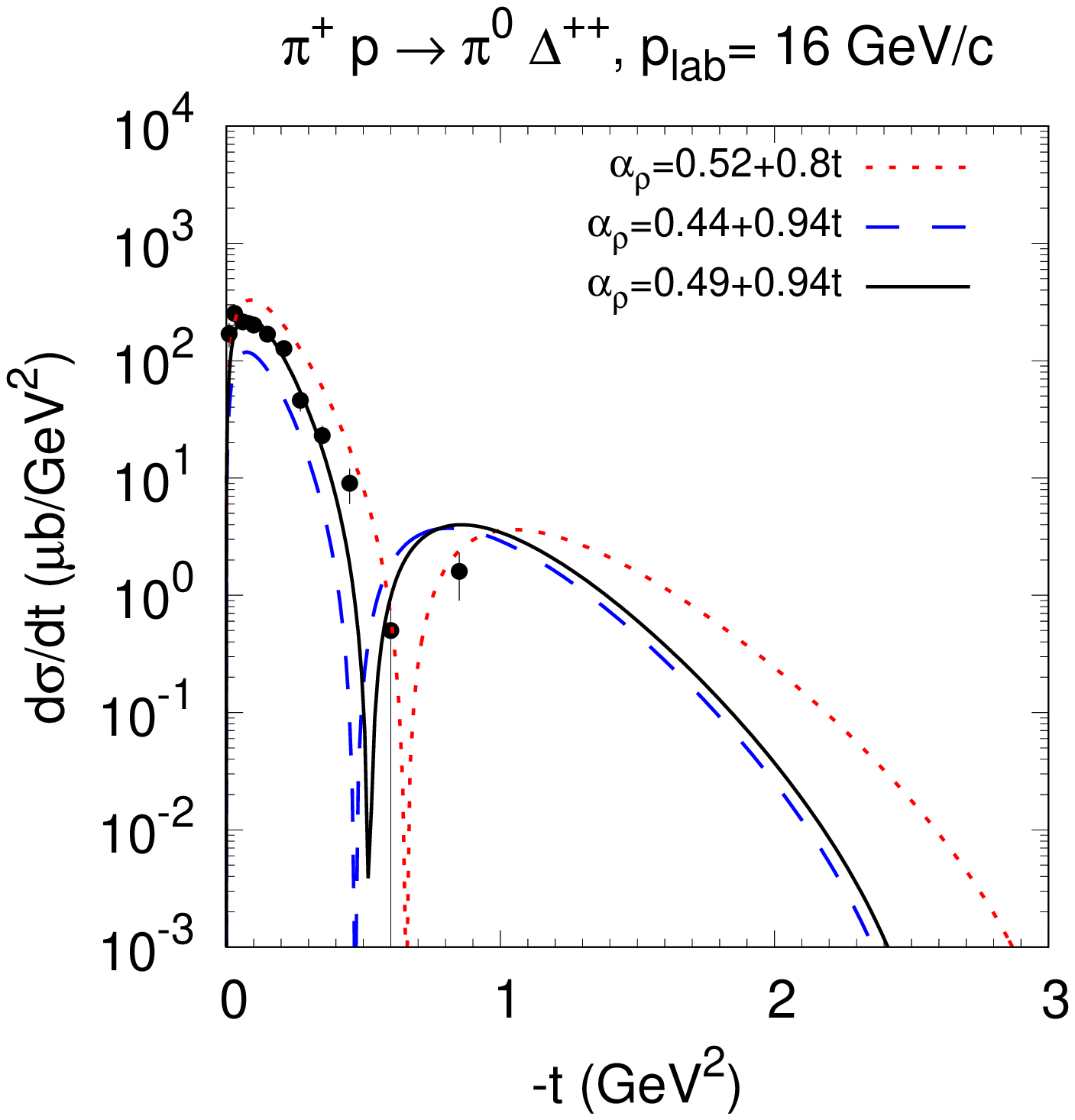}
\end{center}
\caption{\label{fig:piN2piD_CEX} Differential cross section of the $\pi^+ p \to \pi^0 \Delta^{++}$ CEX process.
Different curves represent calculations with different parameters of the $\rho$ meson Regge trajectory as indicated.
Experimental data at 3.6 GeV/c, 5.45 GeV/c, and 16 GeV/c are from ref. \cite{Macnaughton:1977cy}, \cite{Bloodworth:1974ji},
and \cite{Honecker:1977me}, respectively.}
\end{figure}

We have calculated the differential cross section 
\begin{equation}
   \frac{d\sigma}{dt} = \frac{\overline{|M_{\pi N}|^2}}{64 \pi (p_{\rm lab}m_N)^2}~,   \label{dsigma_CEX/dt}
\end{equation}
of the $\pi^+ p \to \pi^0 \Delta^{++}$ process with different parameters of
the $\rho$ Regge trajectory.
As shown in Fig.~\ref{fig:piN2piD_CEX},  the intercept $\alpha_{\rho 0}=0.49$
and slope $\alpha_\rho^\prime=0.94$ GeV$^{-2}$ produce a quite
reasonable description of available experimental data at small $-t$.
Thus, these parameters are used in the calculations of the CEX background
(sec. \ref{CEX}).

\subsection{$\bar N N \to \pi \pi \pi$}

\label{NbarN23pi}

\begin{figure}
\begin{center}
   \includegraphics[scale = 0.60]{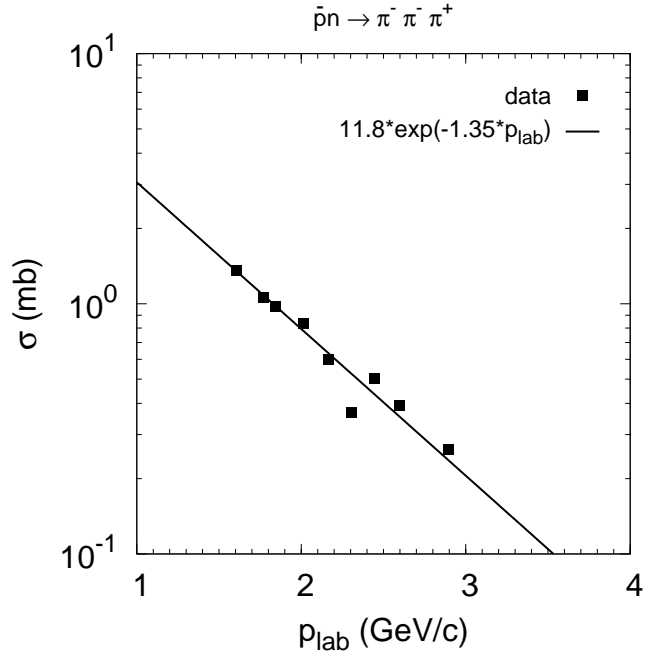}
\end{center}
\caption{\label{fig:sigpbarn23pi} Fit to the experimental data of ref. \cite{Eastman:1973va} on
  the $\bar p n \to \pi^- \pi^- \pi^+$ cross section by the exponential function given in Eq.(\ref{Fit}).}
\end{figure}

For the three-pion annihilation amplitude we assume an $s$-dependent invariant matrix element extracted from the fit
of the $\bar p n \to \pi^- \pi^- \pi^+$ cross section, see Fig.~\ref{fig:sigpbarn23pi}, by the function
\begin{equation}
  \sigma_{\bar p n \to \pi^- \pi^- \pi^+} = 11.8\exp(-1.35\,p_{\rm lab})~,  \label{Fit}
\end{equation}
where $p_{lab}$ is in GeV/c and the cross section in mb.
The invariant matrix element can be estimated as
\begin{equation}
  M_{\bar p n}(s) = \left(\frac{4I_{\bar p n} \sigma_{\bar p n \to \pi^- \pi^- \pi^+}}{(2\pi)^4 \Phi_3(s)}\right)^{1/2}~,   \label{M_pbarn23pi}
\end{equation}
where $I_{\bar p n}=\sqrt{(s/4-m_N^2)s}$ is the M\"oller flux factor and $\Phi_3(s)$ is the 3-body phase space integral
(cf. PDG review \cite{Patrignani:2016xqp}).

\subsection{$\pi N \to \Delta$}

\label{piN2D}

The $\pi N \Delta$ interaction is described by the standard $P$-wave coupling Lagrangian of Eq.(\ref{L_piND}).
The invariant matrix element of the $\pi^+ p \to \Delta^{++}$ transition (see Fig.~\ref{fig:bgr_3pi}) is
\begin{equation}
  i M_{\pi p}(p_\Delta;p_3,p_p) = \frac{f_{\pi N \Delta} F_{\pi N \Delta}(p_3)}{m_\pi} p_{3\mu} \bar u^\mu(p_\Delta,\lambda_\Delta) u(p_p,\lambda_p)~.
                                \label{M_pip}
\end{equation}
The formfactor is chosen in the monopole form
\begin{equation}
     F_{\pi N \Delta}(q)= \frac{\Lambda^2-m_\pi^2}{\Lambda^2+\bvec{q}^2}~,    \label{FF_piND}
\end{equation}
with $\Lambda=1.2$ GeV \cite{Dymarz:1990ac}. Note that the formfactors of Eq.(\ref{F_piND}) and Eq.(\ref{FF_piND})
differ since they are applied in different regimes: the former is valid in the hard while the latter
is valid in the soft regime.

\end{subappendices}

\end{appendices}

\end{document}